\documentclass[3p]{elsarticle}
\usepackage{amssymb,amsmath,graphicx,epsfig,psfrag}
\usepackage{slashed}
\journal{jcp}

\def\Web{\mbox{\text{We}}} 
\def\Fro{\mbox{\text{Fr}}}  
\def\Rey{\mbox{ \text{Re}}}  
\def\Pe{\mbox{\text{Pe}}}
\def\Bi{\mbox{\text{Bi}}}
\def\Da{\mbox{\text{Da}}}

\newcommand{\mbP}{{\mathbb{P}}} 
\newcommand{\be}{{\bf e}}
\newcommand{\mbD}{{\mathbb{D}}}
\newcommand{\mbI}{{\mathbb{I}}} 
\newcommand{\mbS}{{\mathbb{S}}}
\newcommand{\bnu}{\boldsymbol\nu}
\newcommand{\bv}{{\bf v}}
\newcommand{\bw}{{\bf w}}
\newcommand{\bu}{{\bf u}}

\newcommand{\btau}{\boldsymbol\tau}

\newcommand{\bX}{{\bf X}}
\newcommand{\bY}{{\bf Y}}
\newcommand{\dive}{\nabla\cdot}

\usepackage{xcolor,cancel,ifpdf}
\usepackage[normalem]{ulem}

\begin{document}
\begin{frontmatter}
\title{Computations of an impinging droplet with soluble surfactants and dynamic contact angle}

\author{Sashikumaar Ganesan}
\ead{sashi@serc.iisc.in}
\ead[url]{www.serc.iisc.ernet.in/$\sim$sashi/}


\address{Numerical Mathematics and Scientific Computing,  Supercomputer Education and Research Centre, \\Indian Institute of Science, Bangalore  560012, India.}

\begin{abstract}
 An arbitrary Lagrangian--Eulerian (ALE) finite element scheme for computations of soluble surfactant droplet impingement on a horizontal surface is presented. The numerical scheme solves the time-dependent Navier--Stokes equations for the fluid flow,  scalar convection-diffusion equation for the surfactant transport in the bulk phase, and simultaneously,  surface evolution equations for the surfactants on the free surface and on the liquid-solid interface. The effects of surfactants on the flow dynamics are included into the model through the surfactant-dependent surface tension and dynamic contact angle.  In particular, the dynamic contact angle  $(\theta_d)$ of the droplet  is defined in terms of the surfactant concentration at the contact line and the equilibrium contact angle $(\theta_e^0)$  of the clean surface using the nonlinear equation of state for surface tension.  Further, the surface forces are included in the model as the surface divergence of the surface stress tensor that allows to incorporate the Marangoni effects without calculating the surface gradient of the surfactant concentration on the free surface. In addition to a mesh convergence study and   validation of the numerical results with experiments, the effects of adsorption and desorption surfactant coefficients on the flow dynamics in wetting, partially wetting and non-wetting droplets are studied in detail. It is observed that the effect of surfactants are more in wetting droplets than in the non-wetting droplets. Further, the presence of surfactants at the contact line reduces the equilibrium contact angle further when $ \theta_e^0 $  is less than $90^\circ$, and increases it further when $ \theta_e^0 $  is greater than $90^\circ$. The presence of  surfactants has no effect on the contact angle, when $\theta_e^0 =90^\circ$. The numerical study clearly demonstrates that  the  surfactant-dependent contact angle has to be considered,  in addition to the Marangoni effect, in order to study the flow dynamics and the equilibrium states of surfactant droplet impingement accurately. The proposed numerical scheme guarantees the conservation of fluid mass and of the surfactant mass well.
\end{abstract}

\end{frontmatter}

\section{Introduction}
Liquid droplets impinging on a solid substrate is encountered in many applications such as spray cooling,
spray forming, spray coating, ink-jet printing, fuel injecting, etc. Apart form these applications, computations of the impinging droplets are also of a scientific interest for many researches due to the challenges associated with it. Main challenges associated in computations of impinging droplets are to prescribe the boundary condition on the liquid-solid interface, especially at the moving contact line,  and to incorporate the wetting effects, in particular, the inclusion of the contact angle into the model equations. In addition to these challenges, the presence of soluble surfactants in the droplet makes the model more complicate.  

Numerous studies on the choice of the boundary condition on the liquid-solid interface, especially in the vicinity of moving contact line, have been reported in the literature~\cite{DUS76,DUS91,EGG04,HOC77,LAU05,QIA06,REN07,SUI14,THO97,MAR12}.
Using the usual no-slip boundary condition on the liquid-solid interface could induce an unbounded stress singularity at the moving contact line. This singularity is also called kinematic paradox in the literature. Different types of slip boundary conditions have been proposed in the literature~\cite{EGG04,LAU05,MAR12} to alleviate  this singularity. Among all, the Navier-slip boundary condition is widely accepted, but it introduces the so-called slip coefficient. This unknown slip coefficient is also called a momentum transfer coefficient~\cite{HUH71}. Even though a number  of expressions have been proposed for the slip coefficient, it is often determined by comparing the computationally obtained wetting diameter with their corresponding experimental results~\cite{GAN13M}. Based on this approach, an expression for the slip coefficient has recently been proposed in~\cite{SGJV15} for computations of impinging droplets.

Another challenge associated with the computations of the moving contact line flows is the inclusion of the contact angle. This subject has also been studied by several researchers to a great extent, see for example,~\cite{COX86,DEG85,HAL91,HOC76,HOC02,HUH71}. In the  lubrication theory approximations,   the contact angle has been imposed as a boundary condition at the moving contact line, see for example~\cite{HAL91,HOC92}. However, the inclusion of the contact angle is not straight forward in discretization based numerical schemes for computations of moving contact line flows~\cite{FUK95,SG06,GT07,REN01,SPE05,SIK05}. Moreover, the correct choice of the contact angle value in discretization based numerical schemes has also been a topic of research, see for example~\cite{GAN13M}.

One of the main components in a free surface flow solver is the interface capturing/tracking methods, and these methods can be classified into Eulerian and Lagrangian methods. In the Eulerian methods such as Volume-of-Fluid
\cite{HIR81,PIL04,REN01,REN02,RID98,ANN05}, Level set~\cite{GRO06,OS88,JAS96,SUS00,SU94,XU06}, Front  Tracking \cite{TRY01,UNV92}, etc,  the Navier--Stokes equations are solved in a fixed domain with variable material coefficients.  Contrarily, the Navier--Stokes equations are solved in each phase simultaneously with a deforming domain in the Lagrangian methods such as arbitrary Lagrangian--Eulerian \cite{DON83,GAN06E,GAN07,HIR74,NIT05,NOB01} and
pure Lagrangian \cite{FUK95,FUK93}  approaches. Although, a number of numerical studies have been  reported in the literature for free surface and two-phase flows with insoluble surfactants~\cite{BAZ06,EGG01,GAN09JCP,JAM04,KRU04,LAI08,LEE06,LI97,POZ04,REN02,XU06,XYL12}, the effects of soluble surfactants have been considered only in a few recent studies~\cite{ADA10,BOO10,DIE15,GAN12,MUR08,MUR14,TEI11,ZHA06}.
In all these studies,   numerical schemes have been developed for flows with surfactants in closed boundaries, that is, for flows without moving contact lines.

Due to the challenges in handling the moving contact line, only a few works have been reported in the literature for flows with surfactants and moving contact line. The effect of insoluble surfactants on a droplet attached to a plane wall subjected to an over passing Stokes flow has been studied in~\cite{YON99} using a boundary integral method.  The authors  used marker points to track the interface, and assumed that the contact line remains circular and the interfaces having the shape of
sections of a sphere. An immersed boundary method using the Marker-and-Cell method has been proposed in~\cite{MCY010} for computations of two-dimensional (2D) semicircular droplet deformation with insoluble surfactants on a horizontal surface. The authors assumed that the initial velocity in the droplet is zero, and incorporated the surfactant effects on the contact angle as an unbalanced  Young force \cite{DEG85}. Numerical studies for different equilibrium contact angles and surfactant concentrations have been performed in~\cite{MCY010}. Recently, a finite difference scheme using the level-set for computations of 2D semicircular droplet deformation  with insoluble surfactants  on a horizontal surface and for the detachment of a pendant droplet from a wall under gravity has been proposed in~\cite{XU014}. A contact angle condition, which relates the unbalanced Young force and the slip velocity at the contact line, has been used to include the effects of surfactants on wetting. One of the main challenges in the applications of level-set method is the conservation of mass, and an additional mass correction step is needed inorder to conserve the mass~\cite{XU014}. For an overview of moving contact line flows with insoluble surfactants, we refer to~\cite{ZHA14}.

In all the previous studies, a hemispherical/semicircular  droplet subjected to an external or non-equilibrium forces with insoluble surfactants has been considered. To the best of the authors knowledge, numerical studies of impinging droplets with soluble surfactant have not been reported in the literature so far. 
In particular,  a sharp interface model, which is known for conserving the mass without additional correction and suppressing spurious velocities when appropriate  solution spaces are used~\cite{GMT}, has not been reported for computations of moving contact lines with soluble surfactants.

 In this paper, we present a finite element scheme using the arbitrary Lagrangian--Eulerian approach for computations of impinging droplets with soluble surfactants. Since the free surface is resolved by the moving mesh in the ALE approach, the surface forces including Marangoni effects can accurately be
incorporated into the numerical scheme. Moreover,  the surface evolution-equation is approximated on the discrete representation of the free surface directly. In addition, the inclusion of the dynamic contact angle and the  adsorption/desorption balance condition for the surfactant mass transfer are straightforward in the considered sharp interface model. More importantly,  an additional correction is not needed in order to conserve the mass of the fluid   and  of the total surfactants.

The paper is organized as follows. In Section~2, the governing equations of the impinging droplet with soluble surfactants are presented. The dynamic contact angle that depends on surfactants is described in
Section~3. The dimensionless form of the model equations, ALE approach, finite element formulations and mesh handling techniques are presented in Section~4. The mesh convergence of the proposed numerical scheme and the
numerical  results for an impinging droplet with soluble surfactants are given in Section~5 and 6. Finally, in Section~7
we summarize the key observations of this study.

\section{Governing equations}
\begin{figure} 
\begin{center}
\unitlength1cm
\begin{picture}(15, 5.5  )
\put(0,-0.5){{\includegraphics[width=15cm]{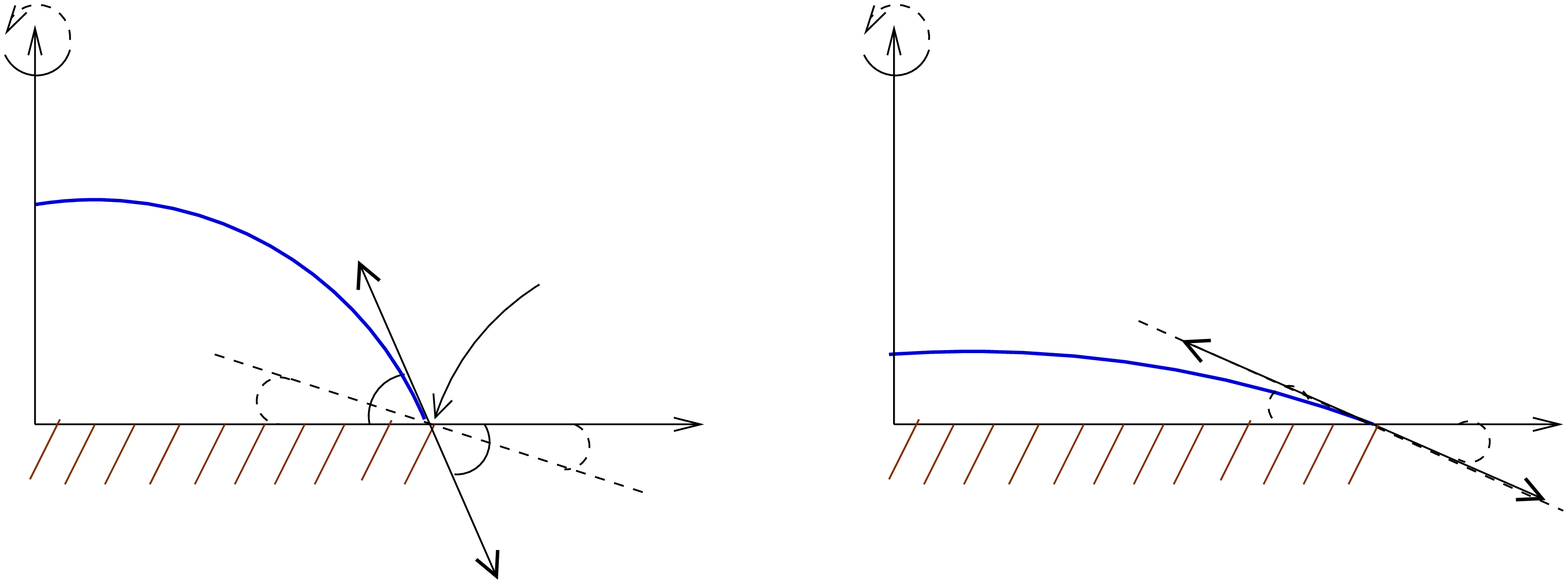}}}
\put(2.25,4.5){$(a)$}
\put(10.75,4.5){$(b)$}
\put(2,1.25 ){$\theta_e^0$}
\put(3.3,1.45){$\theta_d$}
\put(4.65,0.35 ){$\theta_d$}
\put(5.75,0.65){$\theta_e^0$}
\put(10.65,1.12){$\theta_d = \theta_e^0$}
\put(1.25,2){$\Omega(t)$}
\put(1. ,1.15){$\Gamma_2(t)$}
\put(2,3){$\Gamma_1(t)$}
\put(5.25,2.25){${\zeta(t)}$}
\put(4,-.25){$\bnu_{\zeta}$}
\put(5.5,1.2){$\tau_{1,2}$}
\put(6.5,0.75){$r$}
\put(-0.25 ,4.65){$z$}
\end{picture}
\end{center}
\caption{Schematic view of a droplet: $(a)$ during the deformation, and $(b)$ the equilibrium shape of the droplet after impingement.
\label{model}}
\end{figure}

We consider a surfactant liquid droplet impingement on a horizontal solid substrate. 
The computational domain of the droplet is denoted by $\Omega(t)$, whereas $\zeta(t)$ denotes the moving contact line,  $\Gamma_1$ and $\Gamma_2$ denote the free surface and the liquid-solid interface, respectively. Moreover, $\Gamma(t):=\Gamma_1(t)\cup\Gamma_2(t)\cup\zeta(t)$ is the boundary of $\Omega(t)$, and  $\theta_d$ is the  dynamic contact angle. The schematic view of the considered model is shown in Figure \ref{model}. We assume that the liquid is incompressible and the effects  of the surrounding gas on the flow dynamics of the droplet are negligible. The computation starts immediately after the droplet impinges on the solid surface, and it ends at a specified final time, $\text{I}$.

\subsection{Navier--Stokes equations} 
The fluid flow in the droplet  is described by the  time-dependent incompressible Navier--Stokes equations
\begin{align}\label{tnse}
 \nabla \cdot \mathbf u=0, \qquad \frac{\partial \mathbf u}{\partial t}+(\mathbf u\cdot\nabla) \mathbf u-\frac{1}{\rho}\nabla\cdot(\mbS(\mathbf u,p))= g\be &&\mbox{in }\Omega(t)\times(0,\text{I}),
\end{align}
where  the set $\Omega(t)\times (0,I)$ has to be understood as $\{(x,t)\in \mathbb{R}^4\,:\,x\in\Omega(t),\,t\in (0,I)\}$.
Here, $\mathbf u$ is the velocity, $p$ is the pressure, $\rho$ is the density of the fluid, $g$ is the gravitational constant, $t$ is the time, $\be$ is an unit vector in the opposite direction of the gravitational force.  
We assume that the droplet impinges on the solid surface  perpendicularly  with the impact velocity   
\begin{align}
\mathbf u(\cdot, 0) &=  (0,0,-u_{imp}) \quad \mbox{in }\Omega(0),
\end{align}
where $u_{imp}$ is the impact speed of the droplet.
Furthermore,  $\mbS(\mathbf u,p)$ is the   stress tensor,  and for the considered Newtonian incompressible fluid, it is defined as
\begin{align*}
 \mbS(\mathbf u,p):=2\mu \mbD(\mathbf u)-p \mbI ,\quad \mbD(\mathbf u)_{i,j}=\frac{1}{2}\left(\frac {\partial u_i}{\partial x_j}+\frac  {\partial u_j}{\partial x_i}\right),  \quad i,j=1,2,3, 
\end{align*}
where $\mu$ is  the dynamic viscosity, $\mbD(u)$ is the velocity deformation tensor and $\mbI$ is the identity tensor. The Navier-Stokes equations~\eqref{tnse} are closed with the kinematic and force balancing conditions  
\begin{align} \label{bc}
\mathbf u\cdot\bnu_1=\bw\cdot\bnu_1, \qquad
\mbS(\mathbf u,p)\cdot\bnu_1=   \nabla_{\Gamma_1}\cdot\mbS_{\Gamma_1}   &&\mbox{on }\Gamma_1(t),                                                                                                                                   
\end{align}
and the Navier-slip boundary condition
 \begin{align*}
\mathbf u\cdot\bnu_2&=0, \quad
\beta_{\epsilon}(\btau_{i,2}\cdot\mbS(\mathbf u,p)\cdot\bnu_2)=-\mathbf u\cdot\btau_{i,2}, \quad i=1,2,   &&\mbox{on }\Gamma_2(t).
\end{align*}
Here, $\bw$ is the velocity of the computational domain $\Omega(t)$ and $\beta_{\epsilon}$ is a given slip coefficient. Further, $\bnu_1(n_1,n_2,n_3)$ and $\bnu_2$ denote the unit outer normal vector on  $\Gamma_1(t)$ and $\Gamma_2(t)$, respectively.
Moreover, $\btau_{1,2}$ is the scaled projection of 
$\bnu_1$ onto the plane $\Gamma_2(t)$, and $\btau_{2,2}$ is perpendicular to $\btau_{1,2}$ and $\bnu_2$ defined by
\begin{align}\label{tangdef}
\btau_{1,2}:=\frac{\bnu_1-(\bnu_1\cdot\bnu_2)\bnu_2}{\|\bnu_1-(\bnu_1\cdot\bnu_2)\bnu_2\|}, \qquad \btau_{2,2}=\frac{\btau_{1,2}\times \bnu_2}{ \|\btau_{1,2}\times \bnu_2    \|}.
\end{align}
The surface gradient, ${\nabla}_{\Gamma_1}(\cdot)$, of a scalar function $\Psi$, and the surface divergence, $\nabla_{\Gamma_1}\cdot (\cdot)$, of a vector $\bu$ on the surface $\Gamma_1(t)$ are defined by
\[
{\nabla}_{\Gamma_1}  \Psi = \mbP_{\bnu_1}\nabla \Psi, \qquad \nabla_{\Gamma_1}\cdot \mathbf u  = \text{tr }\left( \mbP_{\bnu_1}\nabla \mathbf u\right),
\]
where $\mbP_{\bnu_1} = \mbI -\bnu_1\otimes\bnu_1 $ is the projection onto the tangential plane. The surface stress tensor, $\mbS_{\Gamma_1}$, in the force balance condition~\eqref{bc} is  
modeled by
 
 \[
 \mbS_{\Gamma_1}=\sigma (C_{\Gamma_1})  \mbP_{\bnu_1},
\]
 which is a special case of the Boussinesq-Scriven model 
\[
  \mbS_{\Gamma_1}= \left[\sigma   +(\lambda_{\Gamma_1} - \mu_{\Gamma_1})\nabla_{\Gamma_1}\cdot \mathbf u \right] \mbP_{\bnu_1} + \mu_{\Gamma_1} \mbD_{\Gamma_1}(\mathbf u)
\]
with $\lambda_{\Gamma_1}=0$  and  $\mu_{\Gamma_1} =0$. Moreover,   
\[
  \nabla_{\Gamma_1}\cdot\mbS_{\Gamma_1}  = \text{tr}\, \left( \nabla_{\Gamma_1}  \left( \sigma    \mbP_{\bnu_1}  \right) \right) = \mbP_{\bnu_1} \nabla_{\Gamma_1}  \sigma  + \sigma\, \text{tr}\,\left( \nabla_{\Gamma_1}     \mbP_{\bnu_1}   \right).
\]
Since the surface gradient is in the tangential plane, we have $ \mbP_{\bnu_1} \nabla_{\Gamma_1}  \sigma  =  \nabla_{\Gamma_1}  \sigma$. Further, for $1\le n_j\le 3$,
 \begin{align*}
  \text{tr}\,\left( \nabla_{\Gamma_1}     \mbP_{\bnu_1}   \right) & = \sum_{i=1}^3 \frac{\partial}{\partial x_i} \left( \delta_{i,j} - n_in_j  \right) - \sum_{i,k=1}^3 \frac{\partial}{\partial x_k}\left( \delta_{i,j} - n_in_j  \right) n_kn_i \\
  &= - \sum_{i=1}^3 n_j\frac{\partial n_i}{\partial x_i}  - \sum_{i=1}^3 n_i\frac{\partial n_j}{\partial x_i}
     + \sum_{k=1}^3 n_k\frac{\partial n_j}{\partial x_k}\underbrace{\sum_{i=1}^3n_i^2}_{=1} + \frac{1}{2}\sum_{k=1}^3 n_j n_k \frac{\partial  }{\partial x_k} \underbrace{\left(  {\sum_{i=1}^3n_i^2}  \right)}_{=1} = -\mathcal{K}\bnu_1,
 \end{align*}
 where $\mathcal{K}$ is the sum of the principle curvatures.
Hence, we have
\[
  \nabla_{\Gamma_1}\cdot\mbS_{\Gamma_1}  = \nabla_{\Gamma_1}\cdot\left(\sigma (C_{\Gamma_1})  \mbP_{\bnu_1} \right)= \nabla_{\Gamma_1}  \sigma(C_{\Gamma_1})  - \sigma(C_{\Gamma_1})\mathcal{K}\bnu_1, 
\]
which is the standard form used in the literature to include the Marangoni effects. However, we prefer the surface divergence form, and the advantage is that  it avoids the calculation of $\nabla_{\Gamma_1}  \sigma(C_{\Gamma_1}) $  and the handling of  $\mathcal{K}$  in the variational form, see~\eqref{varsurf}. Next, the surfactant-dependent surface tension, $\sigma(C_{\Gamma_1})>0$, can be defined using the  Henry linear equation of
state
\begin{align} \label{LEOS}
\sigma(C_{\Gamma_1}) = \sigma_{ref} + RT(c_{ref}^\Gamma-C_{\Gamma_1}),
\end{align}
where $\sigma_{ref}$  is the reference surface tension corresponds to the surfactant concentration $c_{ref}^\Gamma$, see for example \cite{JAM04,POZ04}. For instance, if $c_{ref}^\Gamma=0$ then $\sigma_{ref}=\sigma_{0}$,   the surface tension coefficient of the surfactant free (clean)  free surface.  Further, $R$ is the ideal gas constant, $T$ is the absolute temperature.  The linear equation of
state  is valid only for a small variation of the surfactant around their reference value. Moreover,  a
non-linear Langmuir equation of state
\begin{equation} \label{NLEOS}
\sigma(C_{\Gamma_1}) = \sigma_0 + RTC_\Gamma^\infty \ln(1-C_{\Gamma_1}/C_\Gamma^\infty),
\end{equation}
is also used,   see for example \cite{EGG01,KRU04,POZ04}. Here,   $C_\Gamma^\infty$ is the maximum surface packing surfactant concentration.

\subsection{Surfactant transport equations}
In a soluble surfactant model,   surfactant  transports  in the bulk phase (inside the droplet) and on the boundaries of the droplet, that is, on the free surface and on the liquid-solid interface have to be modeled by scalar transport equations. The exchange of surfactants between the bulk phase and the boundaries is modeled by a source term that contains  adsorption and desorption coefficients.
The transport of surfactant concentration in the liquid droplet   is  described \cite{GAN12} by the scalar convection-diffusion equation 
\begin{align} \label{CBULK}
 \frac{\partial C}{\partial t} + \mathbf u\cdot \nabla C =   \nabla\cdot\left( D_c \nabla C\right) &&\mbox{in }\Omega(t)\times(0,\text{I})
\end{align}
with the initial and  boundary conditions
\begin{align*}
 C(\cdot, 0) &=c_{0}&&  \mbox{in }\Omega(0) \\  
 -\bnu_k \cdot \left(D_c\nabla C \right) &=S( C_{\Gamma_k}, C) &&  \mbox{on } \Gamma_k(t)  
\end{align*}
for $k=1,2$.
Here, $C$ is the   surfactant concentration in the bulk phase, $C_{\Gamma_1}$ is the surfactant concentration on  the free surface, $C_{\Gamma_2}$ is the surfactant concentration on the liquid-solid interface,   $D_c$ is the diffusive coefficient of the surfactant in the bulk phase and $c_{0}$ is the initial concentration of surfactants in the bulk phase. The source term $S( C_{\Gamma_k}, C)$ is  given by 
\begin{align} \label{SOURCE}
S(C_{\Gamma_k},C)= K^a_k C\left(C_\Gamma^\infty - C_{\Gamma_k}\right) - K^d_k C_{\Gamma_k},
\end{align}
where  $K^a_k$  and $K^d_k$  are adsorption and desorption coefficients, respectively, on $\Gamma_k(t)$,  $k=1,2.$
The surfactant transport on the moving boundaries $\Gamma_k(t)$ is described \cite{GAN09JCP} by the surface transport equation
\begin{align} \label{FSURF}
 \frac{\partial^{^\Gamma} C_{\Gamma_k}}{\partial t~} + U_\tau\cdot \nabla_{\Gamma_k} C_{\Gamma_k}
+ C_{\Gamma_k} \nabla_{\Gamma_k}\cdot \bw  & =   \nabla_{\Gamma_k}\cdot (D_k \nabla_{\Gamma_k} C_{\Gamma_k}) + S(C_{\Gamma_k}, C) && \mbox{on } \Gamma_k(t),
\end{align}
for $k=1,2$, together with the initial and the continuity condition on the moving contact line, $\zeta(t)$
\begin{align}
 C_{\Gamma_k}(\cdot, 0) &=C_{\Gamma_{k,0}}&&  \mbox{in }\Gamma_k(0) \\  
  C_{\Gamma_1} = C_{\Gamma_2}, \qquad -\bnu_\zeta\cdot \left( D_1 \nabla_{\Gamma_1} C_{\Gamma_1}\right) &= \btau_{1,2}\cdot\left(D_2 \nabla_{\Gamma_2} C_{\Gamma_2}\right) && \mbox{on }\zeta(t) \label{CONT}
\end{align}
Here, $D_1$, and  $D_2$ are the surface diffusive coefficients of  $ C_{\Gamma_1}$ and $ C_{\Gamma_2}$, respectively, $U_\tau$ is the tangential velocity of the free surface/interface, $C_{\Gamma_{k,0}}$, initial concentrations on $\Gamma_k$. Further, the time derivative in \eqref{FSURF} is the normal time derivative of $C_{\Gamma_k}$  following the motion of the free surface along its normal trajectories~\cite{DE13,PAO05}.
In the  continuity condition~\eqref{CONT} at the moving contact line, $\bnu_\zeta$ is the co-normal vector that is normal to $\zeta(t)$ and tangent to $\Gamma_1(t)$, see Figure~\ref{model}. 

\section{Dynamic contact angle at the moving contact line}
\subsection{Equilibrium contact angle}
In thermal, mechanical, and chemical equilibrium state, the equilibrium contact angle, $\theta_e^0$, of a liquid droplet  on a clean, homogeneous, and smooth  surface   satisfies  the Young-Dupre equation
\begin{align}
 \label{young}
 \theta_e^0 =  \cos^{-1}\left(\frac{\sigma^{sg}_0 -  \sigma_0^{ls}}{\sigma_0}\right),
\end{align}
where  $\sigma^{sg}_0 $ and $\sigma^{ls}_0$ are interfacial tension of the clean (surfactant free) solid-gas and liquid-solid interfaces, respectively.  In general, the contact angle and the surface tension are  measured experimentally, and used in computations to describe the wetting behavior.  However, the experimental measurements of  $\sigma^{sg}_0 $ and $ \sigma^{ls}_0$ are seldom available.
The contact angle $\theta_e^0$ in~\eqref{young} is also referred to as the static contact angle, and is unique for the considered gas, liquid and solid material phases at the equilibrium state. Nevertheless, the contact angle deviates from the equilibrium value when the contact line moves, and the difference between the advancing contact angle  and the receding  contact angle  is referred to as a contact angle \textit{hysteresis}.  
 In general, the contact angle that incorporates the hysteresis is called the dynamic contact angle~$\theta_d$. Surface roughness, contamination on surfaces, thermal effects are also some of the reasons for the dynamic behavior of the  contact angle. The  evidence for the dynamic behavior of the contact angle can be found in experiments, see for example,~\cite{DUS79,HOF75}.   The measurement of the contact angle in experiments depends on the resolution of the microscope, and the contact angle is  measured a certain distance away from the contact line. Hence, the experimentally measured angle is referred   as an apparent or macroscopic contact angle.

\subsection{Contact angle in computations}\label{dyca}
Several models have been proposed in the literature for the choice of the contact angle in computations of flows with moving contact lines, see~\cite{GAN13M} for a recent comparative study of different contact angle models in computations of droplet impingement.  In general, the contact angle, irrespective of the contact angle model, is incorporated as a surface force at the contact line in discretization based numerical schemes. Further,   the equilibrium contact angle has been preferred  in  computations when a sharp interface model is used~\cite{GAN13M}. 
In an equilibrium state, the Young-Dupr\'e equation satisfies
\[
  \sigma_0\cos\theta_e^0 =\sigma^{sg}_0 -  \sigma^{ls}_0.
\]
However, an unbalanced Young force~\cite{DEG85}     
\begin{align*}
    F_Y =    \sigma^{sg}_0 -  \sigma^{ls}_0 - \sigma_0\cos\theta_d=  \sigma_0(\cos\theta_e^0 - \cos\theta_d)  
\end{align*}
is induced   at the contact line during the droplet deformation since $\theta_d\ne\theta_e$, see  Figure~\ref{model}. Moreover, we have 
\begin{align*}
   \sigma_0 \cos\theta_d   =   \sigma_0 \cos\theta_e^0  + F_Y.
\end{align*}
We now impose $F_Y=0$ in the model, and it results in
\begin{align} \label{caeq}
     \theta_d   =    \theta_e^0,  
\end{align}
in the numerical scheme~\eqref{weakNSE}.  Thus, the surface force at the contact line becomes unbalanced for the geometry with $\theta_d$, see  Figure~\ref{model}.
The imbalance in the surface force induces a non-zero slip velocity, that is, the surface force is translated  into a kinetic energy. 
Consequently, the slip velocity drives the contact line into the equilibrium position. Moreover, the dynamic contact angle will attain the prescribed equilibrium value  when the contact line attains its   equilibrium position. To incorporate this phenomenon, it is necessary to allow the liquid to slip in the vicinity of the  contact line. Further, the calculated value of $\theta_d$ varies when the contact line moves, see  Figure~\ref{model}. Thus, the slip velocity   directly  influences the dynamics of the contact angle   in   sharp interface models. Therefore, it is necessary to use   an  ``appropriate slip'' in computations of moving contact line flows when a sharp interface  model is used.

Several boundary conditions for the fluid velocity on the liquid-solid interface have been proposed in the literature for moving the contact line problems, see~\cite{EGG04} for an overview. Among all, the Navier-slip boundary condition is widely accepted. However, an   appropriate choice of the slip length (friction coefficient) in   the Navier-slip boundary condition is a main challenge. The experimental evidences show that the slip length varies for different flows at different configurations.
Although, a number of expressions have been proposed for the slip length~\cite{COX86,DUS76,HOC77}, an exact mathematical expression is missing. A more complicate non-linear form of the slip length has been   proposed for a Newtonian liquid in molecular length scale \cite{THO97}.  Recently, an expression for the  numerical slip   as a function of $\Rey$ and $\Web$  has been proposed in \cite{SGJV15} for impinging droplets, and it is used in this paper. 

\subsection{Surfactant-dependent contact angle} \label{surfca}
Suppose that the interfaces are clean, then we have  the balanced Young-Dupr\'e equation \eqref{young}.
However, a nonuniform distribution of surfactants on   interfaces induces  an unbalanced Young force
\begin{align*}
 F_Y = \sigma^{sg}(C_{\Gamma_3}) -  \sigma^{ls}(C_{\Gamma_2}) - \sigma(C_{\Gamma_1})\cos\theta_d
\end{align*}
at the contact line. Here, $C_{\Gamma_3}$ denotes the surfactant concentration  on  the solid-gas interface.
It is impractical to use the above relation since the values of $\sigma^{sg}(C_{\Gamma_3})$ and $\sigma^{ls}(C_{\Gamma_2})$  are still needed in computations, and are seldom available. However, using the nonlinear equation of state~\eqref{NLEOS} for $\sigma^{sg}(C_{\Gamma_3})$ and $\sigma^{ls}(C_{\Gamma_2})$  in the above relation, we get
\begin{align}
 \label{modyoung0}
  F_Y =    \sigma_{0}  \cos(\theta_{e}^0) +  RTC_\Gamma^\infty \ln(M) - \sigma(C_{\Gamma_1})\cos(\theta_d), \qquad M = \frac{C_\Gamma^\infty -C_{\Gamma_3}}{C_\Gamma^\infty -C_{\Gamma_2}}.
\end{align}
 As before,  we impose  $F_Y=0$ in the model, and it results in
\begin{align}
 \label{casurf1}
   \theta_d   =   \cos^{-1}\left(\frac{ \sigma_{0}  \cos(\theta_{e}^0) +  RTC_\Gamma^\infty \ln(M)} {\sigma C_{\Gamma_1}  }\right).
\end{align}
Contrary to the clean case \eqref{caeq}, the dynamic contact angle~\eqref{casurf1} will not attain the equilibrium value $\theta_e^0$  when the contact line attains its equilibrium position. The equilibrium value of  \eqref{casurf1} depends on the surfactant concentrations, $C_{\Gamma_k}$, $k=1,2,3$. 

 An increase in $C_{\Gamma_1}$ reduces $\theta_d$ when $\theta_{e}^0$ is less than $90^\circ$, and increases $\theta_d$ when $\theta_{e}^0$ is greater than $90^\circ$. Furthermore, the $\theta_d$ decreases further when $C_{\Gamma_2}$ is greater than $C_{\Gamma_3}$ and increases  when $C_{\Gamma_2} $ is less than $ C_{\Gamma_3}$ even for $\theta_e^0=90^\circ$.  For instant, the surfactant concentration  on the  liquid-solid interface $C_{\Gamma_2}$ may vary in droplets with  soluble  surfactants, and it will not be equal to the  surfactant concentration  on the  solid-gas interface.  It has   been  observed in  experiments~\cite{CHA97}, where  the measured equilibrium contact angle for pure water on a clean stainless steel was $90^\circ$, and  adding 100 and 1000 ppm surfactants reduced the equilibrium contact angle to $55$ and $20^\circ$, respectively. 

Suppose that the solid-gas interface is clean, that is, $C_{\Gamma_3}=0$, then we have 
\begin{align}
 \label{casurf2}
  \theta_d  =   \cos^{-1}\left(\frac{ \sigma_{0}  \cos(\theta_{e}^0) -  RTC_\Gamma^\infty \ln(1 - C_{\Gamma_2}/C_\Gamma^\infty   )} {\sigma(C_{\Gamma_1})}\right).
\end{align} 
It has to be used in computations when the effects of $C_{\Gamma_2}$ are taken into considerations. Nevertheless, the   relation~\eqref{casurf1} can further  be simplified to 
\begin{align}
 \label{casurf3}
\theta_d =  \cos^{-1}\left(\frac{ \sigma_{0}  \cos(\theta_{e}^0) }{\sigma(C_{\Gamma_1})}\right)= \theta_e ({C_{\Gamma_1}})
\end{align}
 by  assuming     $C_{\Gamma_3}=C_{\Gamma_2}$.
Since   the continuity condition \eqref{CONT}  is imposed at the contact line, it is sufficient to use the relation \eqref{casurf3} in computations. 
Note that the above relation~\eqref{casurf3} is independent of  $C_{\Gamma_3}$ and $C_{\Gamma_2}$  provided that these concentrations are equal at the contact line.  
It is interesting to note that the surfactant concentration $C_{\Gamma_1}$ has no influence on the dynamic contact angle when $\theta_{e}^0=90^\circ$ and  $C_{\Gamma_3} = C_{\Gamma_2} $. A similar relation can also be derived, when   the linear equilibrium of state~\eqref{LEOS} is used.

\section{Numerical scheme}
We  transform  the model equations into an arbitrary Lagrangian-Eulerian form after writing it in a dimensionless form. We then derive variational forms of the Navier--Stokes equations and the surfactant concentration equations. In particular, we derive one-field formulation for the  surfactant concentration equations on the interfaces using the continuity condition of surfactants at the moving contact line. After that, we briefly present the spatial and the temporal discretizations of the model equations. Finally,   the mesh moving technique is presented briefly, and for a detailed description of the mesh handling, we refer to~\cite{GAN09JCP,GAN12}.

\subsection{Dimensionless form}
Let the characteristic length and velocity be $L$ and $U$, respectively. Define the dimensionless variables
\[
\tilde{x}=\frac{x}{L},\quad\! \tilde{\mathbf u}=\frac{\mathbf u}{U},\quad\! \tilde{\bw}=\frac{\bw}{U},\quad\! \tilde{t}=\frac{tU}{L},\quad \! \tilde{\text{I}}=\frac{\text{I}U}{L},\quad
\! \tilde{p}=\frac{p}{\rho U^2}, \quad\! \tilde{C} =\frac{C}{c_\infty} \quad\!\tilde{C}^k_\Gamma = \frac{C_{\Gamma_k}}{C_\Gamma^\infty}.
\]
Applying these variables in the Navier--Stokes equations~\eqref{tnse} in the usual way, and omitting the tilde afterwards, we get the dimensionless form of the Navier--Stokes problem  as 
\begin{align}\label{NTNSE}
\nabla \cdot \mathbf u=0, \qquad \quad \frac{\partial \mathbf u}{\partial t}+(\mathbf u\cdot\nabla) \mathbf u-\nabla\cdot\mbS(\mathbf u,p)&=\frac{1}{\Fro}\be && \!\!\mbox{in }\Omega(t)\times(0,\text{I}),\\
\mathbf u\cdot\bnu_1=\bw\cdot\bnu_1, \quad \mbS(\mathbf u,p)\cdot\bnu_1&=\frac{1}{\Web} \nabla_{\Gamma_1}\cdot(\hat\sigma(C_{\Gamma_1})\mbP_{\bnu_1}) && \!\!\mbox{on }\Gamma_1(t)\times(0,\text{I})\label{tnse3e}\\
\mathbf u\cdot\bnu_2=0, \quad
\btau_{i,S}\cdot\mbS(\mathbf u,p)\cdot\bnu_2&=-\beta\,\mathbf u\cdot\btau_{i,S}&  & \!\!\mbox{on }
\Gamma_2(t)\times(0,\text{I}),\label{tnse3d}\\
\mathbf u(\cdot,0)&=\mathbf u_0/U &&  \!\!\mbox{in }\Omega(0),
\end{align}
with the dimensionless numbers (Reynolds, Weber, Froude and slip, respectively)  
\[          
{\text{Re}}=\frac{\rho U L}{\mu}, \quad {\text{We}}=\frac{\rho  U^2 L}{\sigma_{0}}, \quad \Fro= \frac{ U^2}{L g}, \quad \beta=\frac{1}{ \beta_{\epsilon}\rho U},
\] 
and the  dimensionless stress tensor $\mbS( {\mathbf u},{p})$
\begin{align*}
 \mbS({\mathbf u},{p}) =\frac{2}{\Rey}\mbD(\mathbf u)-p \mbI.
\end{align*}
Here, the scaled surfactant-dependent surface tension in the case of linear equation of state~\eqref{LEOS} becomes
\begin{align} \label{SLEOS}
\hat\sigma(C_{\Gamma_1}) = \frac{\sigma_{ref}}{\sigma_0} + E\left(\frac{C^{ref}}{C_\Gamma^\infty}-C_{\Gamma_1}\right),
\end{align}
where E is the surfactant elasticity defined as $E={RTC_\Gamma^\infty }/{\sigma_{0}}$, and in the case of nonlinear equation of state~\eqref{NLEOS} becomes
\begin{equation} \label{SNLEOS}
\hat\sigma(C_{\Gamma_1}) = 1 +E \ln(1-C_{\Gamma_1}).
\end{equation}

Using the dimensionless variables in Eq.~\eqref{CBULK}, we get the dimensionless form of the surfactant transport problem in bulk phase as
\begin{align} \label{NCSURF}
 \frac{\partial C}{\partial t} + \mathbf u\cdot \nabla C &=   \nabla\cdot\left(\frac{1}{\Pe_c}\nabla C\right)  && \!\!\mbox{in }\Omega(t)\times(0,\text{I}),\\
  C(\cdot, 0) &=\frac{c_{0}}{c_\infty}&&  \mbox{in }\Omega(0), \nonumber \\  
  -\bnu_1 \cdot \left(\frac{1 }{\Pe_c}\nabla C\right) &=   S_c( C_{\Gamma_k}, C) && \mbox{on } \Gamma_k(t)\times(0,\text{I}), \nonumber
\end{align}
for $k=1,2$, where the non-dimensional form of the source term becomes
\[
 S_c( C_{\Gamma_k}, C) = \alpha_k\, C\left(1 - C_{\Gamma_k}\right) - \Bi_k \Da\, C_{\Gamma_k}.
\]
Similarly, the dimensionless form of the surface transport equations become
\begin{align} 
\label{NSURF}
\frac{\partial^{^\Gamma} C_{\Gamma_k}}{\partial t~} + U\cdot \nabla_{\Gamma_k}\;C_{\Gamma_k}
+  C_{\Gamma_k}\;  \nabla_{\Gamma_k}\cdot \bw   &=    \nabla_{\Gamma_k}\cdot\left(\frac{1 }{Pe_k}  \nabla_{\Gamma_k} C_{\Gamma_k} \right) +   S_\Gamma( C_{\Gamma_k}, C) 
 && \mbox{on } \Gamma_k(t)\times(0,\text{I}), \\
 C_{\Gamma_k}(\cdot, 0) &=\frac{C_{\Gamma_{k,0}}}{C_\Gamma^\infty}&&  \mbox{in }\partial\Omega_F(0) \nonumber \\ 
   C_{\Gamma_1} = C_{\Gamma_2}, \qquad -\bnu_\zeta\cdot \left( \frac{1}{\Pe_1} \nabla_{\Gamma_1} C_{\Gamma_1}\right) &= \btau_{1,2}\cdot\left(\frac{1}{\Pe_2}  \nabla_{\Gamma_2} C_{\Gamma_2}\right) && \mbox{on }\zeta(t), \nonumber
\end{align} where
\[
  S_\Gamma( C_{\Gamma_k}, C) = \frac{\alpha_k}{Da}\,  C \left(1-C_{\Gamma_k}\right) -  Bi_k\, C_{\Gamma_k}, \qquad  k=1,2.
\]
The dimensionless numbers (Peclet,  Biot, Damk\"ohler and $\alpha_k$) in
Eqs.~\eqref{NCSURF}, and \eqref{NSURF} are given by 
\[
 \Pe_c = \frac{U L}{D_c},  \quad
 \Pe_k = \frac{U L}{D_k}, \quad
 \Bi_k = \frac{K^d_k L}{U }, \quad 
 \Da=  \frac{  \Gamma_\infty}{ L C_\infty}, \quad 
 \alpha_k  =\frac{K^a_k C_\Gamma^\infty}{U }. 
\] 

\subsection{ALE formulation}
The time-dependent domain is handled by the arbitrary Lagrangian-Eulerian approach using moving meshes, which resolve the free surface and the liquid-solid interface. 
Let   $\hat\Omega$ be a reference domain of $\Omega(t)$. Define a family of ALE mappings 
\begin{align*}
 \mathcal{A}_t:\hat{\Omega} \rightarrow \Omega(t),  \qquad  \mathcal{A}_t(\bY)= \bX(\bY, t), \qquad t\in(0,\text{I}),
\end{align*}
where $\bX$ and    $\bY$ are termed as   Eulerian and ALE coordinates, respectively. 
To derive the ALE form of the model equations, we assume that the mapping  $\mathcal{A}_t$ for all $ t\in(0,\text{I})$ is homeomorphic, that is,   $\mathcal{A}_t$ is bijective, continuous and its inverse  $\mathcal{A}_t^{-1}$ is also continuous. Further, assume that the mappings are differentiable almost everywhere in   $(0,\text{I})$. Consequently, these assumptions impose  that the topology of the domain should remains same.
 For the surfactant concentration $C:\Omega(t)\times (0,\text{I}) \rightarrow \mathbb{R}$, which is defined on the Eulerian frame,   define their corresponding $\hat C$  and its time derivative $\left. \frac{\partial C}{\partial t} \right|_{\hat{\Omega}}$ on the ALE frame by
\begin{align*}
  \hat{C}&: \hat \Omega \times (0,\text{I}) \rightarrow \mathbb{R}, \quad (\bY,t) \mapsto C(\bX(\bY,t),t) = C(\mathcal{A}_t(\bY), t)\\
 \left. \frac{\partial C}{\partial t} \right|_{\hat{\Omega}}&:\Omega(t)\times (0, \rm{I}) \rightarrow \mathbb{R}, \quad
(\bX, t) \mapsto \frac{\partial \hat{C}}{\partial t}(\mathcal{A}_t^{-1}(\bX), t),
\end{align*}
Furthermore, the domain velocity on the ALE frame is defined by
\begin{align*}
 \bw(\bX, t) = \left.\frac{\partial \bX}{\partial t} \right|_{\hat{\Omega}}(\mathcal{A}_t^{-1}(\bX), t),\quad \bX\in \Omega(t).
\end{align*}
Applying the chain rule to the time derivative of $C$ in the ALE frame, we    get
\begin{align}
\label{aletrans}
\left.\frac{\partial C}{\partial t} \right|_{\hat{\Omega}} =  \frac{\partial C}{\partial t}  + \frac{\partial C}{\partial \bX}\left.\frac{\partial \bX}{\partial t}\right|_{\hat{\Omega}}(\mathcal{A}_t^{-1}(\bX), t) = \frac{\partial C}{\partial t} +
 \bw\cdot\nabla C.
\end{align}
The time derivatives of a vector valued functions on the Eulerian frame can also be transformed to the ALE frame component-wise. Note that the time derivative in the ALE form will become a material derivative when the convective velocity $\mathbf u$ and the domain velocity $\mathbf w$ are  same, which is the Lagrangian description of the equations. 
After rewriting the time derivatives in the soluble surfactant droplet impingement model equations using~\eqref{aletrans}, the Navier--Stokes equations~\eqref{NTNSE}, the bulk surfactant concentration equation~\eqref{NCSURF}, and the   surfactant concentration equation on the interface~\eqref{NSURF} become
 \begin{align}\label{tnseale}
 \nabla \cdot \mathbf u=0, \qquad \quad \left.\frac{\partial \mathbf u}{\partial t}\right|_{\hat{\Omega}}+((\mathbf u - \mathbf w)\cdot\nabla) \mathbf u&=\nabla\cdot\mbS(\mathbf u,p) + \frac{1}{\Fro}\be && \!\!\mbox{in }\Omega(t)\times(0,\text{I})\\
 \label{CALE}
  \left.\frac{\partial C}{\partial t}\right|_{\hat{\Omega}} + (\mathbf u - \mathbf w)\cdot \nabla C &=   \nabla\cdot\left(\frac{1}{\Pe_c}\nabla C\right)  && \!\!\mbox{in }\Omega(t)\times(0,\text{I}),\\
  \label{SALE}
\left.\frac{D C_{\Gamma_k}}{D t}\right|_{\hat{\Gamma}_k}
+  C_{\Gamma_k}\;  \nabla_{\Gamma_k}\cdot \bw    &=    \nabla_{\Gamma_k}\cdot\left(\frac{1 }{\Pe_k} \nabla_{\Gamma_k} C_{\Gamma_k} \right) +   S_\Gamma( C_{\Gamma_k}, C) 
 && \mbox{on } \Gamma_k(t)\times(0,\text{I}),
\end{align}
for $k=1,2.$ Note that the free surface and the liquid-solid interface move with the liquid velocity, and therefore the  surface transport equations~\eqref{SALE} are written in the Lagrangian description.

\subsection{Variational formulation}
Let $L^2(\Omega(t))$, $H^1(\Omega(t))$  be the Sobolev spaces, and $(\cdot,\cdot)_\Omega$ be the inner product
in $L^2(\Omega)$ and its vector-valued versions, respectively. Define the functional spaces for the velocity and pressure as
\begin{align*} 
V(\Omega(t))&:=\{\bv\in H^1(\Omega(t))^3 :\bv\cdot\bnu_2=0 \mbox{ on } \Gamma_2(t)\},\quad
Q(\Omega(t)):=L^2(\Omega(t)),
\end{align*}
where the no penetration boundary condition on liquid-solid interface is   incorporated in the velocity space. 
To derive the variational form of the Navier--Stokes equations, we multiply the mass and momentum balance equations~\eqref{tnseale}  by test functions    $q\in Q$  and $\bv\in V$, respectively, and integrate over $\Omega(t)$.
 Applying integration by parts to the stress tensor, we get
 \begin{align*}
-\int_{\Omega(t)}\nabla\cdot \mbS(\mathbf u,p)\cdot\bv ~dx =\frac{2}{\text{Re}}\int_{\Omega(t)}\!\!\mbD(\mathbf u):\mbD(\bv) ~dx-\int_{\Omega(t)}\!\!p\nabla\cdot\bv ~dx - \int_{\Gamma(t)}\bv\cdot\mbS(\mathbf u,p)\cdot\bnu~ d\gamma.
\end{align*}
Rewriting the boundary integral into integral over $\Gamma_1(t)$ and $\Gamma_2(t)$, decomposing the test function $\bv$ into
 \begin{align} \label{decompv}
\bv = (\bv\cdot\bnu_2)\bnu_2 + (\bv\cdot\btau_{1,2})\btau_{1,2} + (\bv\cdot\btau_{2,2})\btau_{2,2},
\end{align}
and after applying the Navier-slip condition, the liquid-solid interface integral becomes
\begin{align*}
\int_{\Gamma_2(t)}\!\!\!\!\bv\cdot\mbS(\mathbf u,p)\cdot &\bnu_2~d\gamma
=   -\beta\sum_{i=1}^2\int_{\Gamma_2(t)}(\mathbf u\cdot\btau_{i,2})(\bv\cdot\btau_{i,2})~d\gamma.
\end{align*}
This integral term will be added on the left hand side of the system, and it improves the stability of  the system. Similarly, the free surface integrate, after incorporating the force balancing condition, will become
\begin{align*}
-\int_{\Gamma_1(t)}\bv\cdot \mbS(\mathbf u,p)\cdot\bnu_F~d\gamma    
&=- \frac{1}{\Web}\int_{\Gamma_1(t)}\bv\cdot \nabla_{\Gamma}\cdot(\hat\sigma(C_{\Gamma_1})\mbP_{\bnu_1})  ~d\gamma  \\ 
&= \frac{1}{\Web}\int_{\Gamma_1(t)}  \hat\sigma(C_{\Gamma_1})\mbP_{\bnu_1} :\nabla_{\Gamma}\bv ~d\gamma
   - \frac{1}{\Web}\int_{\zeta(t)}\hat\sigma(C_{\Gamma_1}) \bnu_{\zeta} \cdot  \bv ~ d\zeta.
\end{align*}
Again using the decomposition~\eqref{decompv} of $\bv$ in the last integral term, we obtain
\begin{align}
-\int_{\Gamma_1(t)}\bv\cdot \mbS(\mathbf u,p)\cdot\bnu_F~d\gamma    
&= \frac{1}{\Web}\int_{\Gamma_1(t)}  \hat\sigma(C_{\Gamma_1})\mbP_{\bnu_1} :\nabla_{\Gamma}\bv ~d\gamma
   - \frac{1}{\Web}\int_{\zeta(t)}\hat\sigma(C_{\Gamma_1}) \cos(\theta_d)     \bv \cdot \btau_{1,2} ~ d\zeta,  \label{varsurf} 
\end{align}
since $\bv\cdot\bnu_2=0$, $ \bnu_{\zeta} \cdot\btau_{2,2} =0$ and  $ \bnu_{\zeta} \cdot\btau_{1,2} =  \cos(\theta_d)  $, see Figure~\ref{model} for a geometrical description.
Note that the Marangoni effects induced by the nonuniform surfactant concentration are included into the numerical scheme without evaluating the surface gradient of the surface tension, which is different from the Laplace-Beltrami operator technique used in~\cite{GRT14,GAN09JCP,GAN12}.  
 Now, using the relation \eqref{casurf3}, the variational form  of the Navier--Stokes equations read:\\
 
 \noindent For given $\mathbf u(0)$, $C_{\Gamma_1}(t)$,  $\theta_d=  { \sigma_{0}  \cos(\theta_{e}^0) }/{\sigma(C_{\Gamma_1})} $ and $\Omega(0)$, find $(\mathbf u, p)  \in V\times Q $ such that
\begin{align}\label{weakNSE}
\left(\frac{\partial \mathbf u}{\partial t},\bv \right)_{\hat\Omega }+a(\mathbf u-\bw;\mathbf u,\bv)-b(p,\bv)+b(q,\mathbf u)=f(\bv)
\end{align}
for all  $(\bv,q) \in V \times Q$, where
\begin{align*}
a(\hat{\mathbf u};\mathbf u,\bv)&=\frac{2}{\text{Re}}\int_{\Omega(t) }\mbD(\mathbf u):\mbD(\bv)+(\hat{\mathbf u}\cdot\nabla)\mathbf u\cdot \bv~ dx +  {\beta}\int_{\Gamma_2(t)}
 \sum_{i=1}^2(\mathbf u\cdot\btau_{i,2})(\bv\cdot\btau_{i,2})~d\gamma,\\ 
b(q,\bv)&=\int_{\Omega(t) } q \,\nabla\cdot \bv\,dx ,\\ 
f(\bv)&=\frac{1}{\Fro}\int_{\Omega(t) } \be\cdot \bv\,dx-\frac{1}{\Web}\int_{\Gamma_1(t)}\hat\sigma(C_{\Gamma_1}(t))\mbP_{\bnu_1} :\nabla_{\Gamma_1}\bv\,d\gamma   + \frac{1}{\Web}\int_{\zeta(t)}\hat\sigma(C_{\Gamma_1}(t))\cos(\theta_d )\;  \mathbf v\cdot \btau_{1,2}\,d\zeta.
\end{align*}
Note that the choice of $\theta_d=  { \sigma_{0}  \cos(\theta_{e}^0) }/{\sigma(C_{\Gamma_1})} $ induce an unbalanced Young force at the contact line, which drives the contact line to   the equilibrium position,   see sections~\ref{dyca} and \ref{surfca}.

Next, the variational forms of the  surfactant concentration equation~\eqref{CALE} is obtained in the usual way.  Let  $G(\Omega(t)):=H^1(\Omega(t))$ and   $M(\Gamma(t)):=H^1(\Gamma(t))$ be the usual Sobolev spaces. Further, in order to write the surfactant concentration equations on the free surface and on the liquid-solid interface  in a one-field formulation, we define
\begin{align*}
 C_\Gamma(x, t) = \left\{ \begin{array}{rcl}
                         C_{\Gamma_1}(x,t)& \text{if} &x\in\Gamma_1(t),\\
                         C_{\Gamma_2}(x,t)& \text{if} &x\in\Gamma_2(t),
                        \end{array}\right. \qquad
\Pe_\Gamma(x) = \left\{ \begin{array}{ccl}
       \Pe_1 &\text{if}& x\in \Gamma_1(t),\\ \displaystyle    \Pe_2  &
       \text{if}& x\in \Gamma_2(t).
        \end{array}\right. \\
     \Bi(x) = \left\{ \begin{array}{rcl}
                         \Bi_1(x,t)& \text{if} &x\in\Gamma_1(t),\\
                         \Bi_2(x,t)& \text{if} &x\in\Gamma_2(t),
                        \end{array}\right. \qquad
\alpha(x) = \left\{ \begin{array}{ccl}
       \alpha_1 &\text{if}& x\in \Gamma_1(t),\\ \displaystyle    \alpha_2  &
       \text{if}& x\in \Gamma_2(t).
        \end{array}\right.    
\end{align*}
Multiplying Eqs.~\eqref{CALE}  and~\eqref{SALE} by test functions $\phi\in G$ and $\psi\in M$, 
integrating over $\Omega(t)$ and $\Gamma_k(t)$, respectively, incorporating the boundary and continuity conditions, we obtain
the coupled problem for the soluble surfactant concentration:\\  

\noindent  For given $(C_{\Gamma, 0}, \bu, \bw)$, find $(C,C_\Gamma)\in G\times M$ such 
that  for all $(\phi,\psi)\in G\times M$
\begin{align} \label{WEAKCOUPCSURF}
\left(\frac{\partial C}{\partial t}, \phi\right)_{\hat\Omega} +
a_c(\bu-\bw;C,\phi) + 
b_c(C,C_\Gamma, \phi) &= s_c(C_\Gamma,\phi),\\ 
\left(\frac{D C_\Gamma}{D t},  \psi\right)_{\hat\Gamma} 
+ a_\Gamma(\bw,C_\Gamma,\psi) + b_\Gamma(C_\Gamma,C,\psi)&=
s_\Gamma(C,\psi), \label{WEAKCOUPSURF} 
\end{align}
   where
\begin{align*}
a_c(\bv;C,\phi) &=   \frac{1}{\Pe_c}\int_{\Omega(t)} \nabla C\cdot
\nabla\phi~dx + \int_{\Omega(t)}(\bv\cdot \nabla) C  \phi ~dx,\\ 
b_c(C_\Gamma, C, \phi) &=    \int_{\Gamma(t)} \alpha (1 - C_\Gamma) C
\,\phi ~d\gamma,\\ 
s_c(C_\Gamma,\phi)  &= \Da\int_{\Gamma(t)} \Bi\, C_\Gamma   \phi ~d\gamma,\\
 a_\Gamma(\bw,C_\Gamma,\psi) &=  \int_{\Gamma(t)} \frac{1}{\Pe_\Gamma}
  \nabla_\Gamma C_\Gamma \cdot \nabla_\Gamma \psi ~d\gamma +
 \int_{\Gamma(t)} C_\Gamma\; \nabla_\Gamma\cdot \bw~\psi ~d\gamma,  \\ 
b_\Gamma(C, C_\Gamma,\psi) &=   \frac{1}{\Da}  \int_{\Gamma(t)}\alpha \,
C~ C_\Gamma ~\psi ~d\gamma +    \int_{\Gamma(t)}\Bi\,C_\Gamma ~\psi ~d\gamma,\\ 
s_\Gamma(C,\phi)  &=  \frac{1}{\Da} \int_{\Gamma(t)} \alpha \, C~\psi ~d\gamma.
\end{align*}

\subsection{Discrete problem}
We first present the temporal discretization of the   coupled system \eqref{weakNSE}-\eqref{WEAKCOUPSURF}, in particular, the application of the fractional-step-$\theta$~scheme is discussed. Further,  a fixed point type iteration  for the nonlinear convective term in the Navier--Stokes equations \eqref{weakNSE} and  a Gauss-Seidel type  iteration for the coupled surfactant equations are presented. The choice of finite elements for the spatial discretization of the system \eqref{weakNSE}-\eqref{WEAKCOUPSURF} is also discussed.

\subsubsection{Temporal discretization}
 
Let $0=t^0<t^1<\dots <t^N=\rm{I}$ be a decomposition of the considered
time interval $[0, \rm{I}]$ and $\delta t= t^{n+1} - t^{n}$, $n=0,\ldots,N-1$, be the uniform time step. Also, we use short    notations $\Omega_n:=\Omega(t^n)$ and $\bu^n=\bu(x,t^n)$  to denote the computational domain and the function value, respectively at time $t^n$. We use the fractional-step-$\theta$~scheme, which is strongly A-stable and of 
second-order convergent on fixed domains~\cite{STK99}, for temporal discretization of the coupled system. The fractional-step-$\theta$~scheme consists three sub-steps in a given time interval $(t^{n},t^{n+1})$. Let  
 \begin{align*}
\vartheta = 1- \frac{\sqrt2}{2}, \quad \tilde\vartheta = 1-2\vartheta, \quad \eta= \frac{\tilde\vartheta}{1-\vartheta}, \quad \tilde\eta= 1-\eta.
\end{align*}
The   three fractional-steps of $(t^{n},t^{n+1})$ are ($t^{n},t^{k_1}$), ($t^{k_1},t^{k_2}$) and ($t^{k_2},t^{n+1}$), where $t^{k_1}=t_n+\vartheta\,\delta t,$ and $t^{k_2}=t_{n+1}-\vartheta\,\delta t$. 
Applying the fractional-step-$\theta$ scheme to the coupled system \eqref{weakNSE}-\eqref{WEAKCOUPSURF}, the first sub-step of the three fractional-steps of the coupled system reads:\\

 \noindent   \textbf{Step 1}:
 For given   $\hat\Omega:=\Omega_n$,   $\bu^n$,  $\bw^n$, $C^n$, $C^n_\Gamma$ and $\theta_d=  { \sigma_{0}  \cos(\theta_{e}^0) }/{\sigma(C_{\Gamma_1}^n)} $, find  $(\bu^{k_1}, p^{k_1})\in V(\Omega_{k_1}) \times Q(\Omega_{k_1})$, $\bw^{k_1}\in  H^1(\Omega_{k_1})$, $C^{k_1}\in G^1(\Omega_{k_1})$ and $C^{k_1}_\Gamma  \in M(\Gamma_{k_1})$ such that for all  $(\bv,q) \in V(\Omega_{k_1}) \times Q(\Omega_{k_1} )$ and $\phi\in G^1(\Omega_{k_1})$ and $\psi \in M(\Gamma_{k_1})$
\\
\begin{align} 
\left(\frac{\bu^{k_1} - \bu^n}{\vartheta \,\delta t},\bv \right)_{\hat\Omega} &+ \eta \,a(\bu^{k_1}-\bw^{k_1};\bu^{k_1},\bv)-b(p^{k_1},\bv) \nonumber \\
&+b(q,\mathbf u^{k_1}) =\eta \,f^{k_1}(\bv) + \tilde\eta \,f^{n}(\bv) - \tilde\eta \,a(\bu^{n}-\bw^{n};\bu^{n},\bv), \label{tdiscnse1}\\
\left(\frac{C^{k_1} - C^n}{\vartheta \,\delta t}, \phi\right)_{\hat\Omega} +
\eta \,a_c( &\bu^{k_1}-\bw^{k_1};C^{k_1},\phi)+ \eta\, b_c(C^{k_1},C_\Gamma^{k_1}, \phi)  \nonumber \\
  &= \eta\, s_c(C^{k_1}_\Gamma,\phi)  + \tilde \eta \,s_c(C^{n}_\Gamma,\phi) - \tilde\eta \, a_c(\bu^{n}-\bw^{n};C^{n},\phi) -  \tilde\eta \,b_c(C^{n},C^{n}_\Gamma, \phi),\label{tdiscCsurf1}\\ 
\left(\frac{C^{k_1}_\Gamma - C^n_\Gamma}{\vartheta \,\delta t},  \psi\right)_{\hat\Gamma} 
+ \eta\,a_\Gamma(&\bw^{k_1},C^{k_1}_\Gamma,\psi) + \eta\,b_\Gamma(C^{k_1}_\Gamma,C^{k_1},\psi) \nonumber \\
&= \eta\,s_\Gamma(C^{k_1},\psi) + \tilde \eta\,s_\Gamma(C^{n},\psi) - \tilde\eta\,a_\Gamma(\bw^{n},C^{n}_\Gamma,\psi) -  \tilde\eta\,b_\Gamma(C^{n}_\Gamma,C^{n},\psi).\label{tdiscSsurf1}
\end{align}
The second and third sub-steps of the fractional-step-$\theta$~scheme are obtained in a similar way~\cite{STK99}.
\subsubsection{Solution  of the nonlinear system} \label{secnonlin}
We discuss the solution procedure for the coupled system in the first sub-step of the fractional-step-$\theta$~scheme, and the same procedure is followed in   other two sub-steps. 
In addition to the nonlinear convection term in the Navier--Stokes equations~\eqref{tdiscnse1}, the unknown computational domain, the domain velocity and the surfactant-dependent surface tension make the computation more challenging.  Since the computational domain, $\Omega_{k_1}$, is part of the Navier--Stokes solution, the  Navier--Stokes equations~\eqref{tdiscnse1} are solved in the previous time-step domain, $\Omega_{n}$. Further the  surfactant-dependent surface tension also treated explicitly, that is, $ C^n_\Gamma$ is used in the source term $f^{k_1}(\bv)$, and it decouples the Navier--Stokes equations~\eqref{tdiscnse1} from the surfactant concentration equation~\eqref{tdiscSsurf1}. Moreover, the curvature term in $f^{k_1}(\bv)$ of~\eqref{tdiscnse1} is treated semi-implicitly, that is,
\begin{align}
 -\frac{1}{\Web}\int_{\Gamma_1^n}\hat\sigma(C^n_{\Gamma_1} )\mbP_{\bnu_1^{k_1}} :\nabla_{\Gamma}\bv \,d\gamma &=  -\frac{1}{\Web}\int_{\Gamma_1^n}\hat\sigma(C^n_{\Gamma_1} )\left[ \mbP_{\bnu_1^{n}} + \vartheta \,\delta t \bu^{k_1} \right]:\nabla_{\Gamma_1}\bv \,d\gamma \nonumber\\
 & = -\frac{1}{\Web}\int_{\Gamma_1^n}\hat\sigma(C^n_{\Gamma_1} ) \mbP_{\bnu_1^{n}}:\nabla_{\Gamma_1}\bv \,d\gamma -\frac{1}{\Web}\int_{\Gamma_1^n}\hat\sigma(C^n_{\Gamma_1} )  \vartheta \,\delta t \bu^{k_1} :\nabla_{\Gamma_1}\bv \,d\gamma. \label{cursemi}
\end{align}
The second integral in~\eqref{cursemi} is symmetric, and it is added to the left hand side of~\eqref{tdiscnse1} that gives   additional stability to the system. The nonlinear convection term in~\eqref{tdiscnse1}  is  handled by a fixed point iteration as in~\cite{GAN07}.  Let $\bu^{k_1}_0:=\bu^{n}$, $\bw^{k_1}_0:=\bw^{n}$, and  replace the form
 $a(\bu^{k_1}-\bw^{k_1};\bu^{k_1},\bv)$   by $a(\bu^{k_1}_{i-1}-\bw^{k_1}_{i-1};\bu^{k_1}_i,\bv)$,  $i=1,2,\dots,$  and iterate until the residual of the   Navier--Stokes equations~\eqref{tdiscnse1}  becomes less than $10^{-8}$. The unknown domain velocity $\bw_i^{k_1}$, $i=1,2,\dots,$ is calculated in each iteration by solving the linear elasticity 
problem for a given displacement virtually obtained using $\bu_i^{k_1}$.

At the end of the fixed point iteration,  we move the domain with the domain velocity $\bw^{k_1}$ to obtain $\Omega_{k_1}$.
We then solve the coupled surfactant equations~\eqref{tdiscCsurf1} and~\eqref{tdiscSsurf1} in $\Omega_{k_1}$ and $\Gamma_{k_1}$, respectively, by a Gauss-Seidel type  fixed point iteration as follows. Let $C^{k_1}_{\Gamma, 0}= C^{n}_{\Gamma}$, $C^{k_1}_{ 0}= C^{n}$, be the initial iterative values. Further, replace  $b_c(C^{k_1},C_\Gamma^{k_1}, \phi) $, $s_c(C^{k_1}_\Gamma,\phi)$
by $b_c(C^{k_1}_i,C_{\Gamma,i-1}^{k_1}, \phi) $, $s_c(C^{k_1}_{\Gamma,i-1},\phi)$  and $b_\Gamma(C^{k_1}_\Gamma,C^{k_1},\psi)$, $s_\Gamma(C^{k_1},\psi)$ by $b_\Gamma(C^{k_1}_{\Gamma,i},C_{i-1}^{k_1},\psi)$, $s_\Gamma(C_{i-1}^{k_1},\psi)$ in~\eqref{tdiscCsurf1} and~\eqref{tdiscSsurf1}, respectively, and iterate until the residual of~\eqref{tdiscCsurf1}  becomes less than $10^{-12}$. 
 
 In computations, the fixed point iteration of the Navier--Stokes satisfies the stopping criteria within two or three iterations for $\delta t$=$5\times 10^{-4}$, and the number of iterations increase when $\delta t$ is increased. In addition to the  dependency  on the time step, the number of Gauss-Seidel type  iteration of~\eqref{tdiscCsurf1} and~\eqref{tdiscSsurf1} depends on $\alpha$ and $\Bi$.

\subsubsection{Finite element discretization}
We assume that the droplet impingement is 3D-axisymmetric, and   we rewrite the volume and surface integrals in \eqref{weakNSE}-\eqref{WEAKCOUPSURF} into area and line integrals using the cylindrical coordinates as described in~\cite{GAN07}. It allows to use two-dimensional finite elements for approximating the velocity, pressure and bulk surfactant concentration on the cross-section and a one-dimensional finite elements for approximating surfactant concentration on free surface and liquid-solid interface.  We triangulate the cross-section with triangles, and   use the inf-sup stable isoparametric Taylor-Hood finite elements, that is, continuous piecewise quadratic polynomials  and continuous piecewise linear polynomials for  the appropriation of the velocity components and  pressure, respectively. Moreover, we use the continuous piecewise quadratic polynomials for the approximation of the  surfactant concentrations in the bulk and on the interfaces.

\subsection{Mesh handling}
The mesh velocity needs to be computed in each fixed point iteration step of the Navier--Stokes equations, see Section~\ref{secnonlin}. To compute the mesh velocity, we first obtain the displacement of the boundary using $\bw=\bu$ on the free surface that satisfies the  kinematic condition $\mathbf u\cdot\bnu_1=\bw\cdot\bnu_1$ given in \eqref{bc}. We then solve the linear elasticity equation for the displacement of the inner mesh points with the obtained boundary displacement as boundary value. For instance, to calculate the mesh velocity $\bw^{n+1}$, let the boundary displacement obtained from the Navier--Stokes equations be $\Upsilon^{n+1}$,  then the displacement $\Psi^{n+1} $ is calculate by solving
\begin{equation}\label{Elast}
\begin{array}{rcll}
\vspace{2mm}
\nabla\cdot\mathbb{T}(\Psi^{n+1}) & = & 0 & \mbox{in } {\Omega}(t_n) \\
\Psi^{n+1}  & = & \Upsilon^{n+1}   & \mbox{on }\Gamma_1\cup\Gamma_2\\
\end{array}
\end{equation}
where 
$
 \mathbb{T}(\phi) = \lambda_1(\dive\phi)\mathbb{I} + 2\lambda_2 \mathbb{D}(\phi),
$
In computations, the Lame constants $\lambda_1$ and $\lambda_2$ are chosen as one. Further,  continuous piecewise linear polynomials are used to approximate each component of the displacement vector.  

During the mesh movement,  vertices on the free surface may accumulate at some part of the boundary due to the tangential movement induced by the Marangoni convection. To avoid remeshing, we verify the ratio of the minimum and maximum edge size on the free surface, and redistribute the vertices using interpolated cubic spline. To incorporate the redistribution, we add the tangential displacement vector that requires to redistribute the vertices in the mesh velocity calculation during the nonlinear iteration of the Navier--Stokes equations. This approach will automatically redistribute the vertices on the free surface during the mesh update.  However, the free surface vertices may become a part of the liquid-solid interface due to rolling motion. In this case, the free surface boundary  condition has to be replaced with the slip with friction boundary condition, and the free surface vertex become the wetting point. Consequently, the finite element spaces have to be reconstructed. Also, the surface meshes of $\Gamma_1$ and $\Gamma_2$ change during this process, and need new finite element spaces.  The entire process is handled automatically by mapping the old solution to the new finite element spaces without remeshing, as the number of finite element degrees of freedom (unknown solution coefficients) do not change during the change in boundary description. Moreover, the minimum angle of the triangular mesh is calculated at every time step, and a remeshing will be done when the minimum angle of the mesh is less than $10^\circ$.  During the remeshing, the old solutions are interpolated to the new mesh. To minimize the interpolation error, the Navier--Stokes equations are solved with the interpolated velocity as an initial guess and $\bw=\mathbf 0$  before advancing to the next time step. Note that the remeshing is not necessary at every time step, as the inner mesh points are moved using the elastic mesh update.

\section{Validation}
Simulations  of impinging droplets without surfactants (clean droplets) using the proposed ALE finite element method have been compared with the experimental results in our previous studies~\cite{SG06,GAN13M,GRT14}. Further, the numerical scheme for transport of surfactants in bulk and on the interface/free surface has been validated with analytical solutions in~\cite{GAN09JCP,GAN12}. Numerical studies of surfactant droplet impingement using the proposed numerical scheme with the surfactant-dependent contact angle model are carried out here. We first perform a mesh convergence study using the impinging droplet configuration, in particular, with  surfactant-dependent contact angle. Computational results of the clean droplets are also compared with the experimental results. Moreover the effects of adsorption and desorption coefficients on the flow dynamics of wetting and non-wetting droplets are studied. In these numerical studies, the relative mass fluctuation
of the droplet  over time is given by 
\begin{align*}
 \delta_V(t) =   \frac{|\Omega(t)| -
   |\Omega(0)|}{|\Omega(0)|}, \qquad
 |\Omega(t)|={\int_{\Phi(t)} r\,dr\,dz}, 
\end{align*}
where $r,~z$ are radial and axial coordinates in axisymmetric domain, and $\Phi(t)$ is the axisymmetric meridian domain of $\Omega(t)$.
The above integrals are evaluated in the axisymmetric configurations~\cite{GAN07}. Further, the  relative fluctuation of the total surfactant mass  is
computed by 
\begin{align*}
\delta^\Gamma_c(t) =  \frac{ M(t) - M(0)}{M(0)}, \qquad M(t) = \int_{\Phi(t)} C\, r\,dr\,dz ~+~ Da \int_{\Gamma_1(t)\cup\Gamma_2(t)} C_\Gamma \,r\,ds.
\end{align*}
In addition, the sphericity   of the droplet is calculated using
\[
\text{sphericity} = \displaystyle\frac{\text{surface area of the volume-equivalent sphere} }{\text{surface area of the droplet}}.
\]
It implies that the sphericity will be one when the droplet is in spherical shape, and the sphericity will be less than one when   the droplet deforms. The sphericity gives a quantitative measure of the droplet deformation. Further, the kinetic energy in the droplet is calculated  using
\begin{align*}
\text{kinetic energy} =\frac{\displaystyle\int_{\Phi(t)}~r~\bu \cdot\bu ~dr~dz}{\displaystyle\int_{\Phi(t)} ~r~dr~dz}.
\end{align*}
Note that the dimensionless velocity is used in the above definition, and it has to be multiplied with $U^2$ in order to get the kinematic energy in the dimensional form.

\subsection{Mesh convergence}
A mesh convergence study is performed  for the numerical scheme presented in the previous section. We consider a hemispherical droplet of diameter $d_0$=$1.29\times 10^{-3} $~m on a horizontal surface. The initial surfactant concentration   on the free surface is assumed to be uniform, that is, $C_{\Gamma_1}(\cdot,0)=0.5$, and the no flux is imposed at the moving contact line. Further, the liquid-solid interface is assumed to be clean, that is,  $C_{\Gamma_2} =0$, and remains clean during computations due to the no flux condition at the contact line. Moreover, we assume that $C=0$  and there is no transport of surfactants between the free surface and the bulk phase, that is, no adsorption or desorption of surfactants, $\alpha_1=0$ and $\Bi_1=0$. Using $L$=$d_0$  and $U$=$1.18$~m/s  as characteristic values, we get $\text{Re}$=$1522$, $\Web$=$25$ and $\Fro$=$110$. Further, we used $\Pe_1=1$,  $\delta t =0.00025$ and $\beta=0.55/h_E$, where $h_E$ is the edge size of the liquid-solid interface. The imposed   contact angle of the clean free surface,  $\theta_e^0=110^\circ$. Since $\theta_e^0$ is more than $90^\circ$, the presence of surfactant on the free surface will increase the equilibrium contact angle further, see~\eqref{casurf3}. The initial mesh level (L0) contains 25 vertices on the free surface with  $h_E=0.06282152$, and the successive mesh levels are obtained by uniformly refining the initial mesh, that is,  L1, L2 and L3 meshes contain 50, 100 and 200 vertices on the free surface, respectively.
\begin{figure}[t]
\begin{center}
\unitlength1cm
\begin{picture}(15, 5.5 )
\put(-0.5,-0.5){{\includegraphics[width=7.5cm]{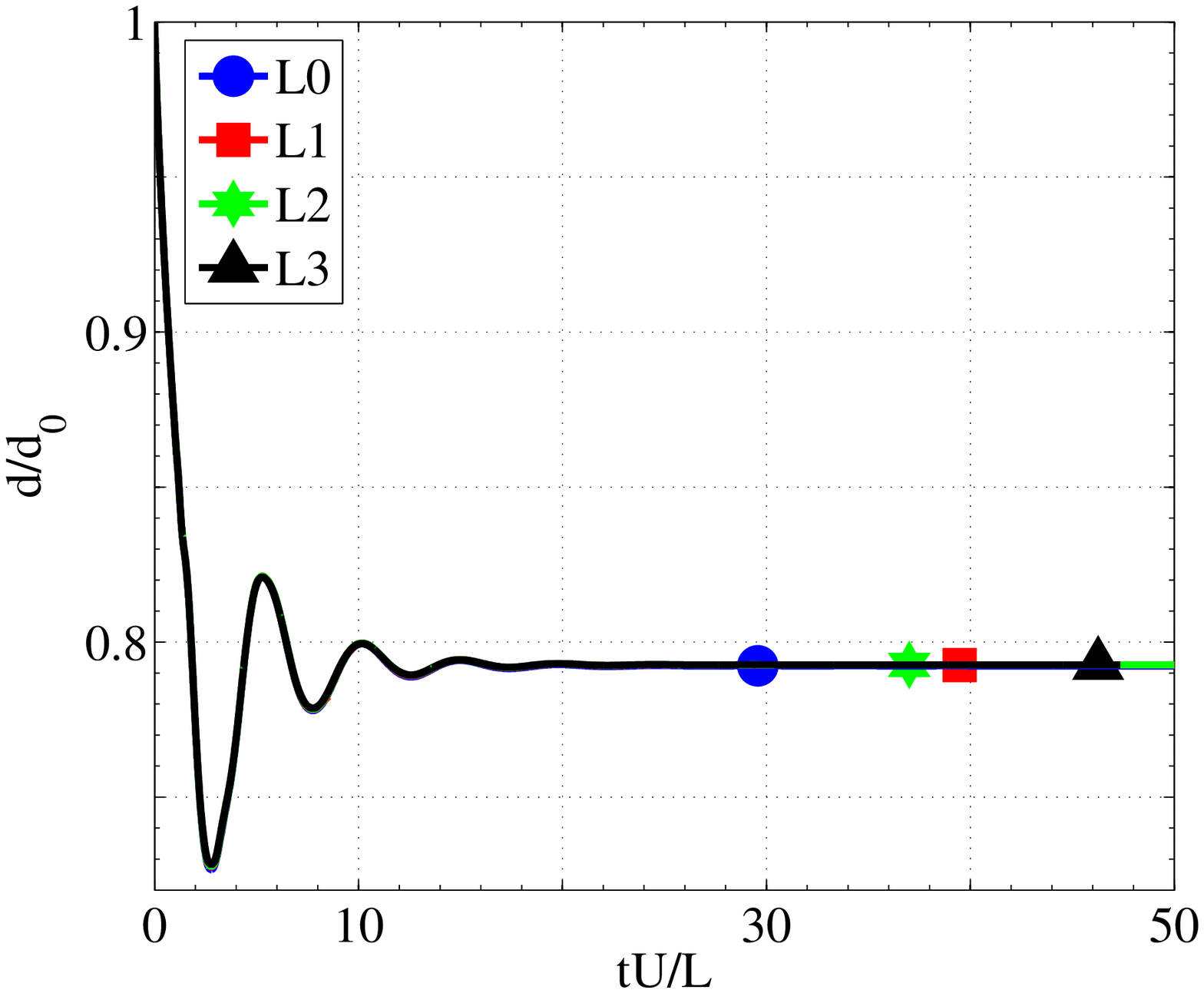}}}
\put(2.25 , 1.85){{\includegraphics[width=4. cm]{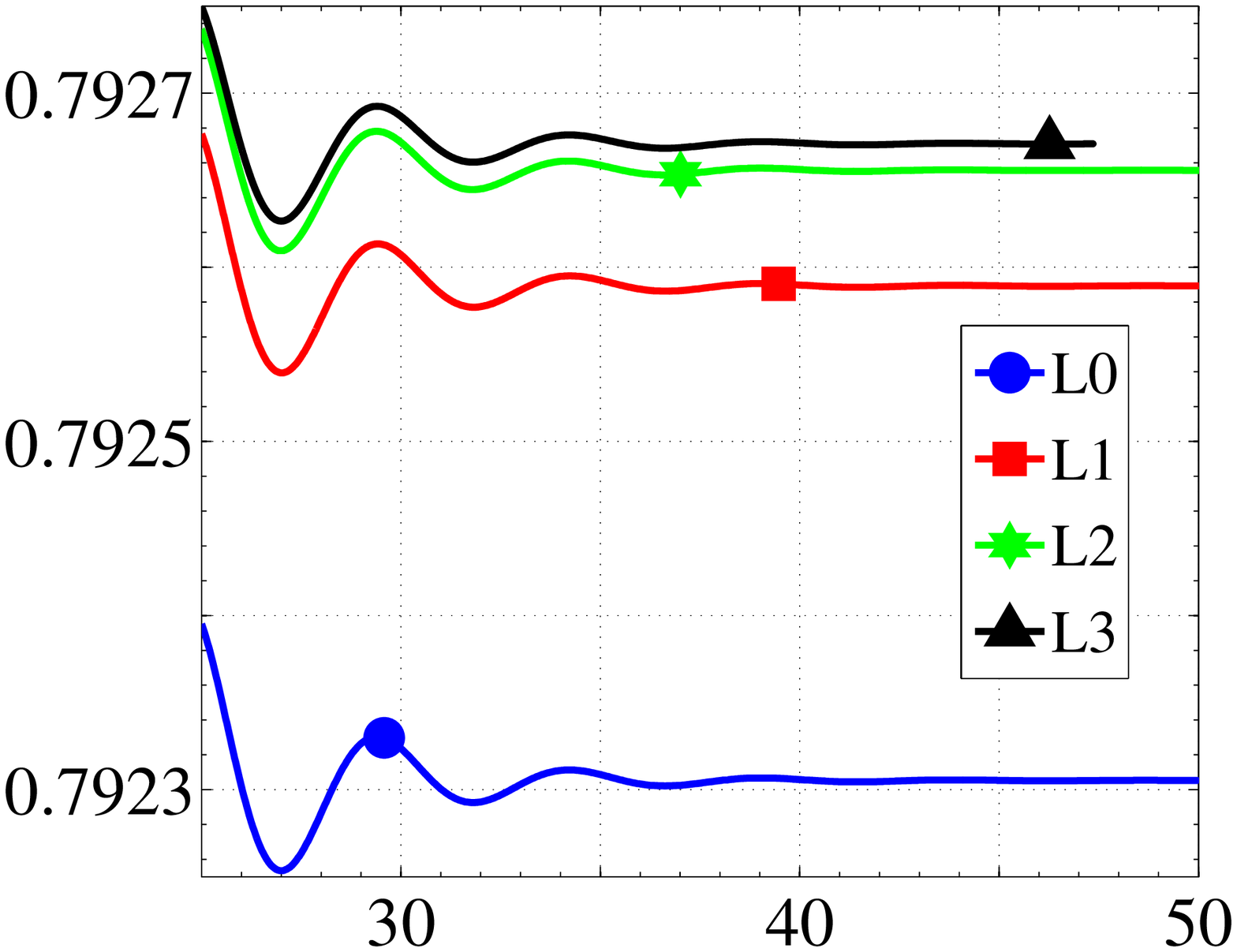}}}
\put(7.75,-0.5){{\includegraphics[width=7.5cm]{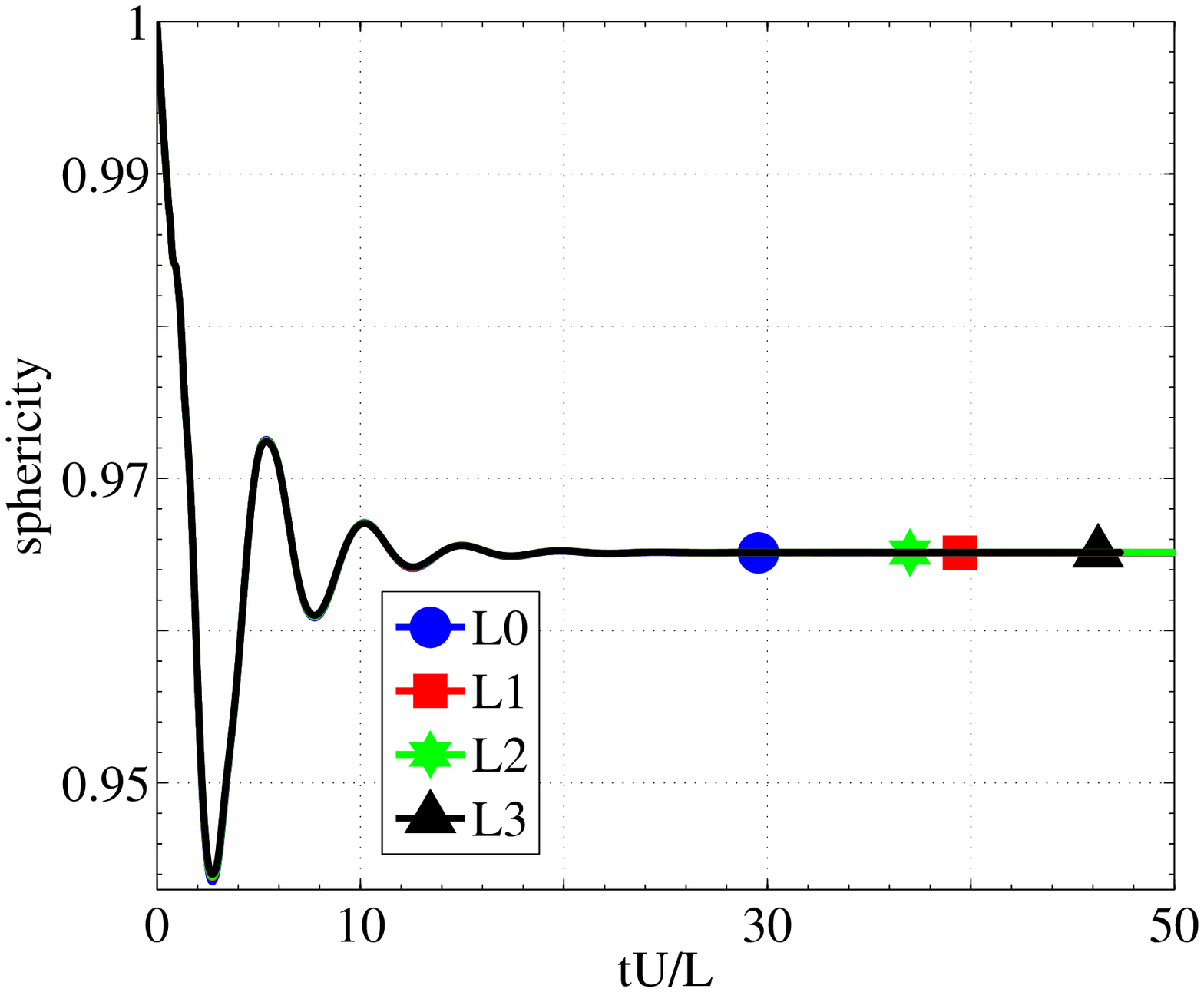}}}
\put(11., 2.25){{\includegraphics[width=3.5cm]{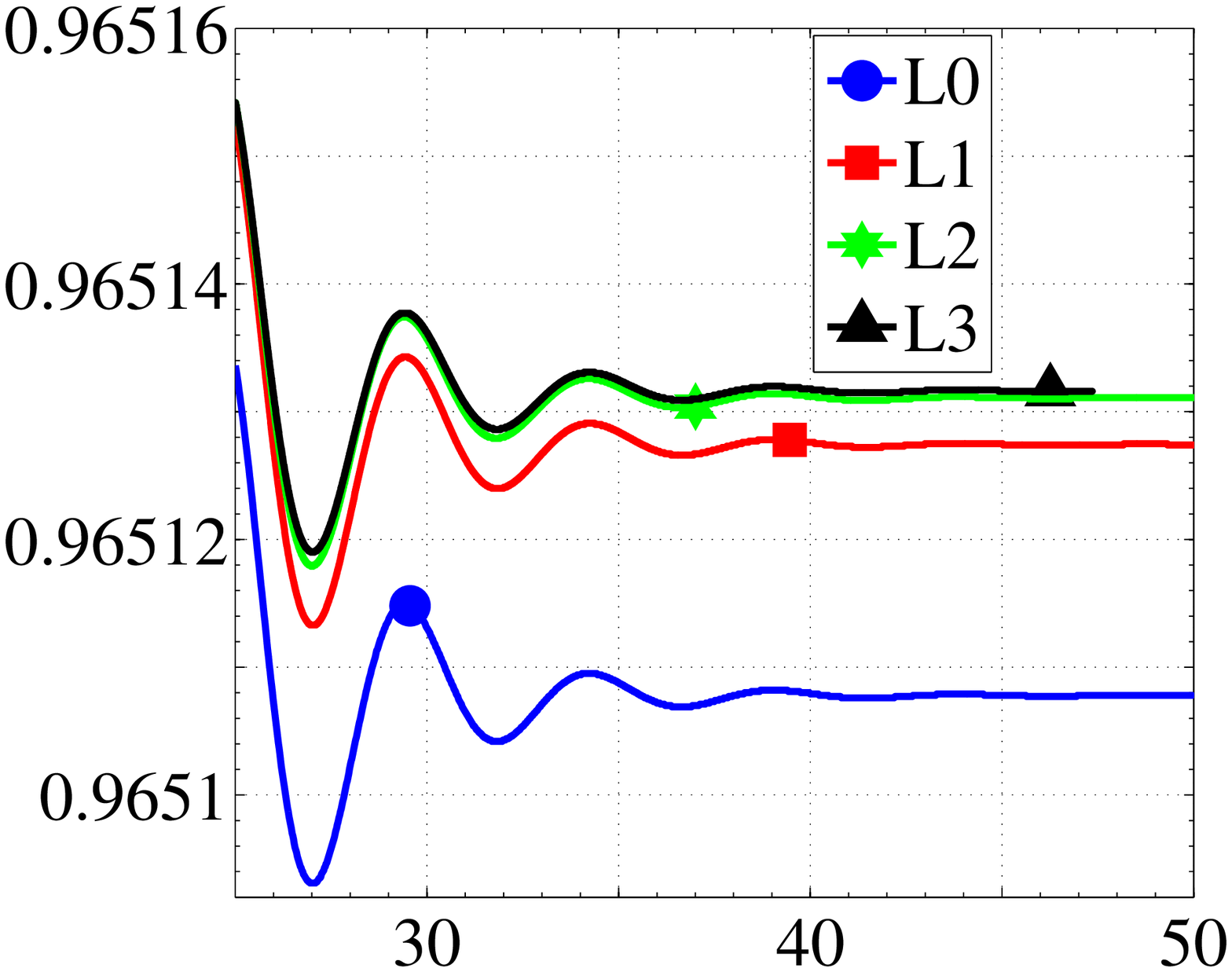}}}
\end{picture}
\end{center}
\caption{Wetting diameter (left) and the sphericity (right) of the droplet in different mesh levels of the mesh convergence study. $\text{Re}$=$1522$, $\Web$=$25$, $\Fro$=$110$, $\alpha_1=0$, $\Bi_1=0$, $\Pe_1=1$ and  $\theta_e^0$=$110^\circ$.
\label{convera}}
\end{figure}

\begin{figure}[t!]
\begin{center}
\unitlength1cm
\begin{picture}(15, 5.5 )
\put(-0.5,-0.5){{\includegraphics[width=7.5cm]{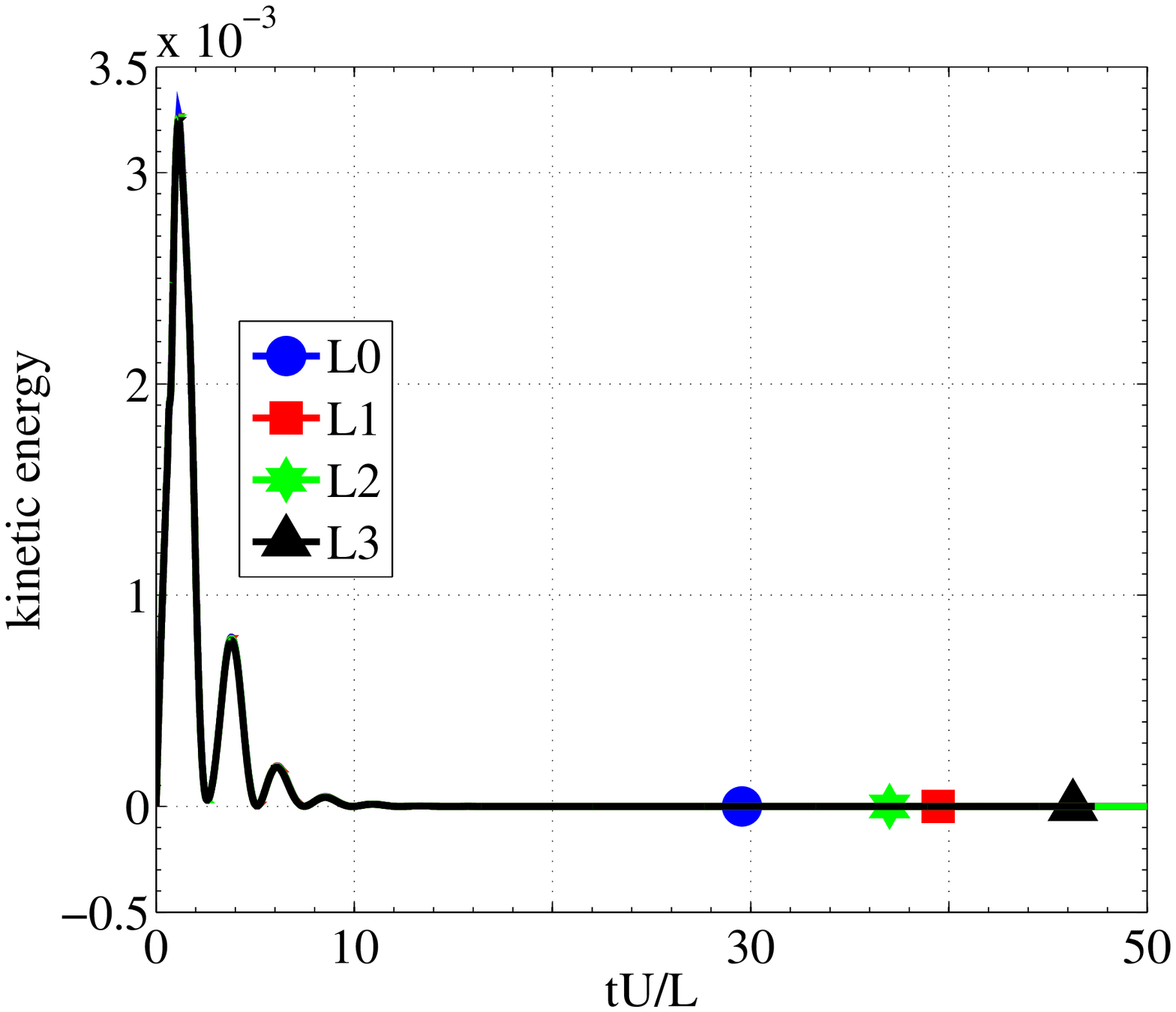}}}
\put(2.25 , 1.75){{\includegraphics[width=4. cm]{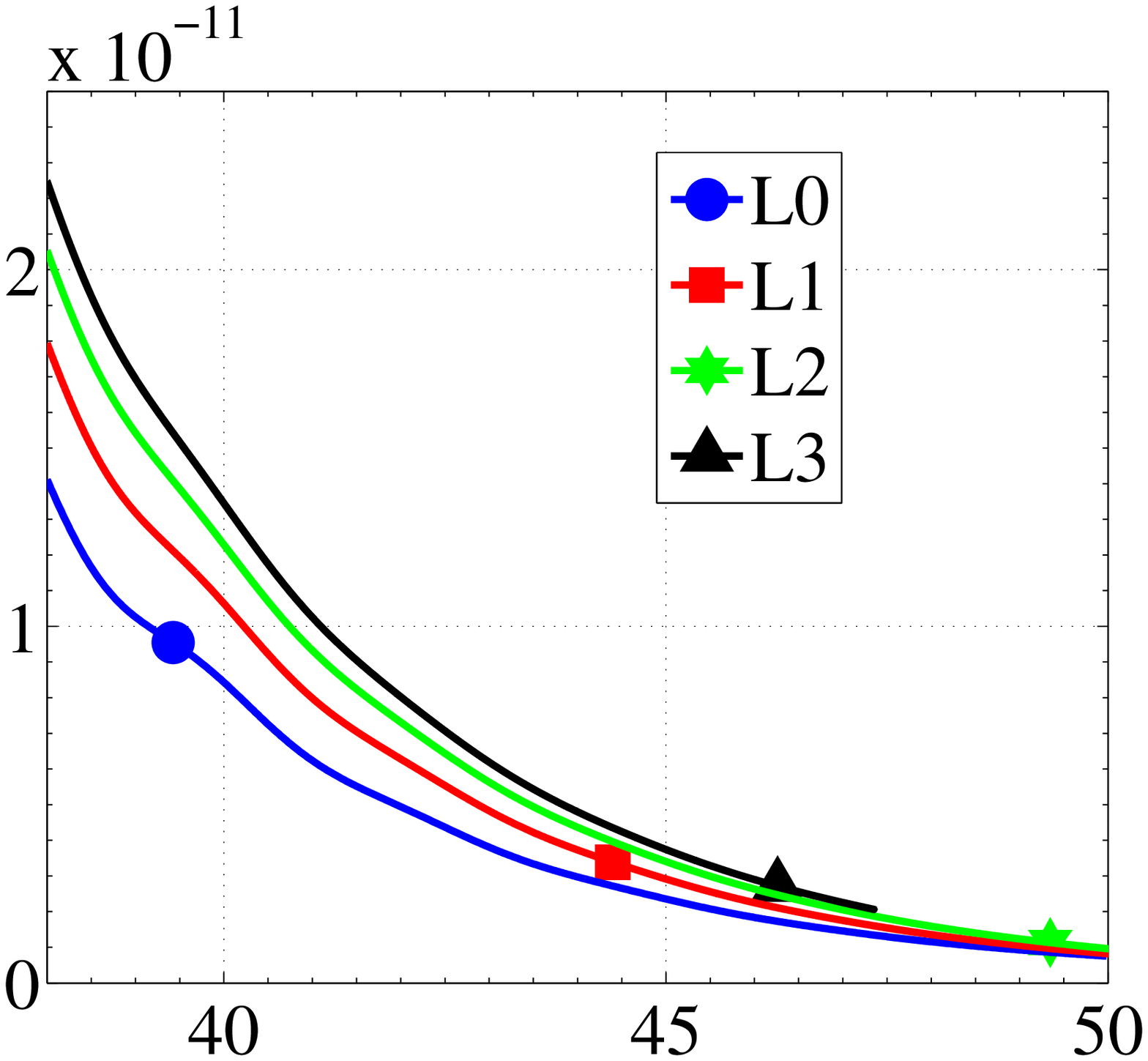}}}
\put(7.75,-0.5){{\includegraphics[width=7.5cm]{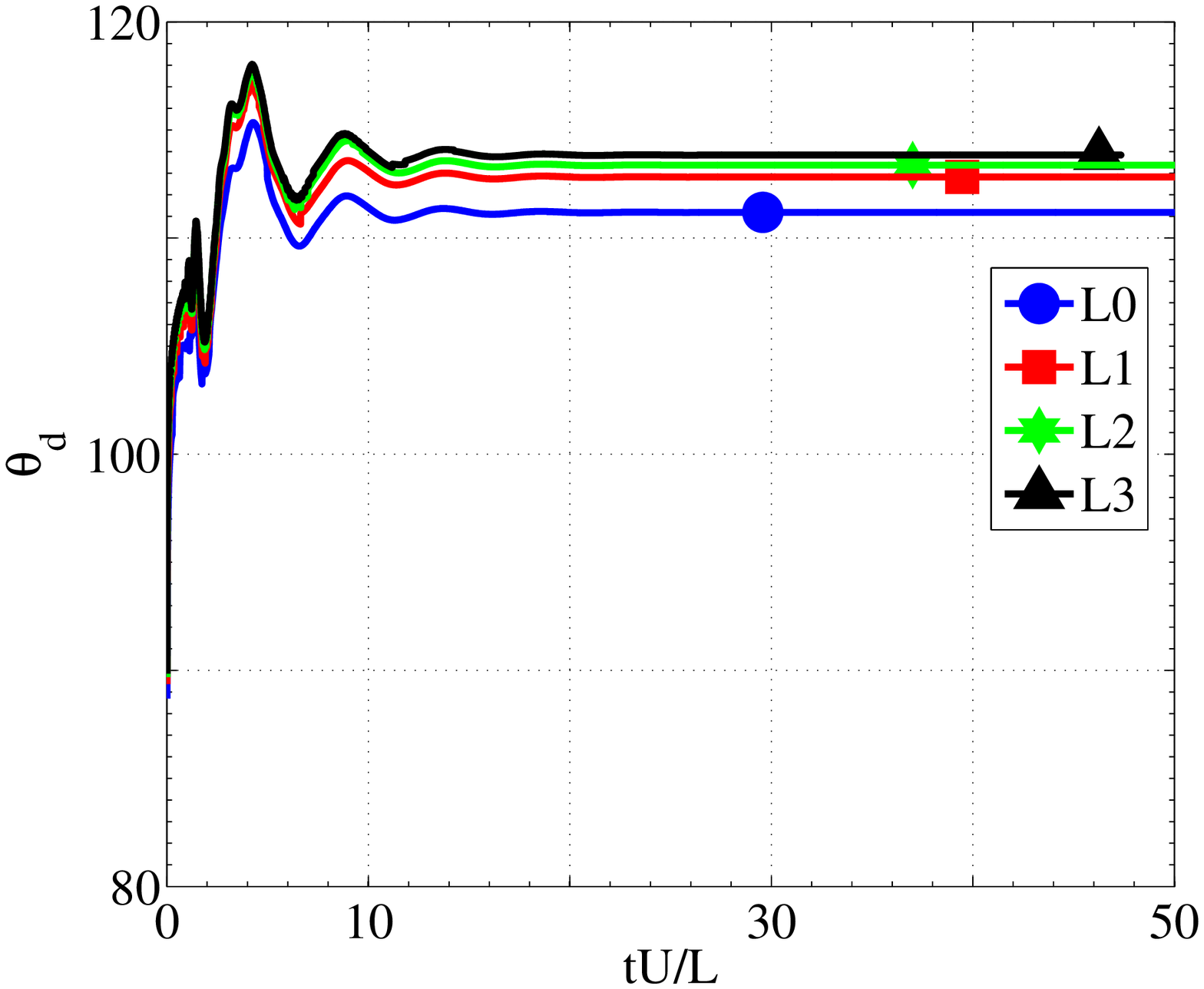}}}
\put(9.25, 0.25){{\includegraphics[width=4cm]{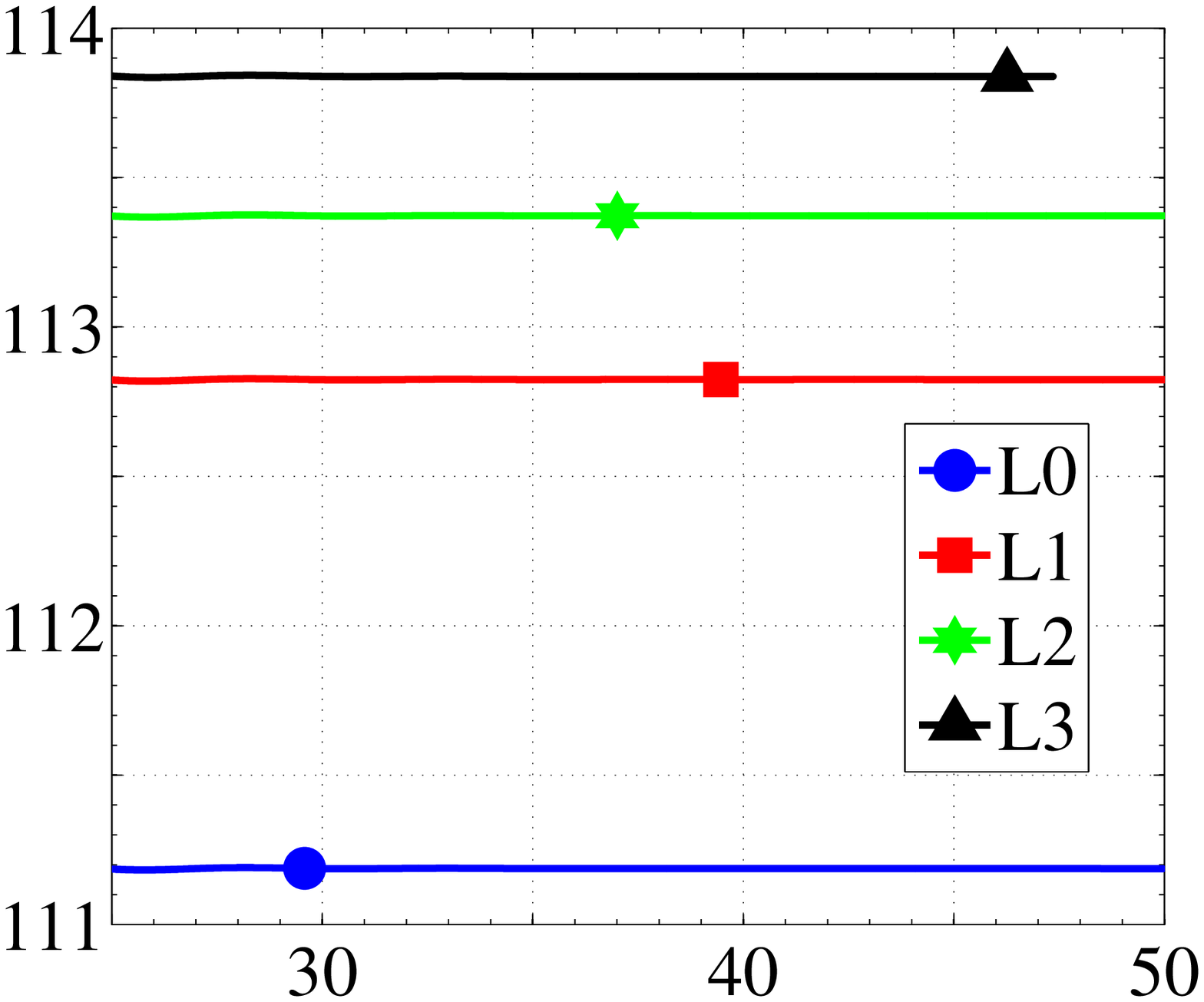}}}
\end{picture}
\end{center}
\caption{Kinetic energy (left) and the dynamic contact angle (right) of the droplet in different mesh levels of the mesh convergence study. $\text{Re}$=$1522$, $\Web$=$25$, $\Fro$=$110$ $\Bi=0$, $\Pe_1=1$ and  $\theta_e^0$=$110^\circ$.
\label{converb}}
\end{figure}

\begin{figure}
\begin{center}
\unitlength1cm
\begin{picture}(15, 5.5 )
\put(-0.5,-0.5){{\includegraphics[width=7.5cm]{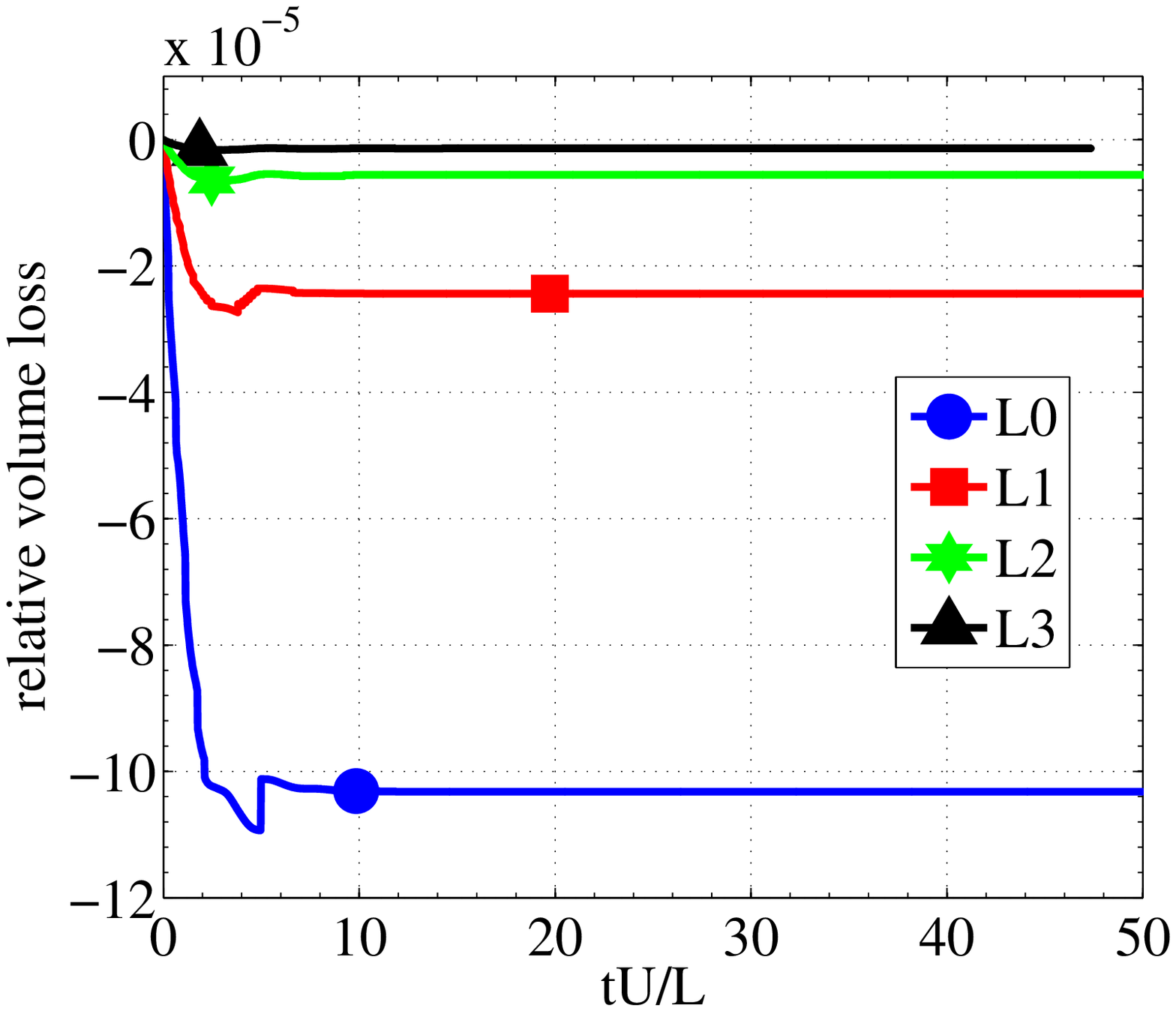}}}
\put(7.75,-0.5){{\includegraphics[width=7.5cm]{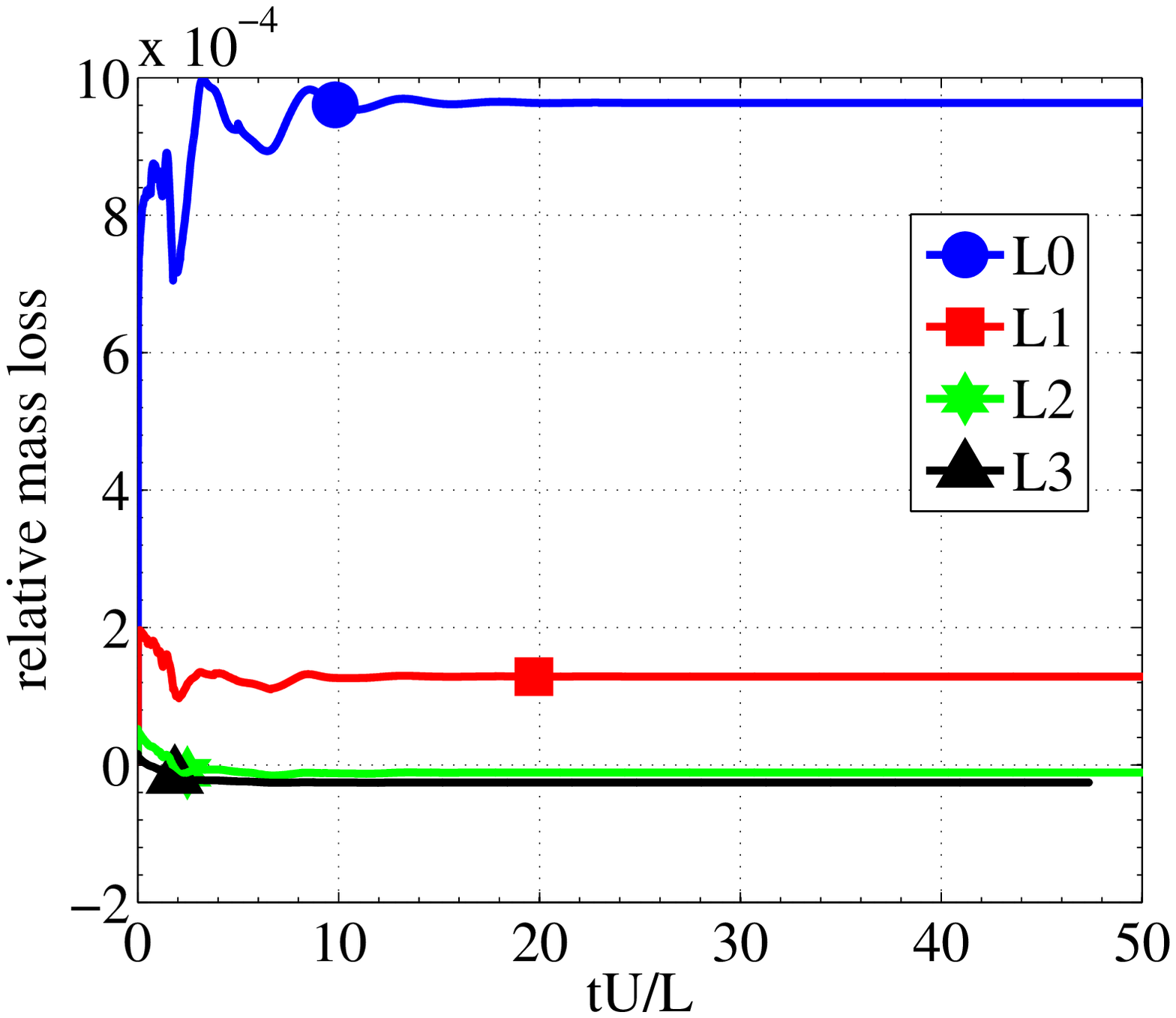}}}
\end{picture}
\end{center}
\caption{Relative mass loss of droplet (left) and the relative mass loss of surfactants (right) in different mesh levels of the mesh convergence study. $\text{Re}$=$1522$, $\Web$=$25$, $\Fro$=$110$ $\Bi=0$, $\Pe_1=1$ and  $\theta_e^0$=$110^\circ$.
\label{converc}}
\end{figure}

Initially, the dynamic contact angle and the wetting diameter are not in the equilibrium state for the chosen parameters, and thus the droplet starts to deform and   attains its equilibrium state after a sequence of recoiling and  spreading. Figure~\ref{convera} shows the wetting diameter (also the position of the contact line, since $d/d_0 = r/ r_0$) and the sphericity of the droplet  till the droplet attains its equilibrium state. The values obtained with different meshes are almost identical, and it shows that the free surface with the L0 mesh is enough to obtain a mesh independent   wetting diameter and   sphericity. 
However, a close view in the sub-figures shows the convergence behavior clearly, and the solution obtained with L2 and L3 are very similar.
The dynamics of the kinetic energy and the  contact angle of the droplet during the sequence of recoiling and spreading are presented in Figure~\ref{converb} for all mesh levels. The dynamic phenomenon observed in computations of dynamic  contact angle supports the earlier discussion, see sections~\ref{dyca} and \ref{surfca}. The initial dynamic  contact angle is $90^\circ$, which  is different from  the equilibrium value for the given droplet configuration. The computationally obtained dynamic  contact angle increases initially to a maximum value ($\simeq 118^\circ$), and oscillates around its equilibrium value before attaining it. Further, the kinetic energy attains a maximum value   when the dynamic  contact angle differs from its equilibrium value, see Figure~\ref{converb}. The dynamic  contact angle value obtained with L0 mesh is different from the values obtained with other meshes. It clearly shows that a mesh with at least $h_E=0.03141076$ (L1 mesh) on free surface is needed for a mesh independent solution, see the sub-figures in  Figure~\ref{converb}. Finally, the observed relative mass fluctuation  of the liquid droplet  and   of the surfactants  during the computations are presented in Figure~\ref{converc} for all mesh levels. Except the L0 mesh, the mass fluctuations in all other mesh levels are similar and very less. Further, we can observe the convergence behavior clearly. Based on this mesh convergence study, we use the L2 mesh with $h_E=0.01570538$ on the free surface in all computations of droplet  impingement. 


\section{Computational examples of soluble  surfactant droplet impingement}
Effects of soluble surfactants on the flow dynamics of impinging droplets are studied in this section. In particular, numerical studies on the influence of adsorption and desorption coefficients on the flow dynamics of wetting, partially wetting and non-wetting droplets are performed.

\subsection{Influence of adsorption in wetting droplets ($\theta_e^0=46^\circ$)}\label{seccase01}
 \begin{figure}[t!]
\begin{center}
\unitlength1cm
\begin{picture}(20,11.5)
\put(10,-2){\makebox(3,6){\includegraphics[trim=1.5cm 0cm 0.5cm 10cm, clip=true,width=7cm]{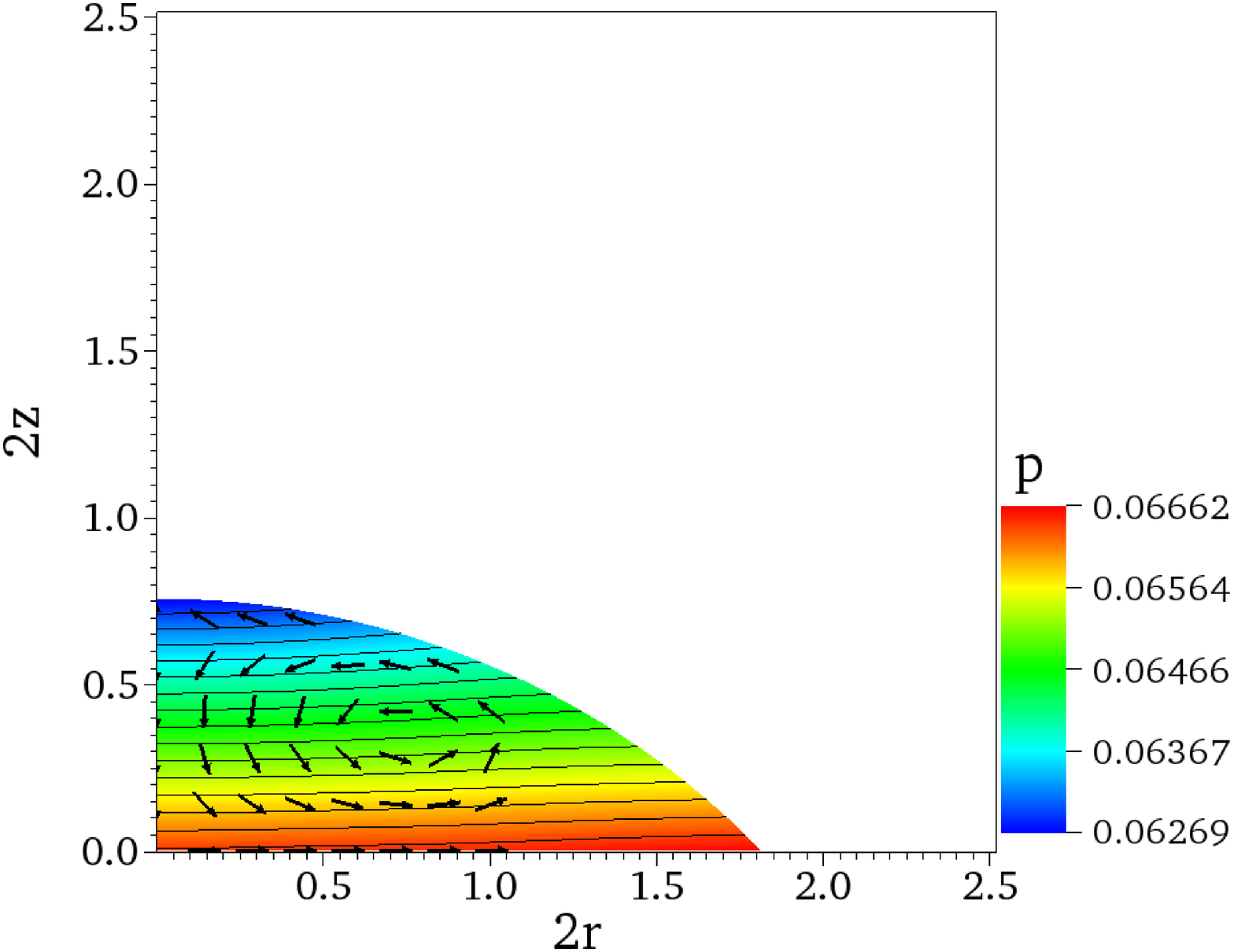}}}
\put(10,2){\makebox(3,6){\includegraphics[trim=1.5cm 0cm 0.5cm 10cm, clip=true,width=7cm]{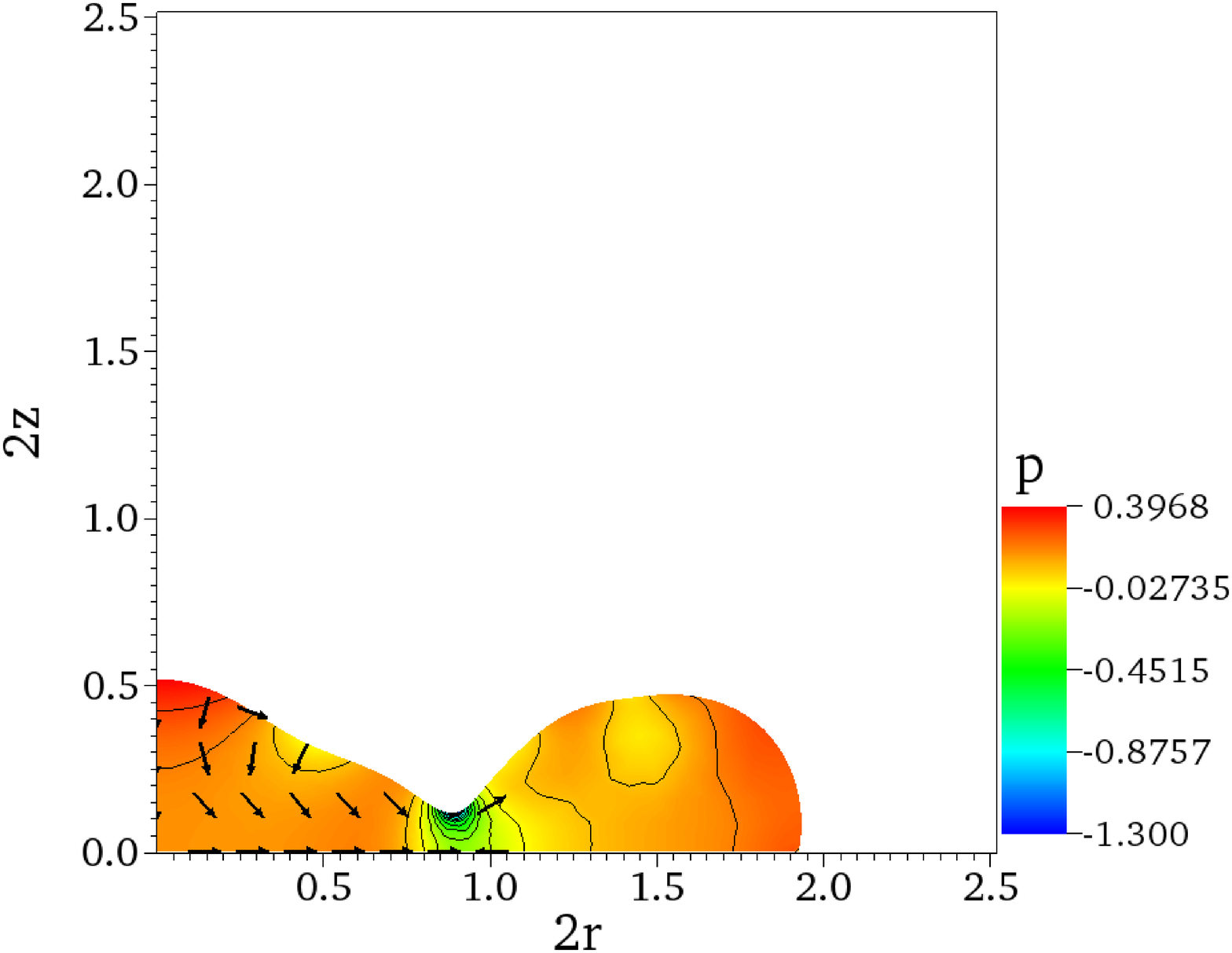}}}
\put(10,6.){\makebox(3,6){\includegraphics[trim=1.5cm 0cm 0.5cm 10cm, clip=true,width=7cm]{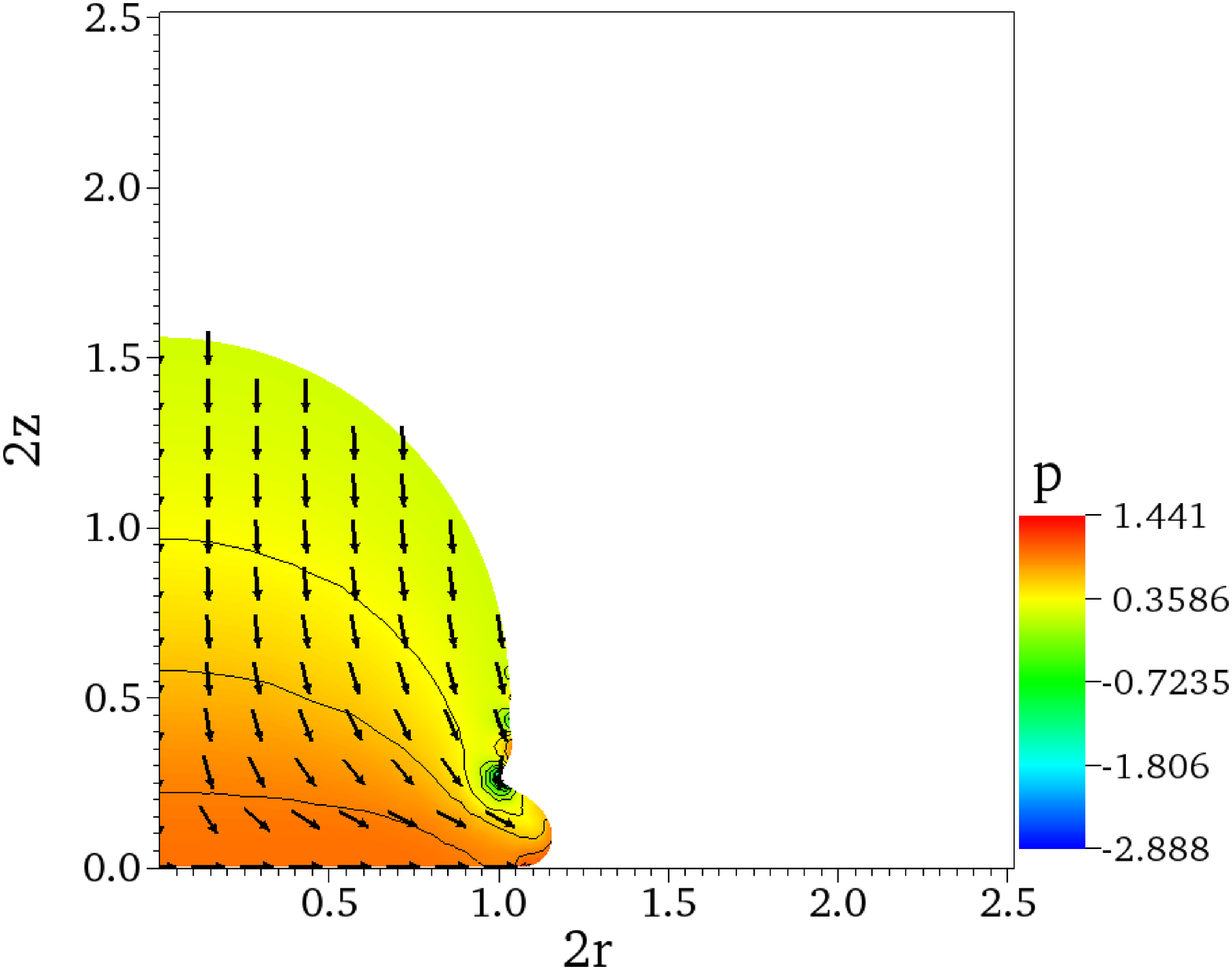}}}
\put(2,-2){\makebox(3,6){\includegraphics[trim=1.5cm 0cm 0.5cm 10cm, clip=true,width=7cm]{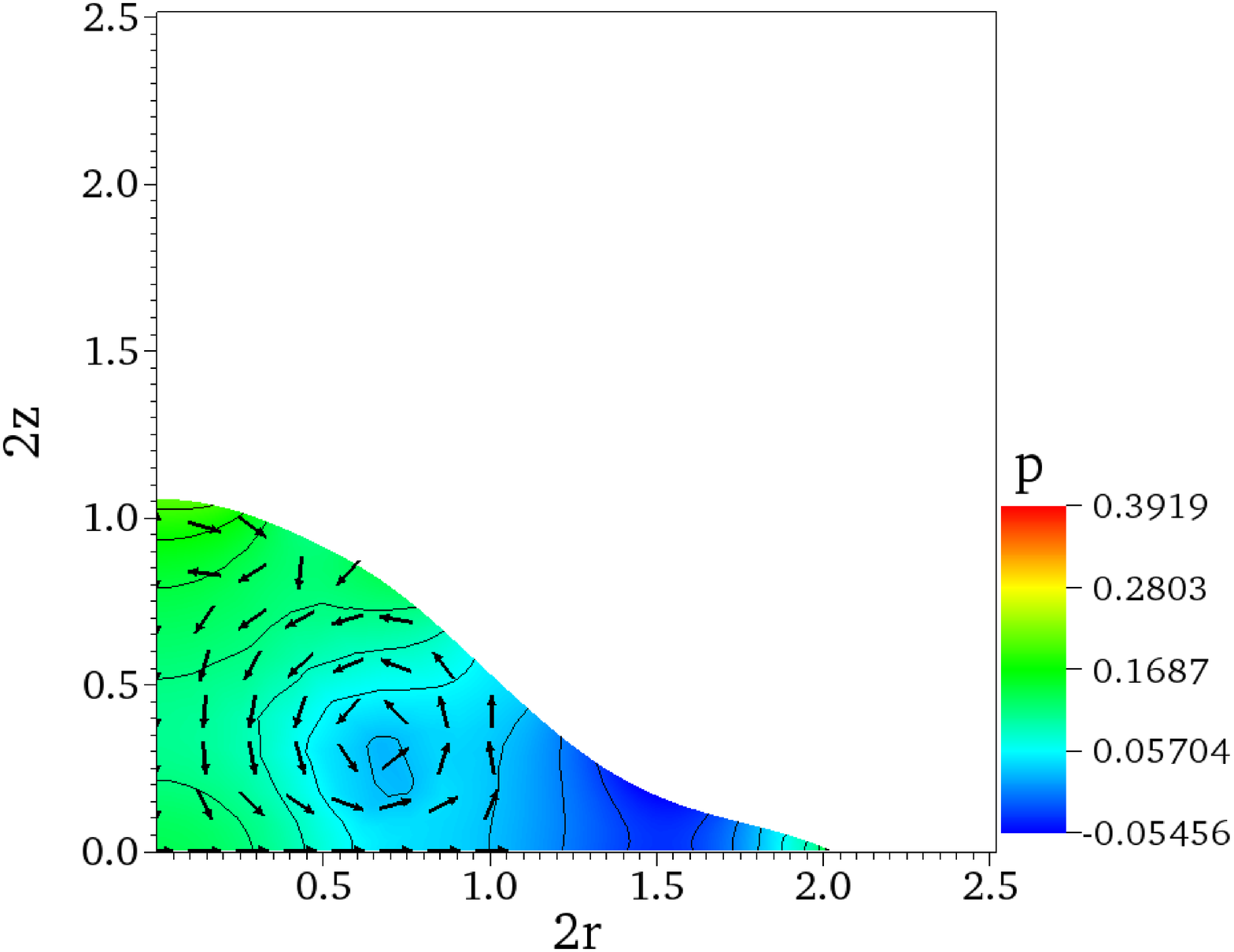}}}
\put(2, 2){\makebox(3,6){\includegraphics[trim=1.5cm 0cm 0.5cm 9cm, clip=true,width=7cm]{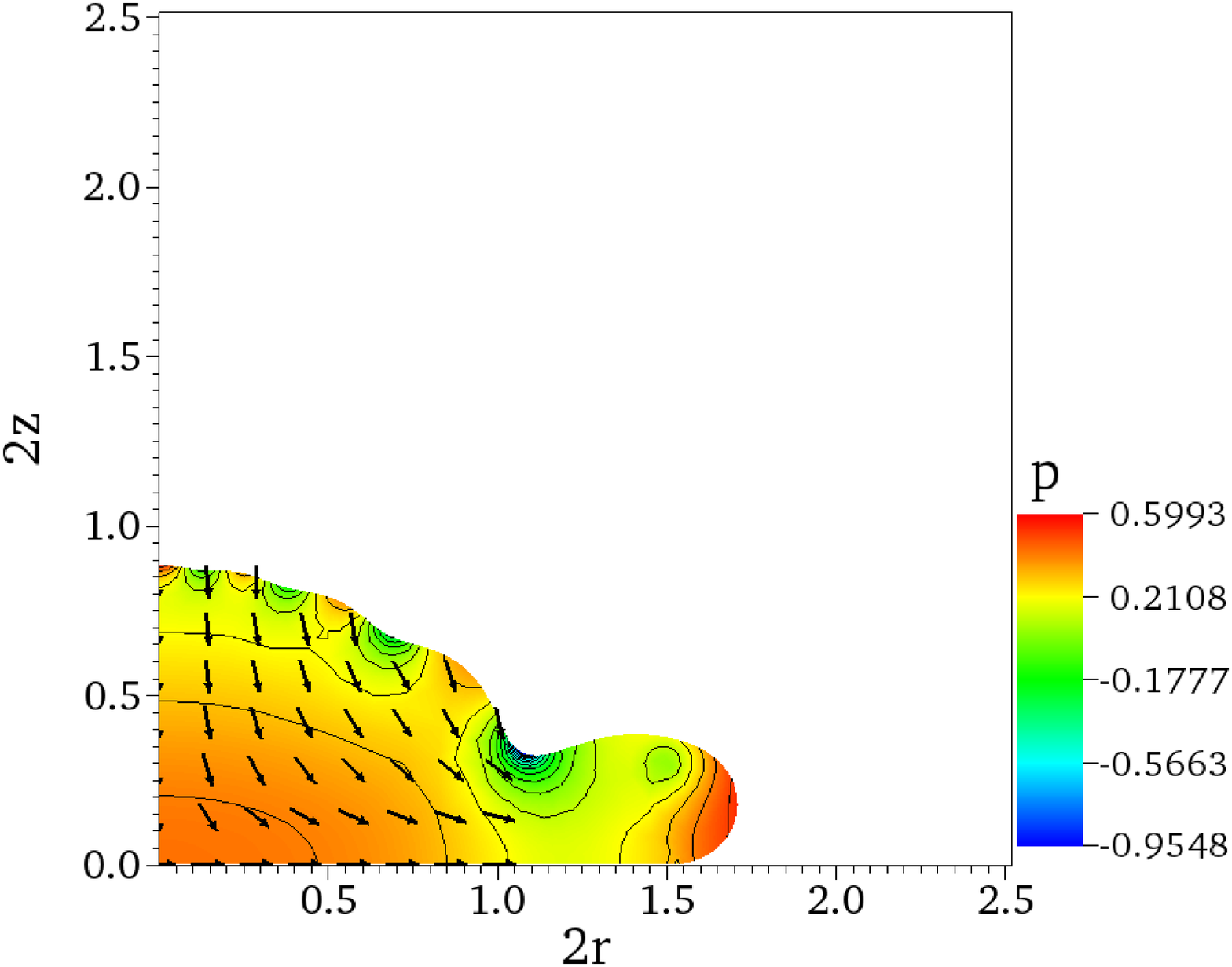}}}
\put(2, 6.){\makebox(3,6){\includegraphics[trim=1.5cm 0cm 0.5cm 7cm, clip=true,width=7cm]{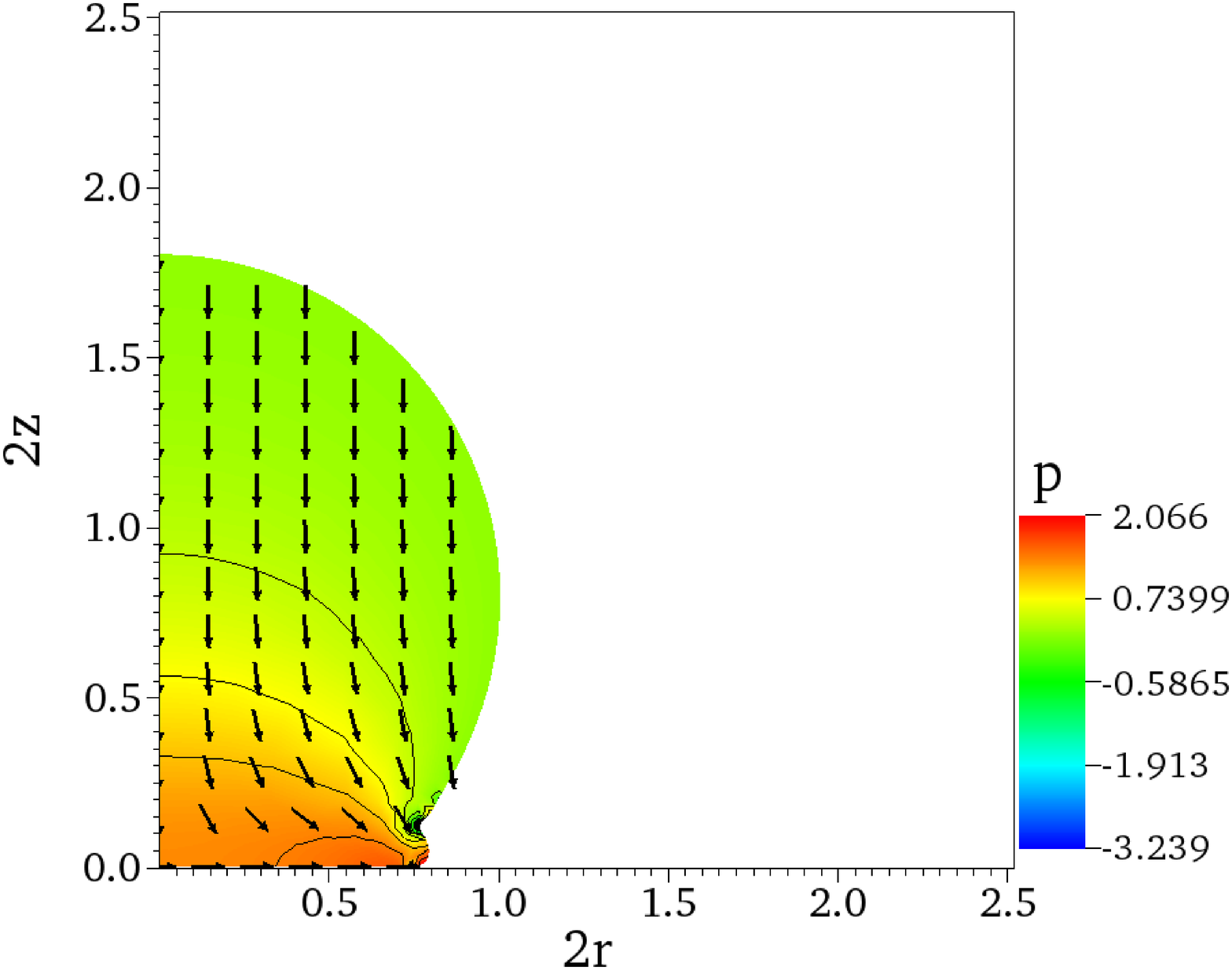}}}
\put(3,10.5){$t=0.1$}
\put(3,6.5){$t=0.625$}
\put(3,2.5){$t=3.75$}

\put(11,10.5){$t=0.215$}
\put(11,6.5){$t=0.125$}
\put(11,2.5){$t=25$}

\end{picture}
\end{center}
\caption{Pressure contours and shapes of the impinging droplet at dimensionless time  $t$=0.1, 0.215, 0.625, 1.25, 3.75 and 25.  $\text{Re}$=$1522$, $\Web$=$25$, $\Fro$=$110$, $\Pe_c=2$,  $\Pe_\Gamma=2$, $\Bi=0$, $\Da=775$, $\alpha=100$ and  $\theta_e^0$=$46^\circ$.}
\label{pressure}
\end{figure}
We first study the influence of the adsorption coefficient on the flow dynamics of a soluble surfactant droplet impingement. We consider a spherical water droplet of diameter $d_0$=$1.29\times 10^{-3}$~m impinging  with the pre-impact speed   $u_{imp}$ =$1.18$~m/s, and initial surfactant concentrations, $c_0=2000$ and $C_\Gamma(x,0)=0$.  An equilibrium contact angle  $\theta_e^0$=$46^\circ$  has been observed in the experimental study~\cite{SHE10} for a  clean water droplet on a polished silicon surface, and it is used here.     Using $L$=$d_0$  and $U$=$1.18$~m/s as characteristic values, we get $\Rey$=$1522$, $\Web$=$25$ and $\Fro$=$110$. In this example, we consider the following three variants: (1)~$\alpha=1$,  (2)~$\alpha=10$ and  (3)~$\alpha=100$. The   computations of all variants are performed with  $\Pe_c=2$,  $\Pe_\Gamma=2$, $\Bi=0$, $\Da=775$, $\delta t =0.00025$ and  $\beta=1.476776/2h_E$. Even though the initial surfactant concentration on the liquid-solid interface is zero, it increases due to the adsorption of surfactants from the bulk phase and  the transport of surfactant from the free surface. The transport of surfactants from the free surface into the liquid-solid interface may occur due to the imposed continuity condition at the contact line \eqref{CONT} and the rolling motion of the droplet during spreading. Nevertheless, the effects of $C_{\Gamma_2}$ will only be on the  dynamic contact angle~\eqref{casurf1}. Since the   condition, $C_{\Gamma_1}=C_{\Gamma_2}$,  is imposed at the contact line, the dynamic contact angle~\eqref{casurf1} will further simplified to the form~\eqref{casurf3}.

\begin{figure}[t!]
\begin{center}
\unitlength1cm
\begin{picture}(15, 11.5)
\put(-0.5,5.75){{\includegraphics[width=7.5cm]{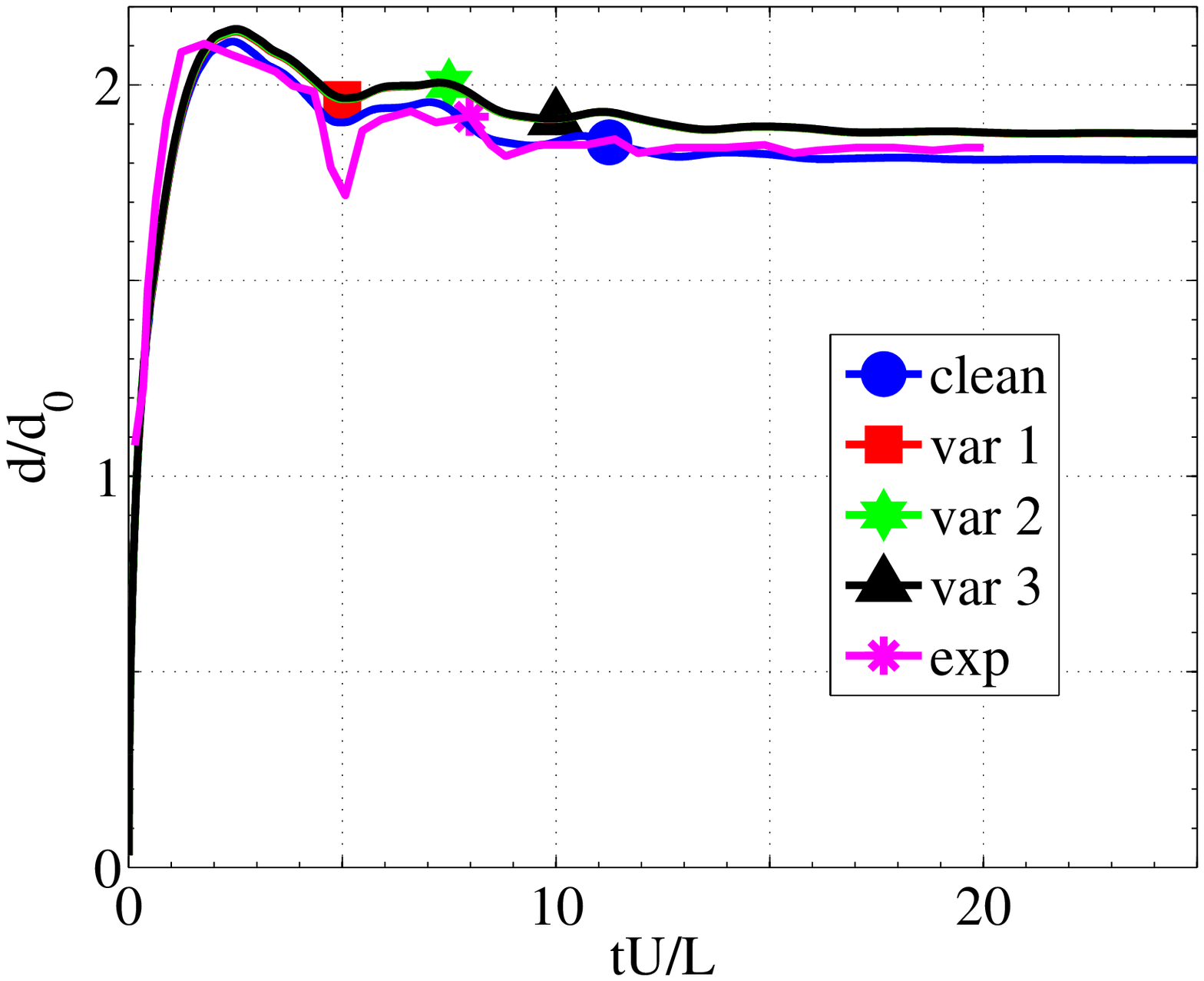}}}
\put(7.75,5.75){{\includegraphics[width=7.5cm]{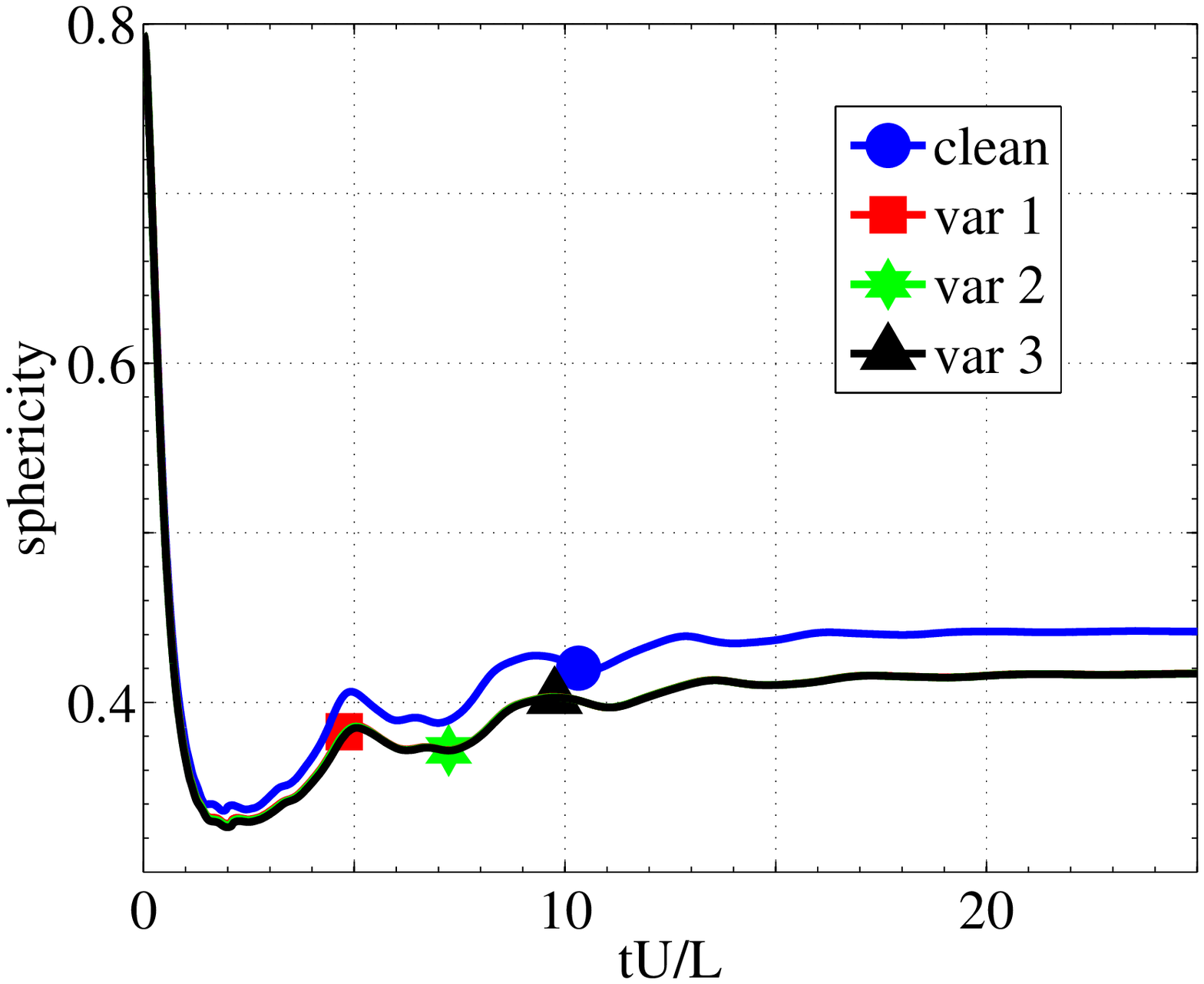}}}
\put(-0.5, -0.5){{\includegraphics[width=7.5cm]{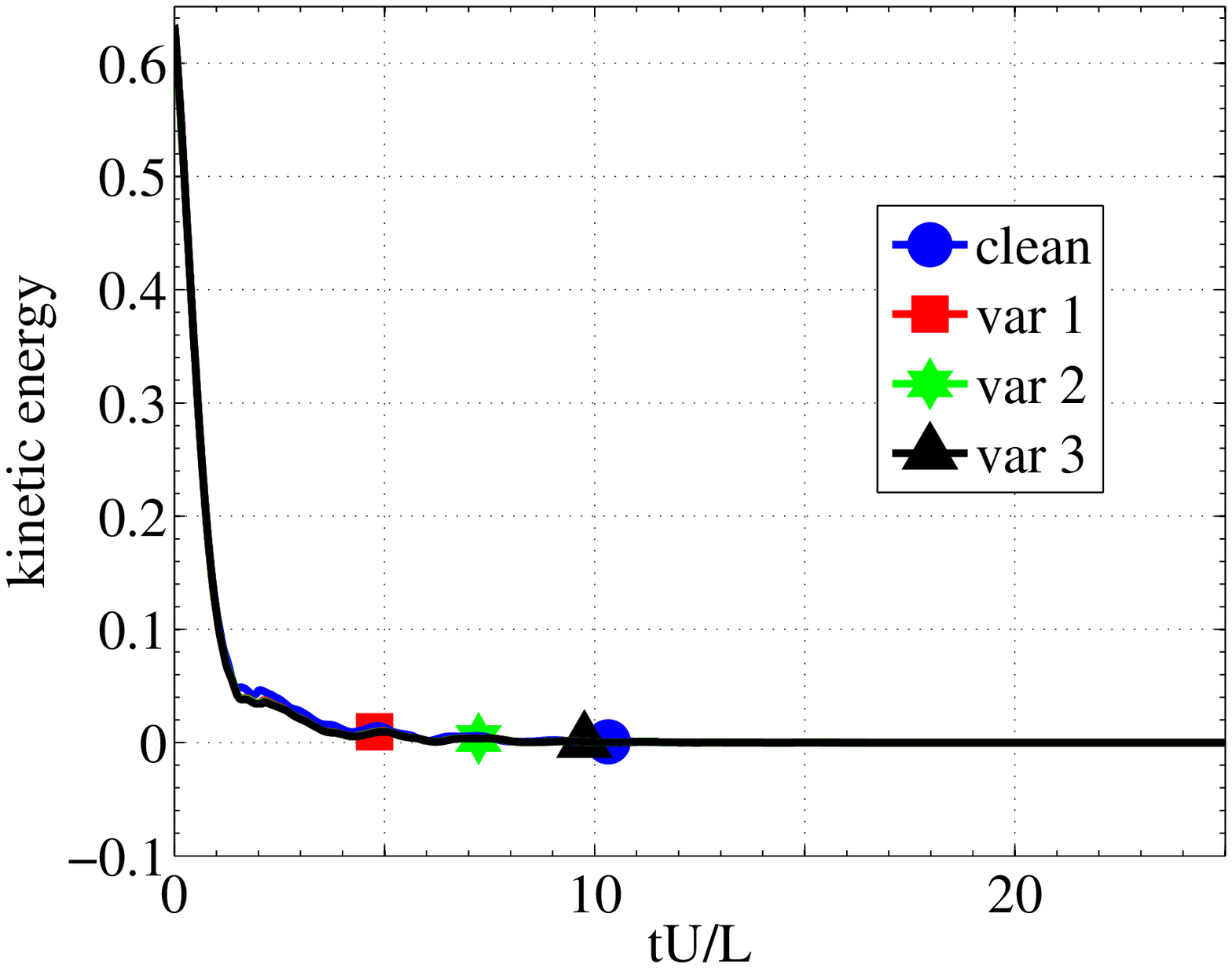}}}
\put(7.75, -0.5){{\includegraphics[width=7.5cm]{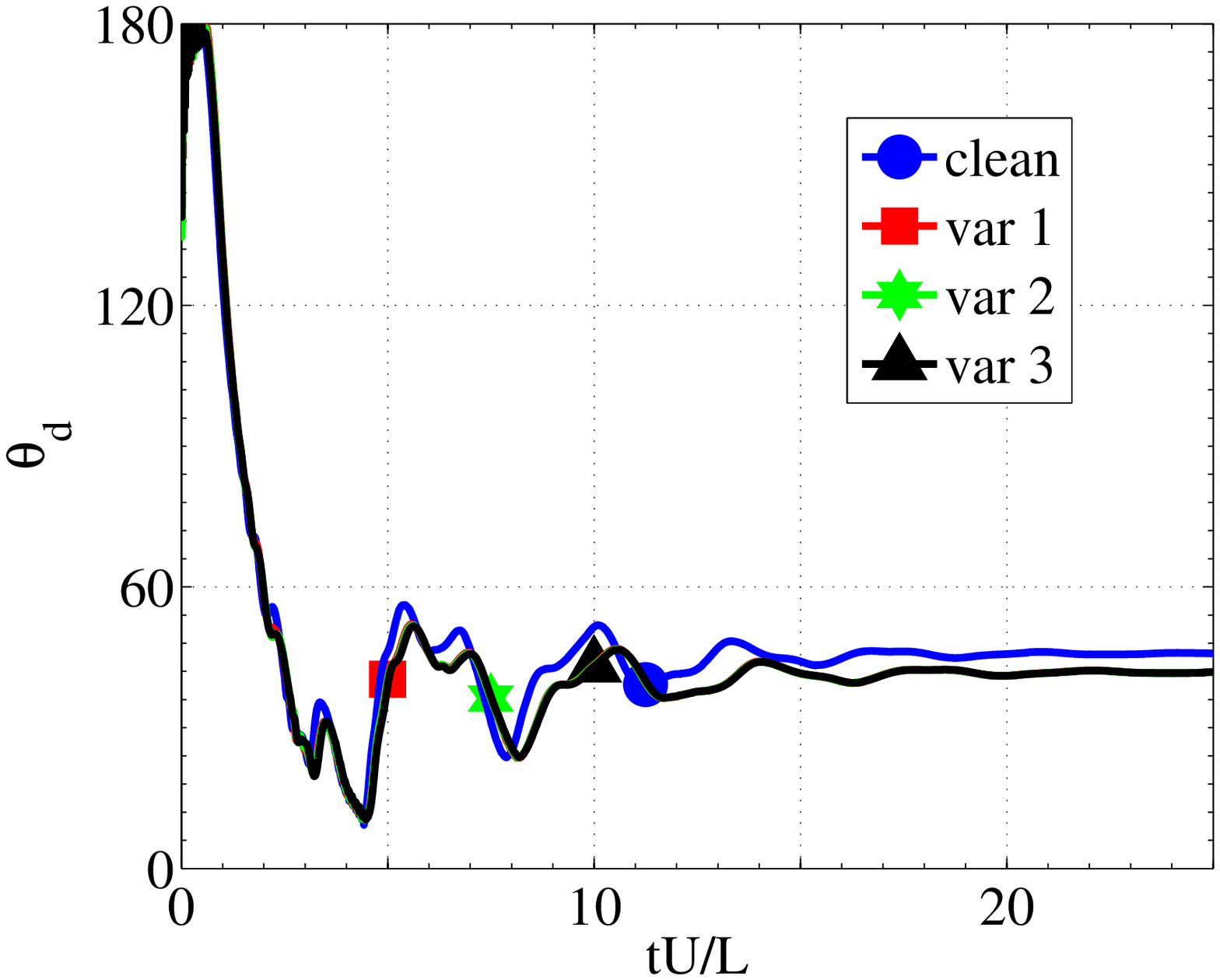}}}
\put(3.25,11.35){$(a)$}
\put(11.75,11.35){$(b)$}
\put(3.25, 5){$(c)$}
\put(11.75, 5){$(d)$}
\end{picture}
\end{center}
\caption{Effects of surfactant adsorption coefficient on the wetting diameter $(a)$, the sphericity $(b)$, the kinetic energy $(c)$, and the dynamic contact angle $(d)$  of an impinging droplet with $\text{Re}$=$1522$, $\Web$=$25$, $\Fro$=$110$, $\Bi=0$ and  $\theta_e^0$=$46^\circ$. Var~1: $\alpha=1$,  Var~2: $\alpha=10$ and  Var~3: $\alpha=100$.
\label{case01}}
\end{figure}
\begin{table}
\caption{Comparison of flow and geometric  parameters  of droplet impingement simulations for    different     examples considered in section~\ref{seccase01} with I=25.} 
\begin{center}
\begin{tabular}{lcccccc} 
\hline \vspace*{-2mm}\\
Variants &   $\displaystyle\max_{t\in(0,\text{I}]}|\delta_V(t)|$  & $\displaystyle\max_{t\in(0,\text{I}]}|\delta^\Gamma_c(t)|$ &    $\displaystyle\max_{t\in(0,\text{I}]}\frac{d(t)}{d_0}$ &   $\underset{\text{at } t=\text{I}}{\text{sphericity}}$ & $\underset{\text{at } t=\text{I}}{\text{kinetic energy}}$ & $\underset{\text{at } t=\text{I}}{ \theta_d(C_{\Gamma_1})}$ \vspace*{2mm}\\ 
\hline 
Clean   & 0.0141  & -      &    2.1109 & 0.4418 & 2.19 $\times 10^{-6}$  & 45.83  \\
Var.~1  & 0.0151  & 0.0127 &    2.1361 & 0.4170 &  6.09 $\times 10^{-7}$ & 41.85 \\
Var.~2  & 0.0151  & 0.0101 &    2.1385 & 0.4169 &  6.06 $\times 10^{-7}$ & 41.82\\
Var.~3  & 0.0144  & 0.0089 &    2.1430 & 0.4169 &  5.90 $\times 10^{-7}$ & 41.87\\
\hline 
\end{tabular}
\end{center}
\label{tabcase01}
\end{table}
The pressure contours and shapes of the impinging droplet at different instances (dimensionless time) $t$=0.1, 0.215, 0.625, 1.25, 3.75 and 25 are depicted  in Fig.~\ref{pressure} for the variant  $\alpha=100$. Initially, the pressure variation is large in the vicinity of the contact line, and becomes almost uniform when the droplet attains its equilibrium state, see the last snapshot. During the deformation, the pressure variation induces a capillary wave over the free surface, and it can clearly be seen in second and third snapshots. Further, the arrows in the snapshots show the flow directions in the droplet.
The computationally obtained wetting diameter, sphericity, kinetic energy and dynamic contact angle for all considered surfactant variants and for the clean droplet case ($c_0=0$ and $C_\Gamma(x,0)=0$) are presented in Figure~\ref{case01}. The wetting diameters of different variants are compared with the experimentally  observed wetting diameter in Fig.~\ref{case01} $(a)$, and  the clean droplet case matches very well, both qualitatively and quantitatively, with the  experiment results presented in~\cite{SHE10}. Since $\theta_e^0$ is less than $90^\circ$ in this example, the surfactant-dependent dynamic contact angle model~\eqref{casurf3} reduces the equilibrium contact angle further when the concentration of surfactant increases. Consequently, the maximum wetting diameter and the equilibrium wetting diameter of the droplet also increase, see Fig.~\ref{case01} $(a)$. The increase in the wetting diameter reduces the sphericity of the droplet, see Fig.~\ref{case01} $(b)$. These observations show that the surfactant-dependent dynamic contact angle is incorporated into the numerical scheme precisely. Initially, the kinetic energy is very high due to the pre-impact velocity and non-equilibrium surface force,  and it approaches to zero when the droplet attains the equilibrium state. The effects of surfactants on the kinetic energy of the droplet is negligible, Fig.~\ref{case01} $(c)$.  Initially, say until $\tilde t=2$, the computationally obtained dynamic contact angle in both surfactant and clean cases are similar, since $C_\Gamma(x,0)=0$. However, the   dynamic contact angle of the surfactant droplets becomes small when the surfactants are transported  to the free surface from the bulk phase, say after  $\tilde t=2$. Though the effects of surfactants on the flow dynamics of the droplet are clearly observed, the influence of the adsorption coefficient is negligible for the considered surfactant droplet configuration.

To  compare different flow and geometric parameters of the droplet impingement simulations quantitatively, the sphericity, the kinetic energy, and the dynamic contact angle of the droplet  are given in Table~\ref{tabcase01}. Further, the  maximum mass fluctuations and the  maximum wetting diameter obtained in all variants are also presented in the table.  The dynamic contact angle, as expected, is less in surfactant droplets   in comparison with the clean droplet case, and it will attain its   equilibrium value when the kinetic energy becomes zero.  The maximum fluctuations in the droplet's volume and in the surfactant mass are less than $1.52\%$ and $1.3\%$, respectively, see Table~\ref{tabcase01}.


\subsection{Influence of desorption in wetting droplets ($\theta_e^0=46^\circ$)}\label{seccase02}
\begin{figure}[t!]
\begin{center}
\unitlength1cm
\begin{picture}(15, 11.5)
\put(-0.5,5.75){{\includegraphics[width=7.5cm]{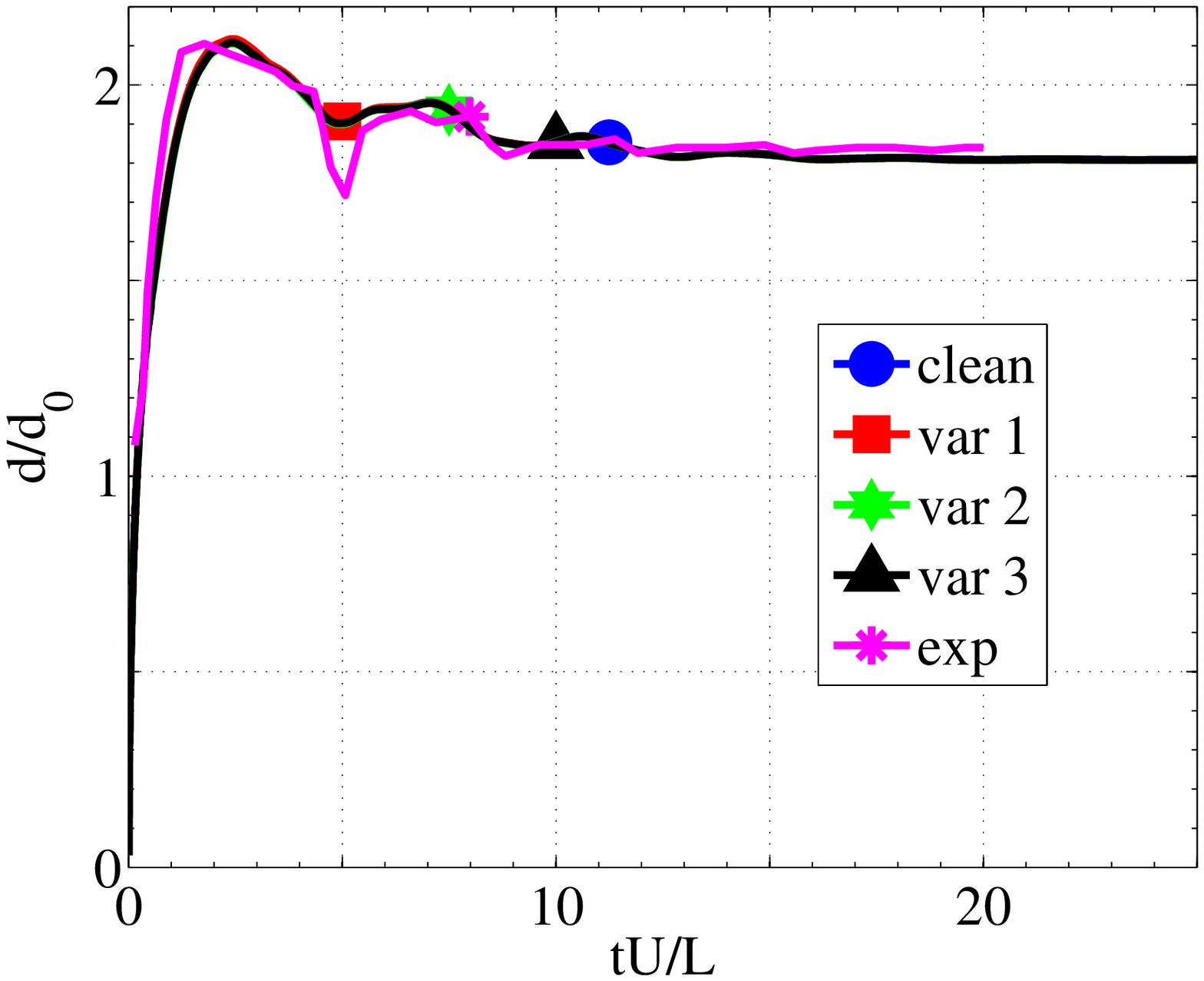}}}
\put(7.75,5.75){{\includegraphics[width=7.5cm]{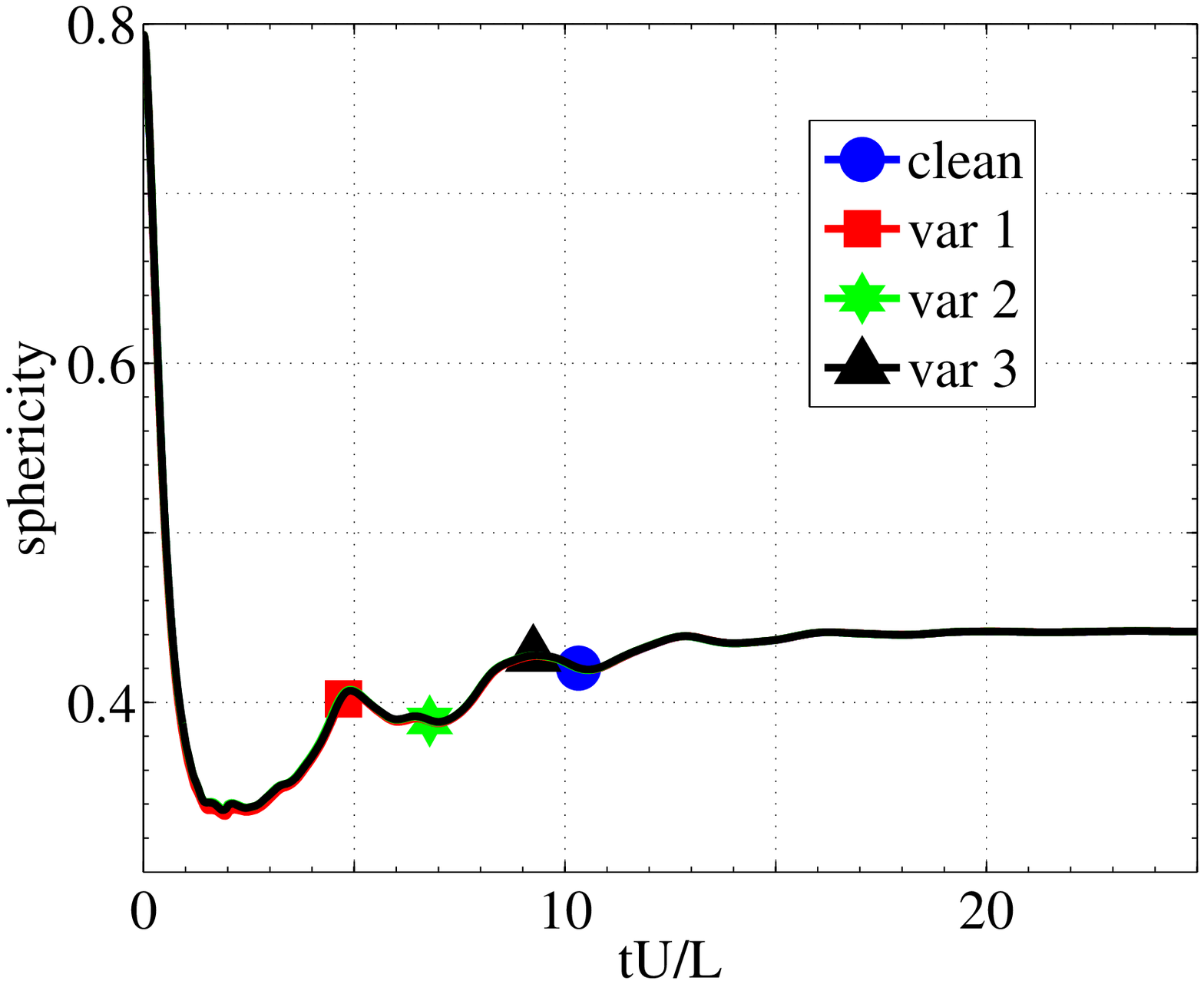}}}
\put(-0.5, -0.5){{\includegraphics[width=7.5cm]{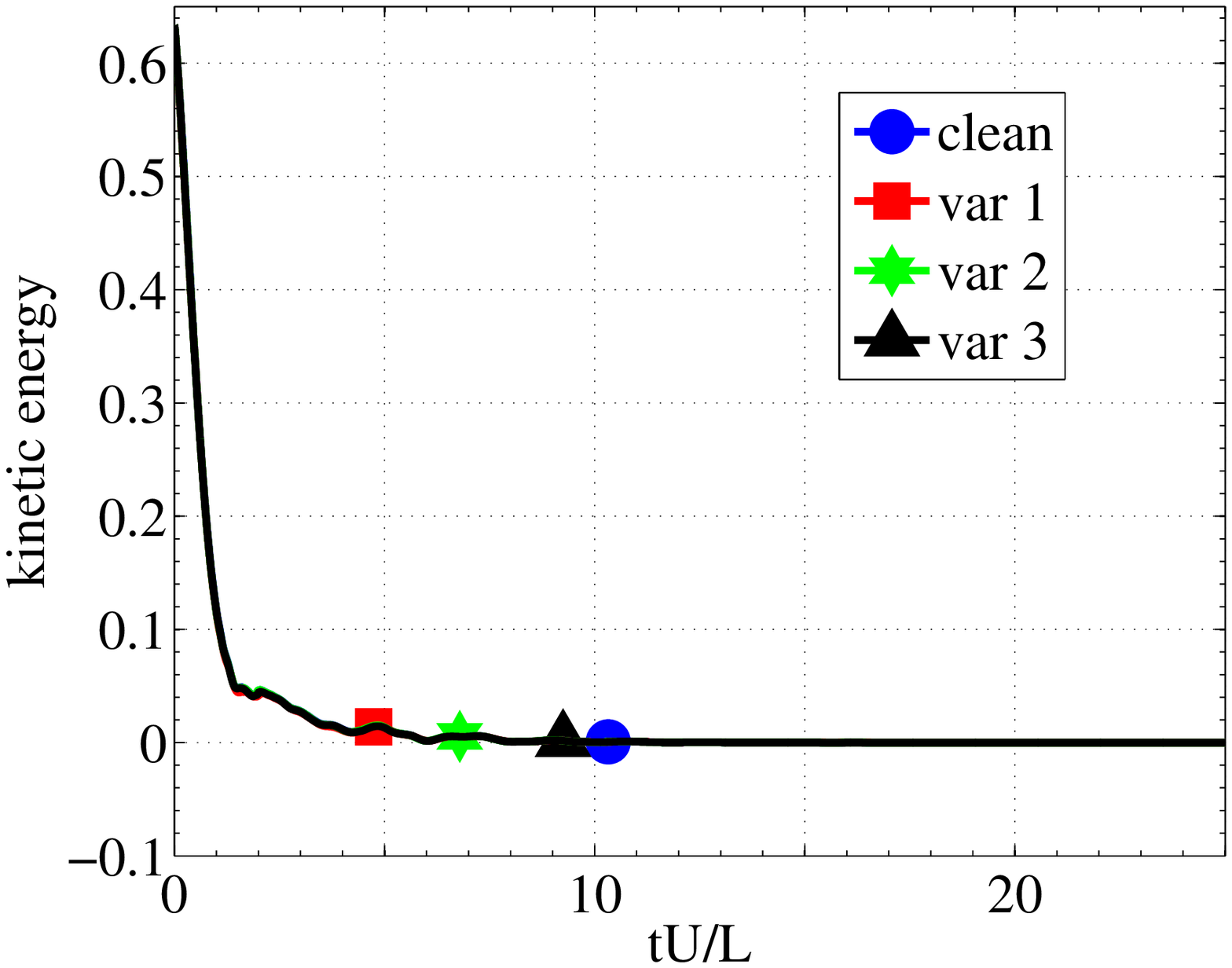}}}
\put(7.75, -0.5){{\includegraphics[width=7.5cm]{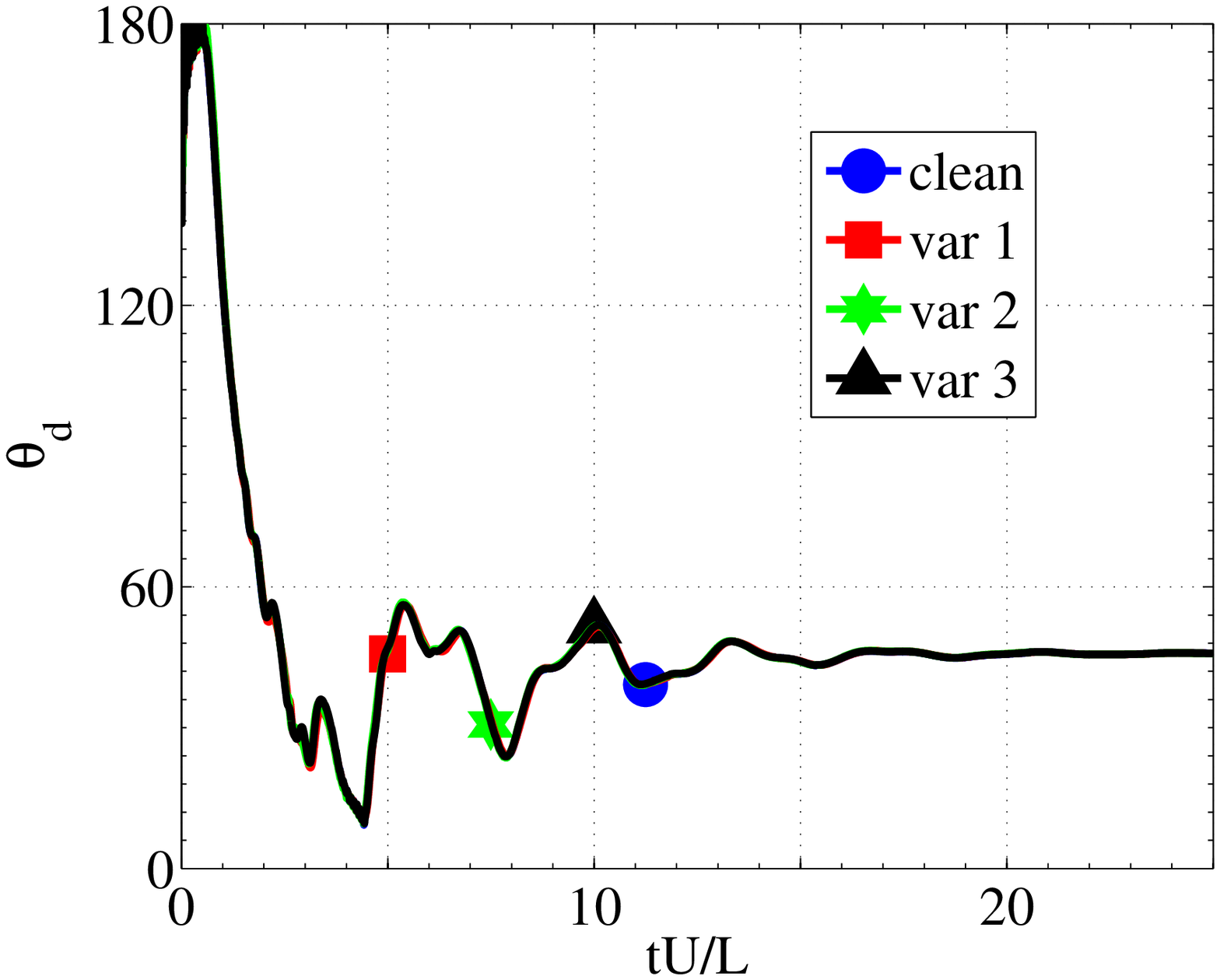}}}
\put(3.25,11.35){$(a)$}
\put(11.75,11.35){$(b)$}
\put(3.25, 5){$(c)$}
\put(11.75, 5){$(d)$}
\end{picture}
\end{center}
\caption{Effects of surfactant desorption coefficient on the wetting diameter $(a)$, the sphericity $(b)$, the kinetic energy $(c)$, and the dynamic contact angle $(d)$  of an impinging droplet with $\text{Re}$=$1522$, $\Web$=$25$, $\Fro$=$110$, $\alpha=0$ and  $\theta_e^0$=$46^\circ$. Var~1: $\Bi=1$,  Var~2: $\Bi=5$ and  Var~3: $\Bi=10$.
\label{case02}}
\end{figure}
\begin{table}
\caption{Comparison of flow and geometric  parameters  of droplet impingement simulations for    different     examples considered in section~\ref{seccase02} with I=25.} 
\begin{center}
\begin{tabular}{lcccccc} 
\hline \vspace*{-2mm}\\
Variants &   $\displaystyle\max_{t\in(0,\text{I}]}|\delta_V(t)|$  & $\displaystyle\max_{t\in(0,\text{I}]}|\delta^\Gamma_c(t)|$ &    $\displaystyle\max_{t\in(0,\text{I}]}\frac{d(t)}{d_0}$ &   $\underset{\text{at } t=\text{I}}{\text{sphericity}}$ & $\underset{\text{at } t=\text{I}}{\text{kinetic energy}}$ & $\underset{\text{at } t=\text{I}}{ \theta_d(C_{\Gamma_1})}$ \vspace*{2mm}\\ 
\hline 
Clean        &  0.0141  & -       & 2.1109    & 0.4418  &   2.19$\times 10^{-6}$ & 45.83 \\
Var.~1   &  0.0147  & 0.0061  & 2.1176    & 0.4418  &   2.52$\times 10^{-6}$ & 45.82 \\
Var.~2  &  0.0153  & 0.0096  & 2.1056    & 0.4418  &   2.15$\times 10^{-6}$ & 45.82 \\
Var.~3 &  0.0146  & 0.0092  & 2.1082    & 0.4418  &   1.99$\times 10^{-6}$ & 45.83\\
\hline 
\end{tabular}
\end{center}
\label{tabcase02}
\end{table}

To study the influence of desorption of surfactants on the flow dynamics of the droplet impingement, we consider the same droplet configurations and flow parameters as in the previous section. However, we take $\alpha=0$, the initial surfactant concentrations,  $c_0=0$ and $C_\Gamma(x,0)=0.5$ in this numerical test. Further, the following three variants,  (1)~$\Bi=1$,  (2)~$\Bi=5$ and  (3)~$\Bi=10$, are considered. The computations are performed until the dimensionless time I$=25$.

The wetting diameter, sphericity, kinetic energy and dynamic contact angle obtained in computations of all variants are presented in Fig.~\ref{case02}. Since the Biot number is nonzero in all variants, the surfactants on the interfaces are transported into  the bulk phase, and eventually the free surface and the liquid-solid interface become clean after some time. Further, we observed that the influence of surfactants on the flow dynamics is negligible, as effects of the Marangoni convection and the surfactant-dependent dynamic contact angle become negligible when the interface becomes clean. These observations can be seen in each picture of Fig.~\ref{case02}, where the curves of different variants are almost identical. It shows that the impurities do not affect the flow dynamics of the droplet much, when the impurities are transported into the bulk phase. To support this observation quantitatively, the parameters obtained in all variants are given in Table~\ref{tabcase02}. The tabulated values are almost identical in all variants. Moreover, the maximum fluctuations in the droplet's volume and in the surfactant mass are less than $1.53\%$ and $1\%$, respectively.   

Since the influence of surfactants in the desorption case is negligible on the flow dynamics of the droplet impingement, only the adsorption cases are studied in the subsequent  sections.


\subsection{Influence of adsorption in non-wetting droplets ($\theta_e^0=100^\circ$)}\label{seccase03}
 \begin{figure}[t!]
\begin{center}
\unitlength1cm
\begin{picture}(20,11.5)
\put(10,-2.25){\makebox(3,6){\includegraphics[trim=0.5cm 0cm 0.5cm 10cm, clip=true,width=7cm]{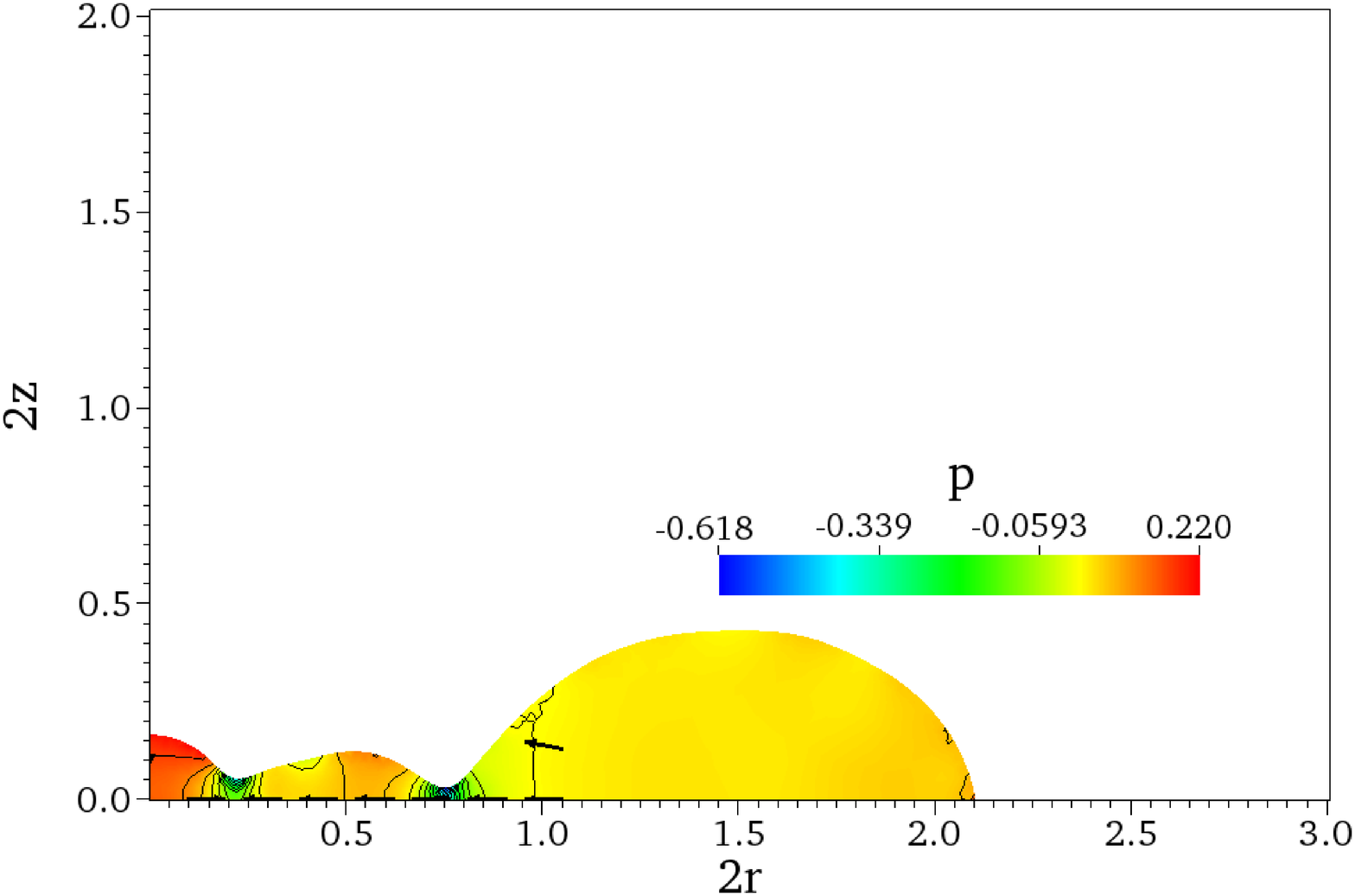}}}
\put(10,1.5){\makebox(3,6){\includegraphics[trim=0.5cm 0cm 0.5cm 5cm, clip=true,width=7cm]{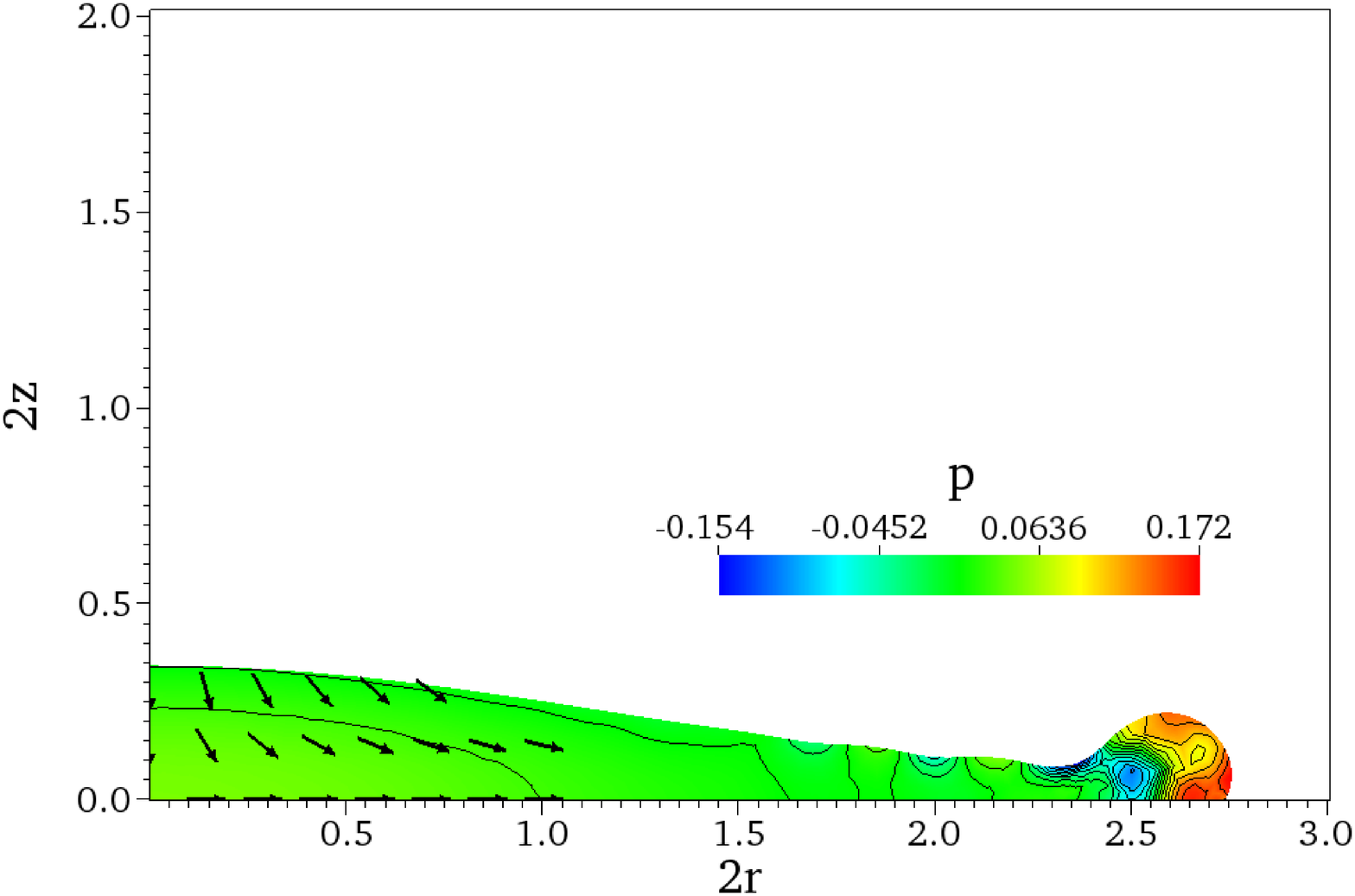}}}
\put(10,6.){\makebox(3,6){\includegraphics[trim=0.5cm 0cm 0.5cm 3cm, clip=true,width=7cm]{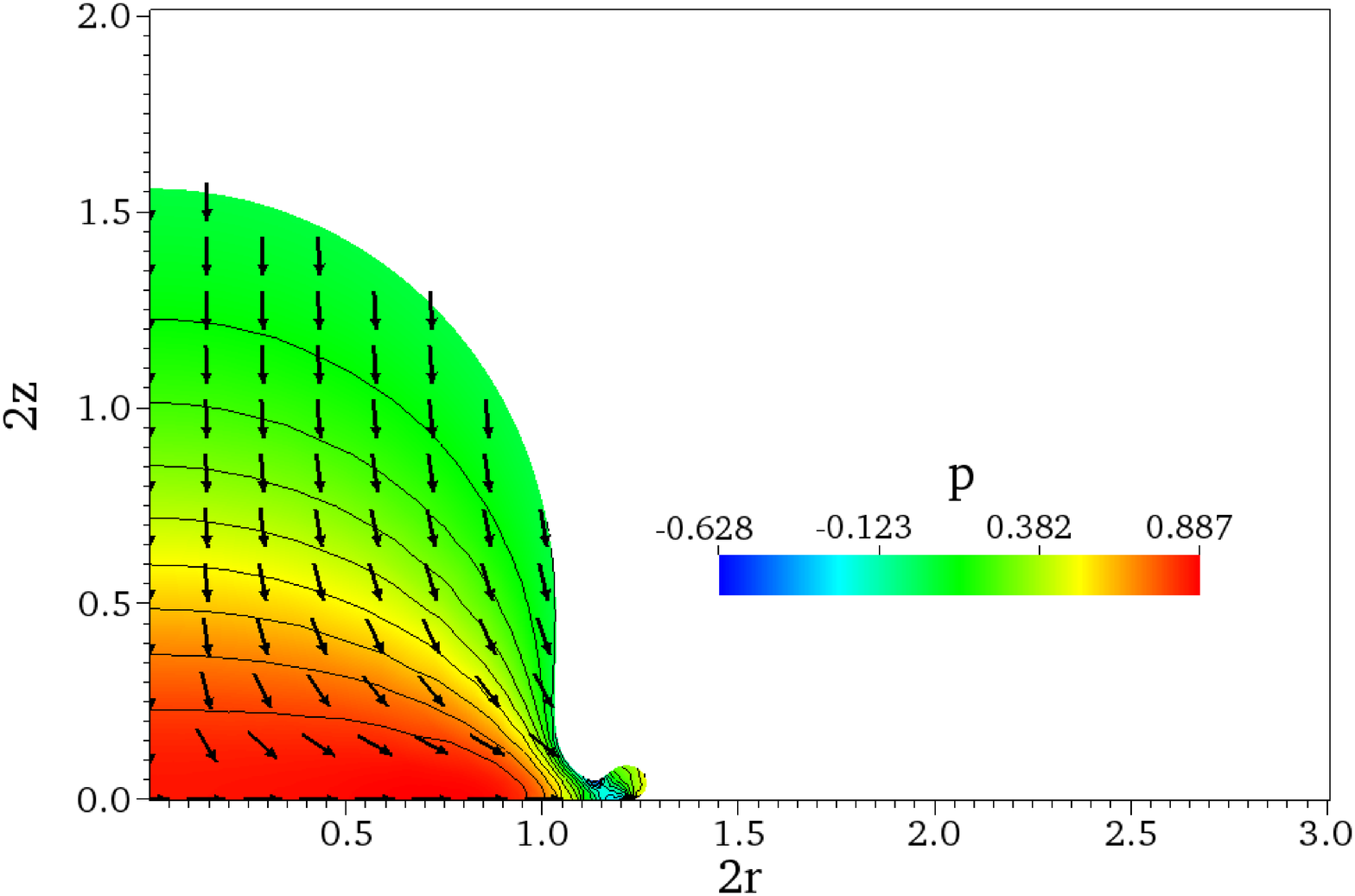}}}
\put(2,-2.25){\makebox(3,6){\includegraphics[trim=0.5cm 0cm 0.5cm 10cm, clip=true,width=7cm]{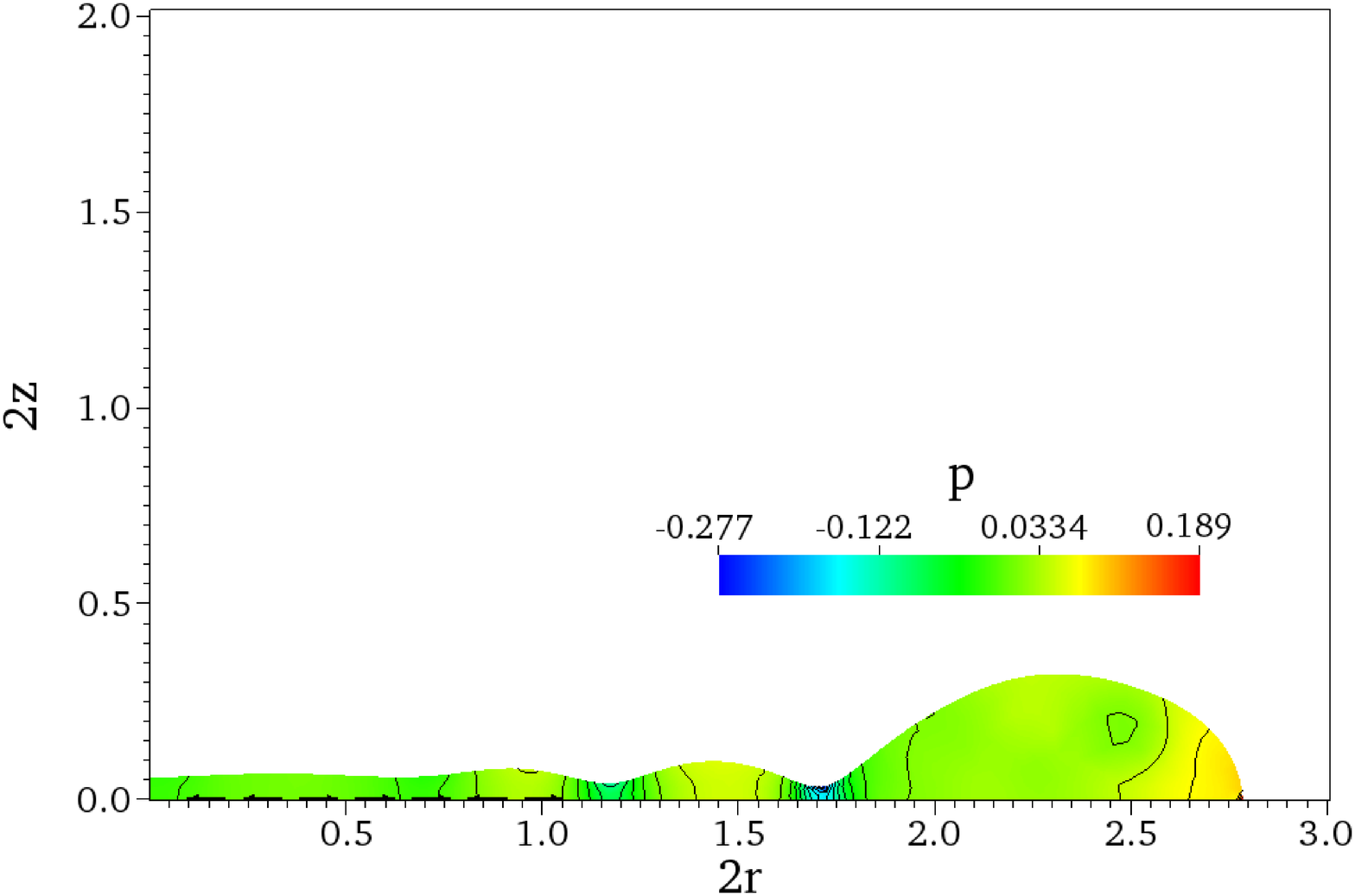}}}
\put(2, 1.5){\makebox(3,6){\includegraphics[trim=0.5cm 0cm 0.5cm 5cm, clip=true,width=7cm]{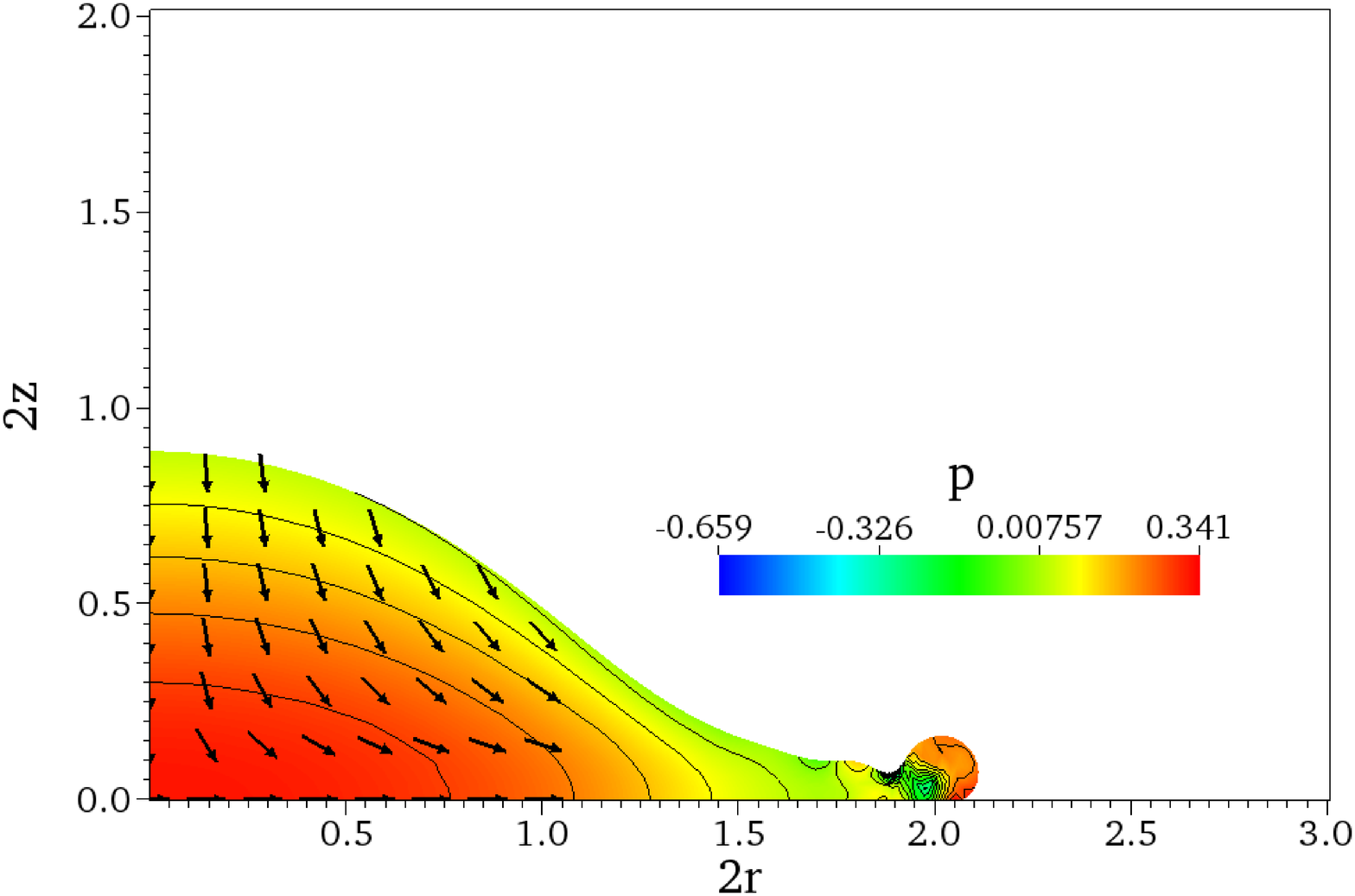}}}
\put(2, 6.){\makebox(3,6){\includegraphics[trim=0.5cm 0cm 0.5cm 3cm, clip=true,width=7cm]{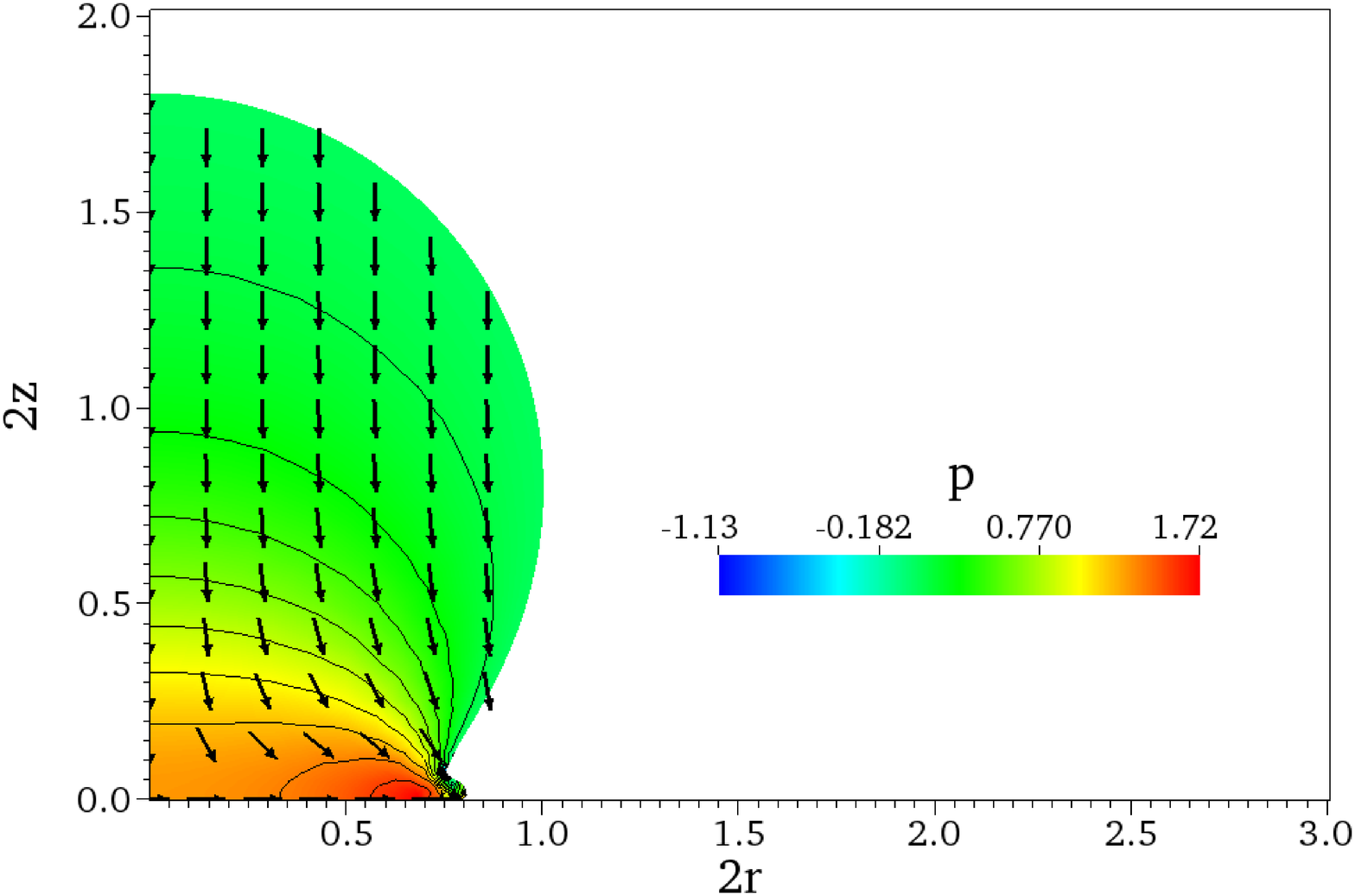}}}
\put(3,10.5){$t=0.1$}
\put(3,6.){$t=0.625$}
\put(3,2.){$t=3.75$}

\put(11,10.5){$t=0.215$}
\put(11,6.){$t=0.125$}
\put(11,2.){$t=25$}

\end{picture}
\end{center}
\caption{Pressure contours and shapes of the impinging droplet at dimensionless time  $t$=0.1, 0.215, 0.625, 1.25, 3.75 and 5. $\Rey$=$4212$, $\Web$=$90$, $\Fro$=$92$, $\Pe_c=2$,  $\Pe_\Gamma=2$,  $\Bi=0$,  $\Da=370$,  $\alpha=10$  and  $\theta_e^0$=$100^\circ$.}
\label{pressureCa100}
\end{figure}

\begin{figure}[t!]
\begin{center}
\unitlength1cm
\begin{picture}(15, 11.5)
\put(-0.5,5.75){{\includegraphics[width=7.5cm]{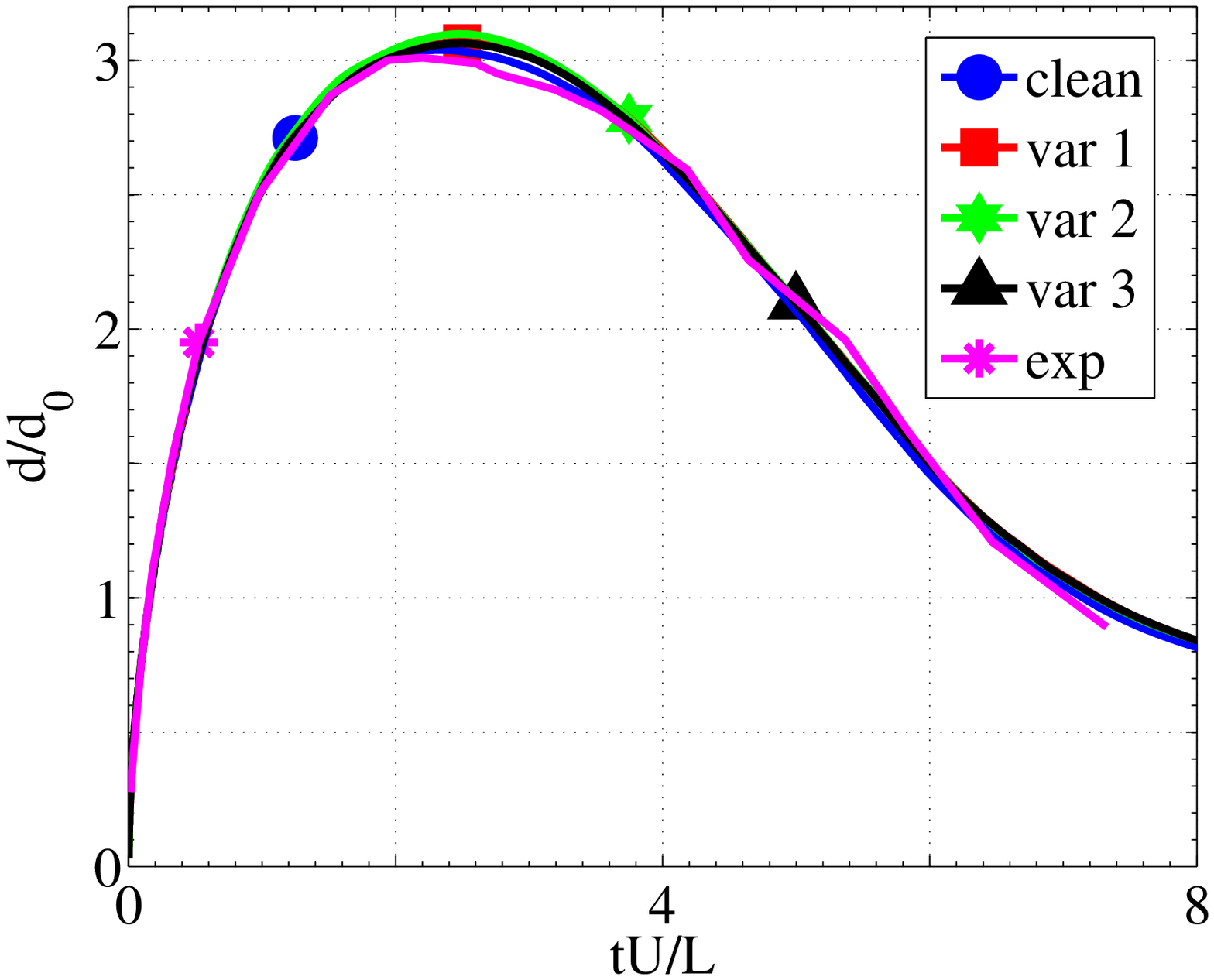}}}
\put(1.,6.75){{\includegraphics[width=3.25cm]{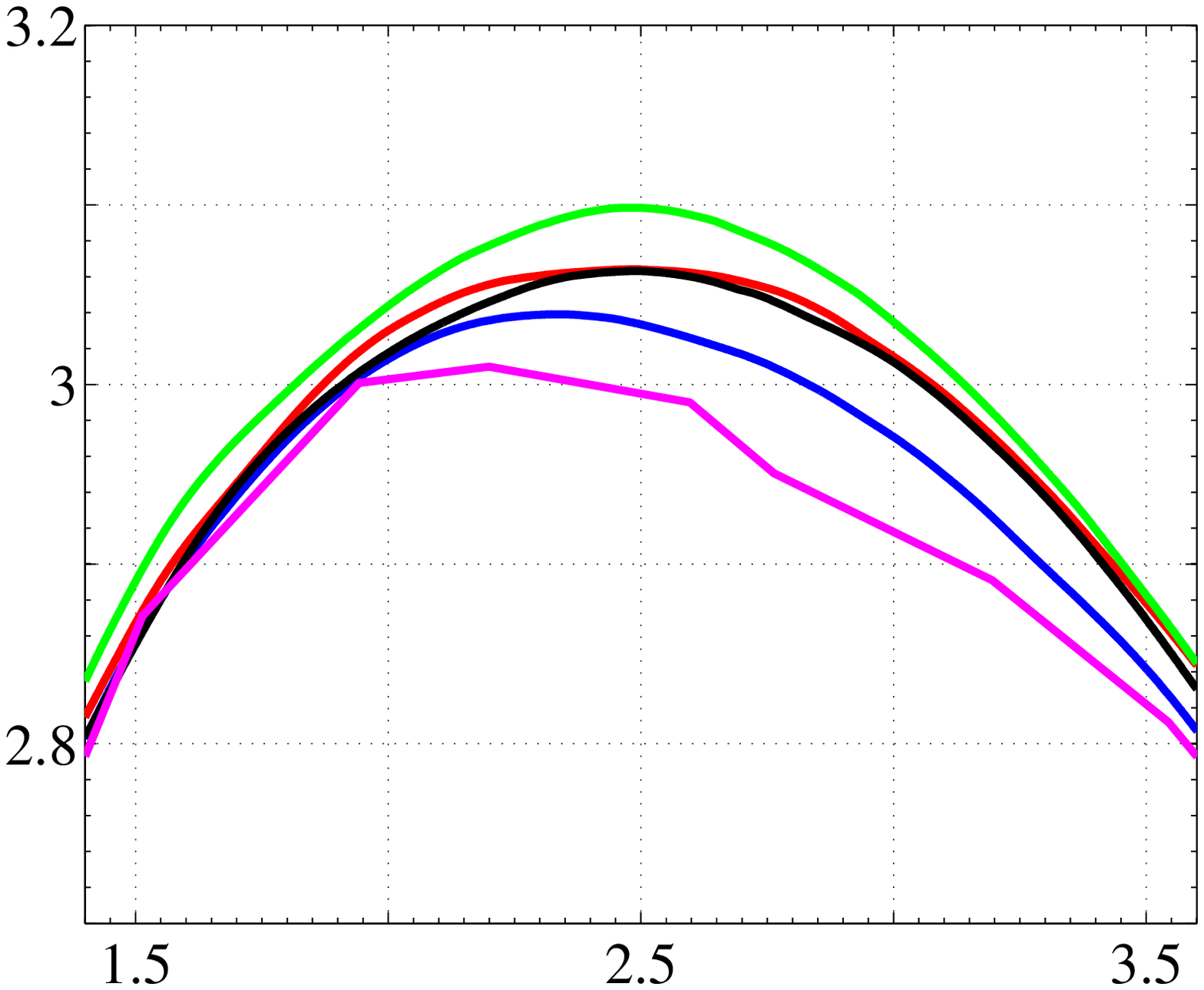}}}
\put(7.75,5.75){{\includegraphics[width=7.5cm]{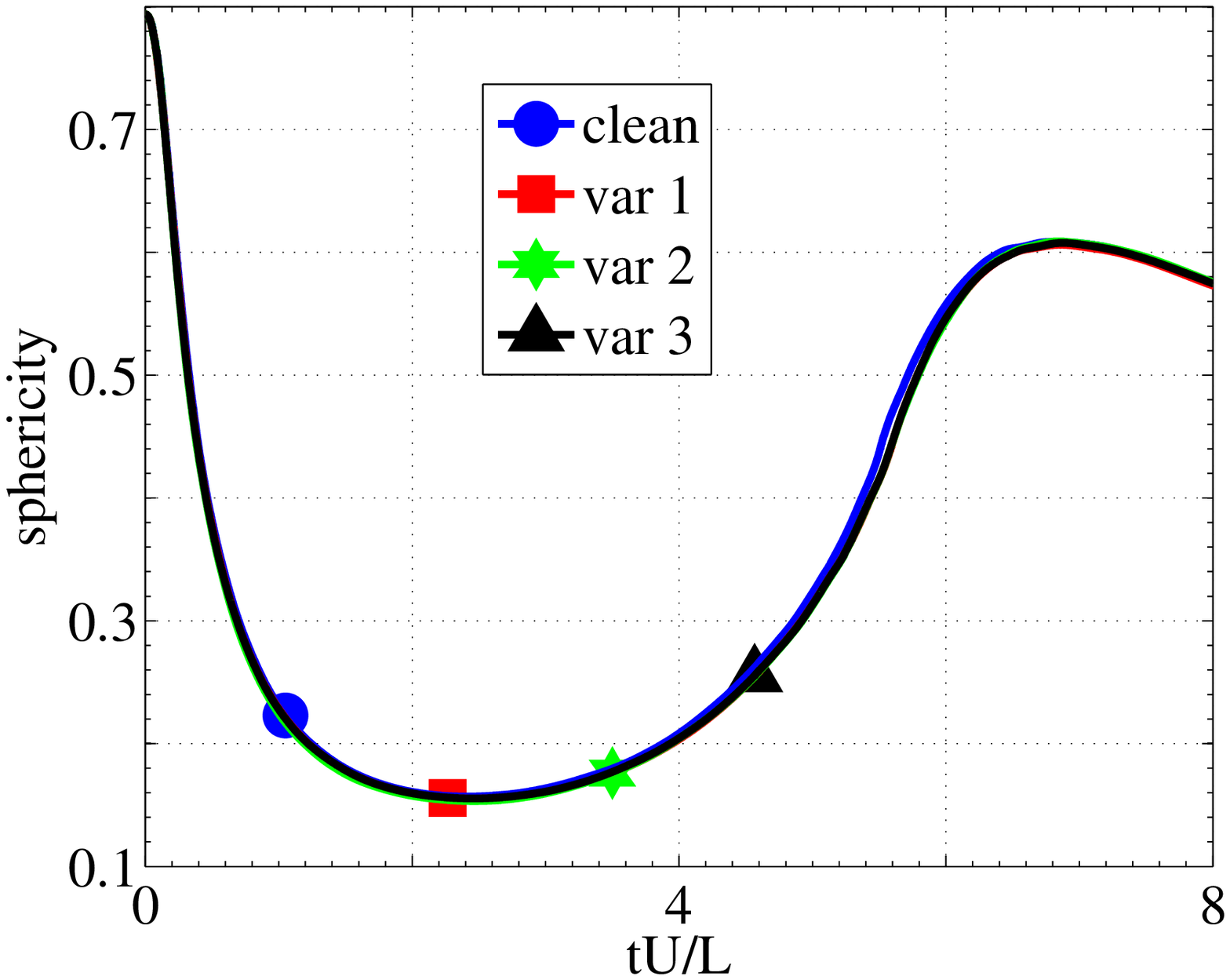}}}
\put(-0.5, -0.5){{\includegraphics[width=7.5cm]{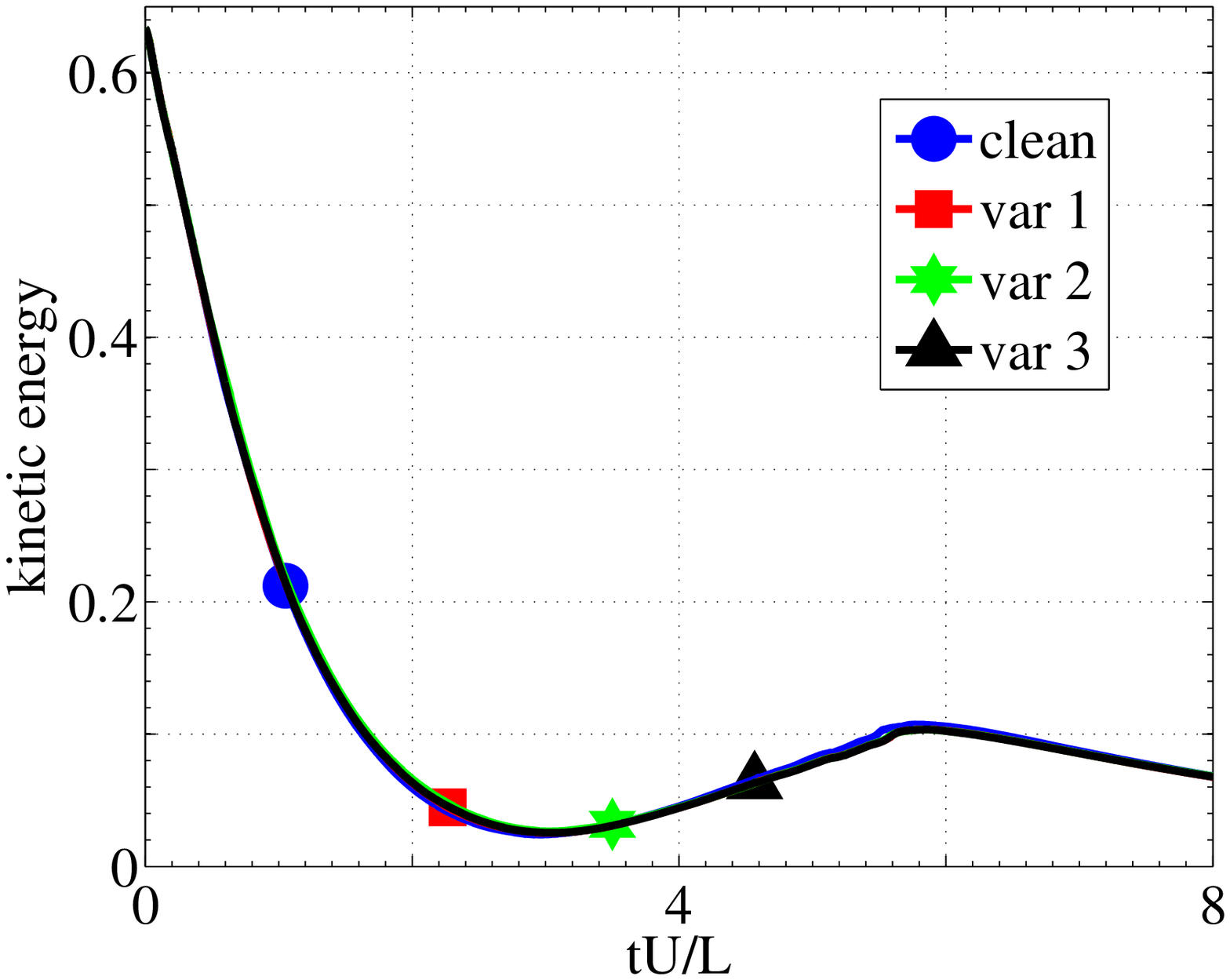}}}
\put(7.75, -0.5){{\includegraphics[width=7.5cm]{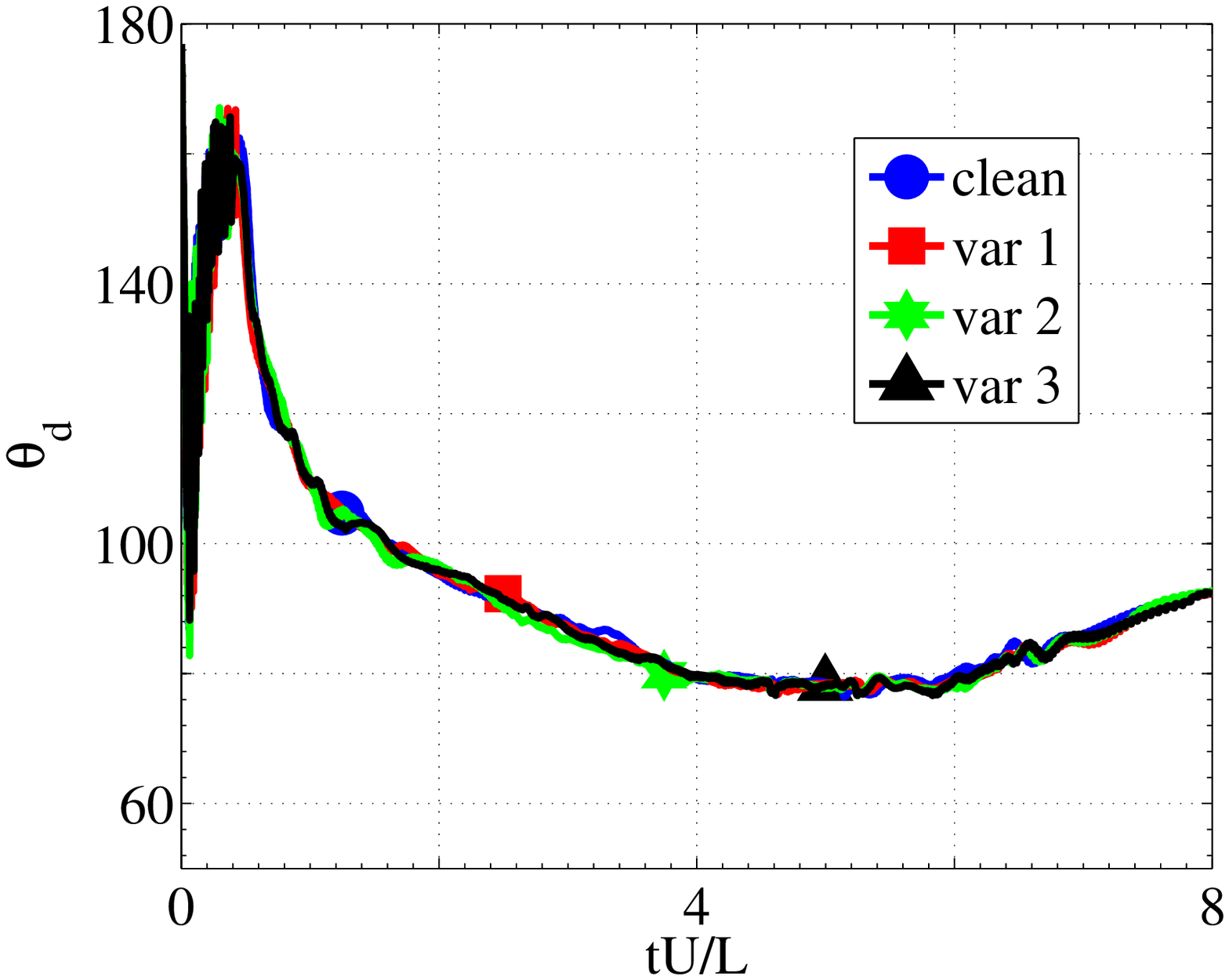}}}
\put(3.25,11.35){$(a)$}
\put(11.75,11.35){$(b)$}
\put(3.25, 5){$(c)$}
\put(11.75, 5){$(d)$}
\end{picture}
\end{center}
\caption{Effects of surfactant adsorption coefficient on the wetting diameter $(a)$, the sphericity $(b)$, the kinetic energy $(c)$, and the dynamic contact angle $(d)$  of an impinging droplet with $\Rey=4212$, $\Web$=$90$, $\Fro$=$92$, $\Bi=0$, and  $\theta_e^0$=$100^\circ$. Var~1: $\alpha=1$,  Var~2: $\alpha=10$ and  Var~3: $\alpha=100$.
\label{case03}}
\end{figure}
\begin{table}
\caption{Comparison of flow and geometric  parameters  of droplet impingement simulations for    different     examples considered in section~\ref{seccase03}.} 
\begin{center}
\begin{tabular}{lccccccc} 
\hline \vspace*{-2mm}\\
Variants &  $\underset{\text{(I)}}{\text{End time}}$   & $\displaystyle\max_{t\in(0,\text{I}]}|\delta_V(t)|$  & $\displaystyle\max_{t\in(0,\text{I}]}|\delta^\Gamma_c(t)|$ &    $\displaystyle\max_{t\in(0,\text{I}]}\frac{d(t)}{d_0}$ &   $\underset{\text{at } t=\text{I}}{\text{sphericity}}$ & $\underset{\text{at } t=\text{I}}{\text{kinetic energy}}$ & $\underset{\text{at } t=\text{I}}{ \theta_d(C_{\Gamma_1})}$ \vspace*{2mm}\\ 
\hline 
Clean    & 8.49 & 0.0344  & -       & 3.039     & 0.5526  &   0.0592 & 93.53 \\
Var.~1   & 8.03 &  0.0306  & 0.008   & 3.0641    & 0.5717  &   0.0670 & 92.41 \\
Var.~2  &  8.65 & 0.0354  & 0.0239  & 3.0984    & 0.5501  &   0.0583 & 93.65 \\
Var.~3 &  8.46 & 0.0353  & 0.0221  & 3.0632    & 0.5564  &   0.0606 & 93.50\\
\hline 
\end{tabular}
\end{center}
\label{tabcase03}
\end{table}

We now consider an impinging water droplet of diameter $d_0$=$2.7\times 10^{-3}$~m impinging  with the pre-impact speed   $u_{imp}$ =$1.56$~m/s and the equilibrium contact angle $\theta_e^0=100^\circ$. The experimental results of the considered
 clean droplet case are presented in~\cite{SIK02}. The resulting dimensionless numbers are $\Rey$=$4212$, $\Web$=$90$ and $\Fro$=$92$. Further, $\Pe_c=2$,  $\Pe_\Gamma=2$, $\Bi=0$, $\Da=370$, $\delta t =0.00025$ and  $\beta=0.75/2h_E$ are used in computations. In general, the non-wetting droplets with surfactants are more interesting to study, as the presence of surfactants on the interface increases the Weber number by reducing the surface tension that eventually increase the wetting diameter. Contrarily, the capillary effect reduces the wetting diameter when  the dynamic contact angle increases due to an increase in the surfactant concentration. Since the $\Web$ has no influence at the equilibrium  state, the wetting diameter will be less and the   contact angle will be more in the surfactant droplet when compared to the clean droplet. To study this behavior in detail, the following three variants, (1)~$\alpha=1$,  (2)~$\alpha=10$ and  (3)~$\alpha=100$, are considered.

The pressure contours and shapes of the impinging droplet at different instances (dimensionless time) $t$=0.1, 0.215, 0.625, 1.25, 3.75 and 5 are depicted  in Fig.~\ref{pressure} for the variant $\alpha=10$. As in the previous cases, the pressure variation is large in the vicinity of the contact line initially. However, the wetting diameter is large in this example due to a high Reynolds number. The arrows in the snapshots of the droplets indicate the flow directions.
The obtained wetting diameter, sphericity, kinetic energy and dynamic contact angle in computations of all variants are presented in Fig.~\ref{case03}. The obtained wetting diameter in computations of all variants are compared in Fig.~\ref{case03} $(a)$ with the experimentally observed wetting diameter of the clean droplet presented in~\cite{SIK02}. The computational results  of the clean droplet case are in good agreement with the experiment result.
Since the considered equilibrium contact angle is close to $90^\circ$, the influence of surfactant-dependent dynamic contact angle on the flow dynamics is less. Unlike the previous test case considered in Section~\ref{seccase01}, a topological change (breaking/splashing) is observed in this example due to high  $\Rey$=$4212$. Since the topological changes are not modeled in the numerical scheme,  the computations break down when the distance between the interfaces and/or boundaries are less than the mesh size. The time at which each variant of droplet breaks down (End time) is presented in Table~\ref{tabcase03}. All other parameters such as the mass fluctuations, maximum wetting diameter, sphericity, kinematic energy and the dynamic contact angle obtained in all variants are comparable, see Table~\ref{tabcase03}. Despite the strong deformation due to high Reynolds number, the maximum fluctuations of the droplet's volume and the surfactant mass are less than $3.6\%$ and $2.4\%$, respectively,  in all computations.

\subsection{Influence of adsorption in an impinging droplet with $\theta_e^0=90^\circ$}\label{seccase04}

\begin{figure}[t!]
\begin{center}
\unitlength1cm
\begin{picture}(15, 11.5)
\put(-0.5,5.75){{\includegraphics[width=7.5cm]{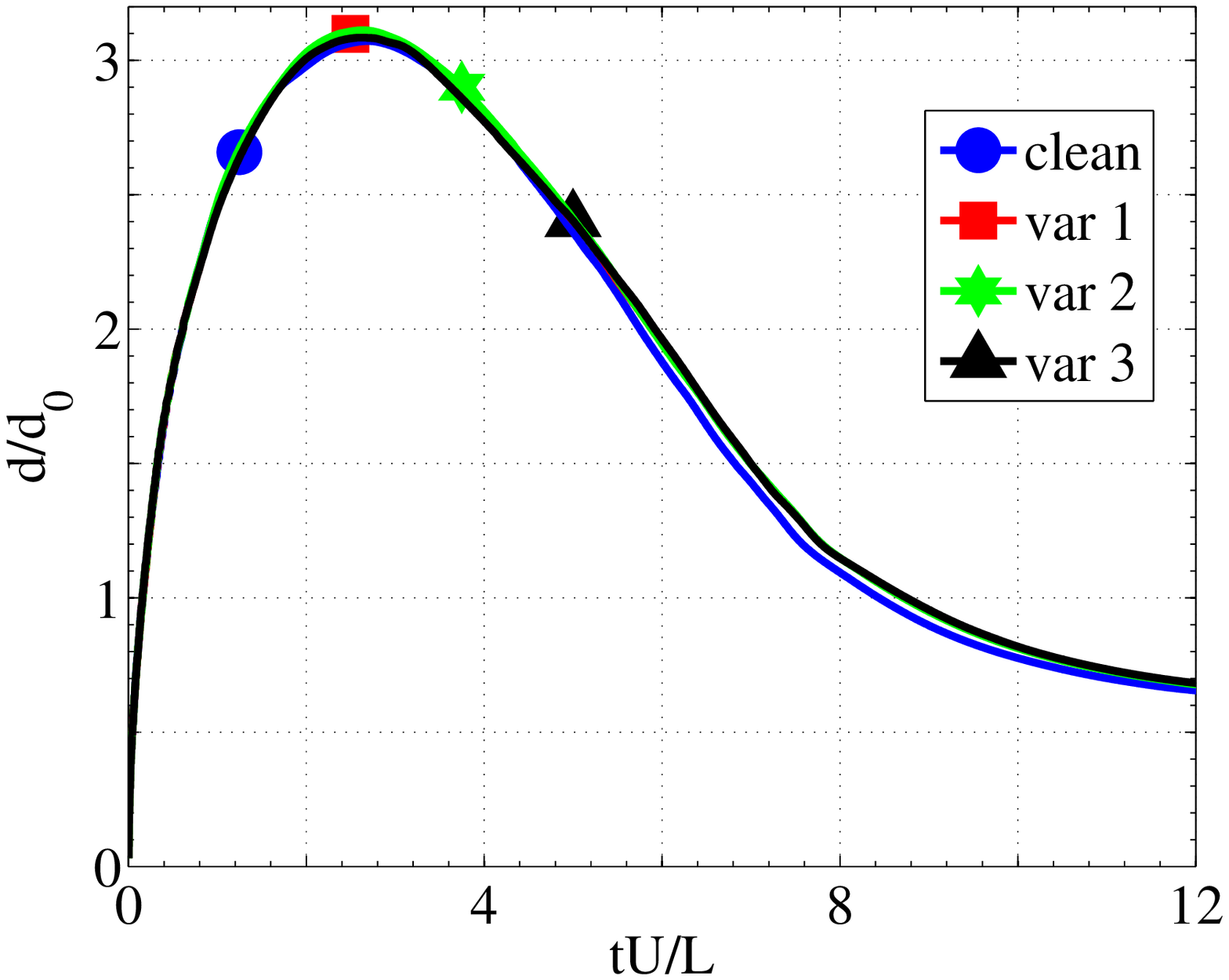}}}
\put(1.,6.75){{\includegraphics[width=3.25cm]{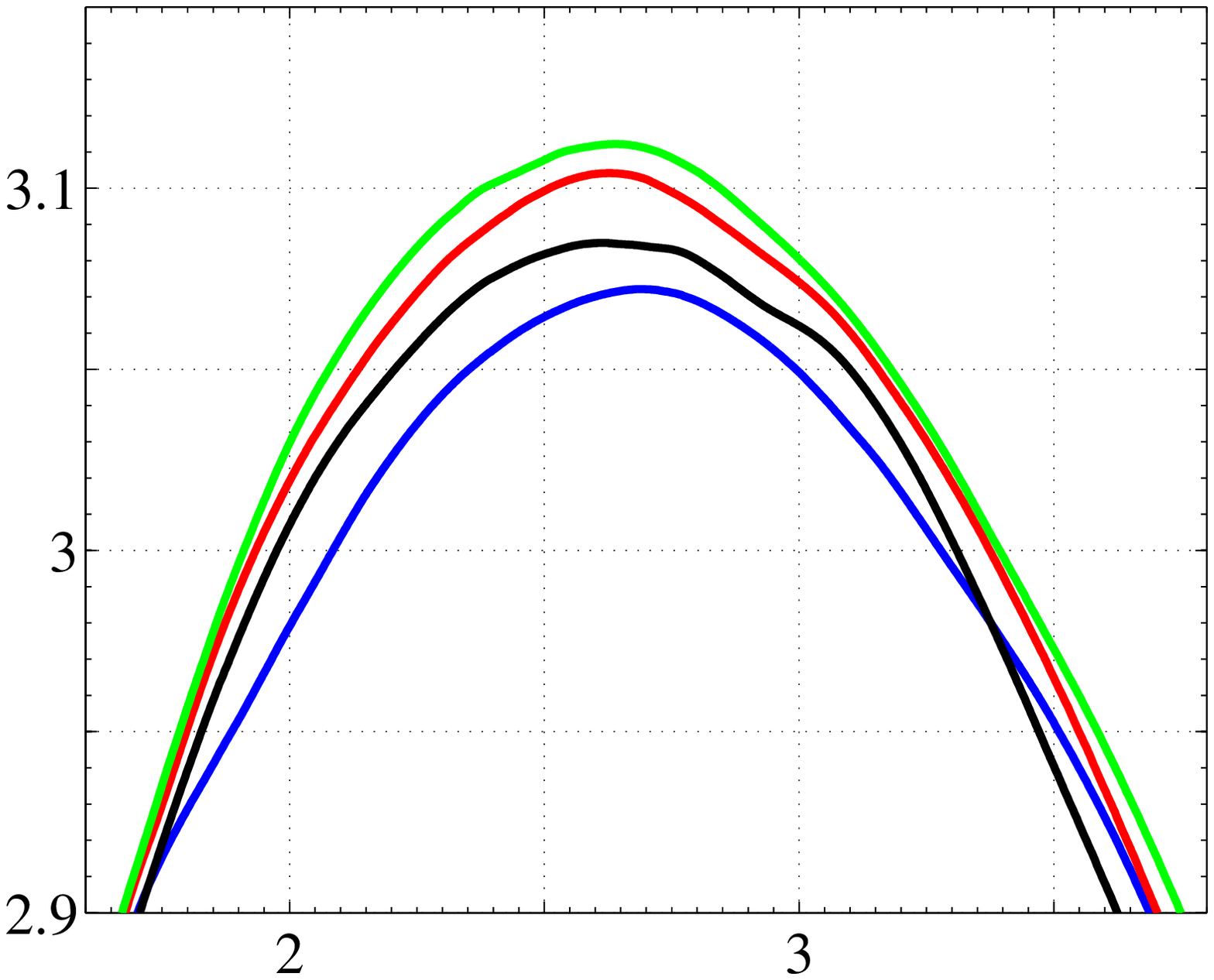}}}
\put(7.75,5.75){{\includegraphics[width=7.5cm]{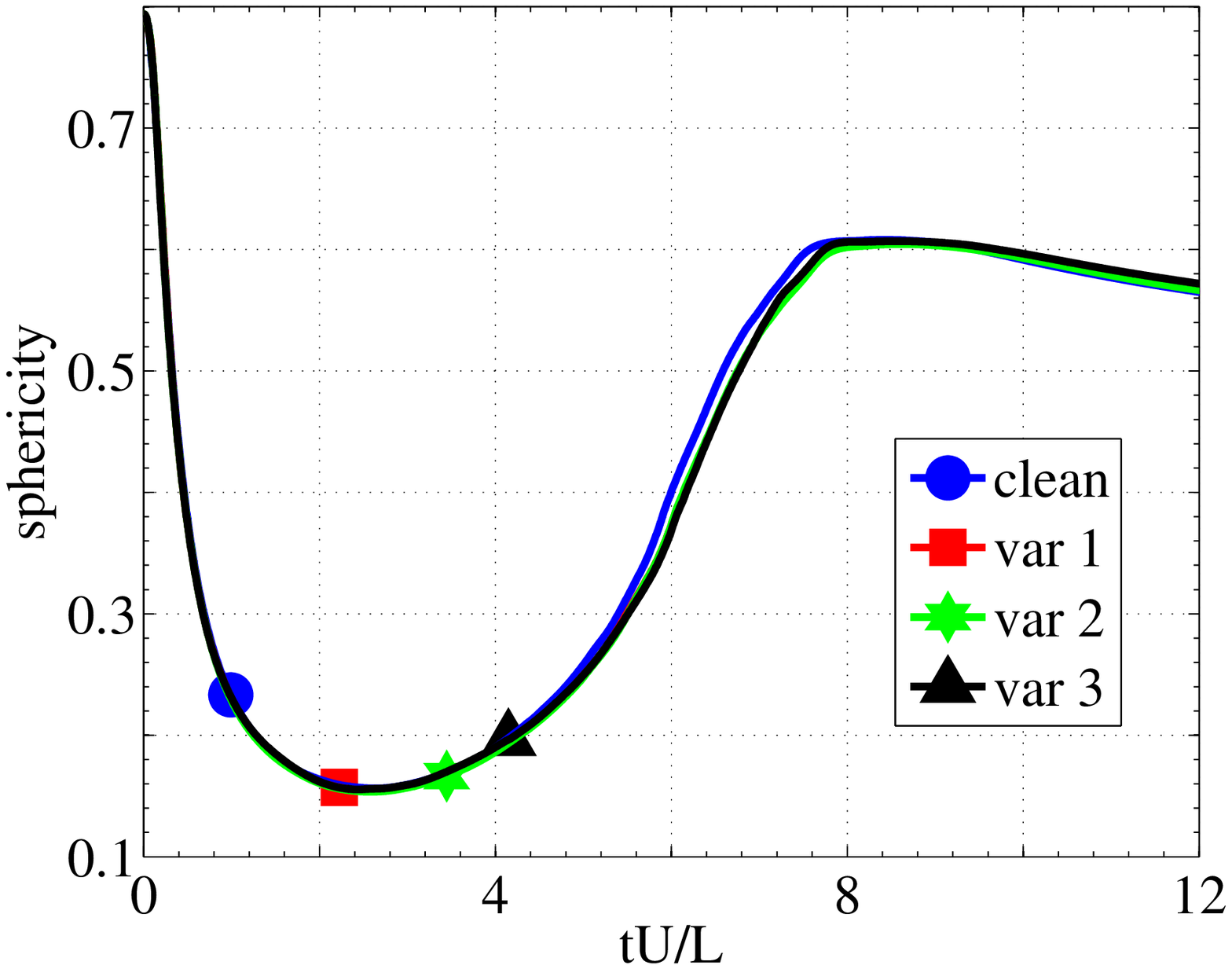}}}
\put(-0.5, -0.5){{\includegraphics[width=7.5cm]{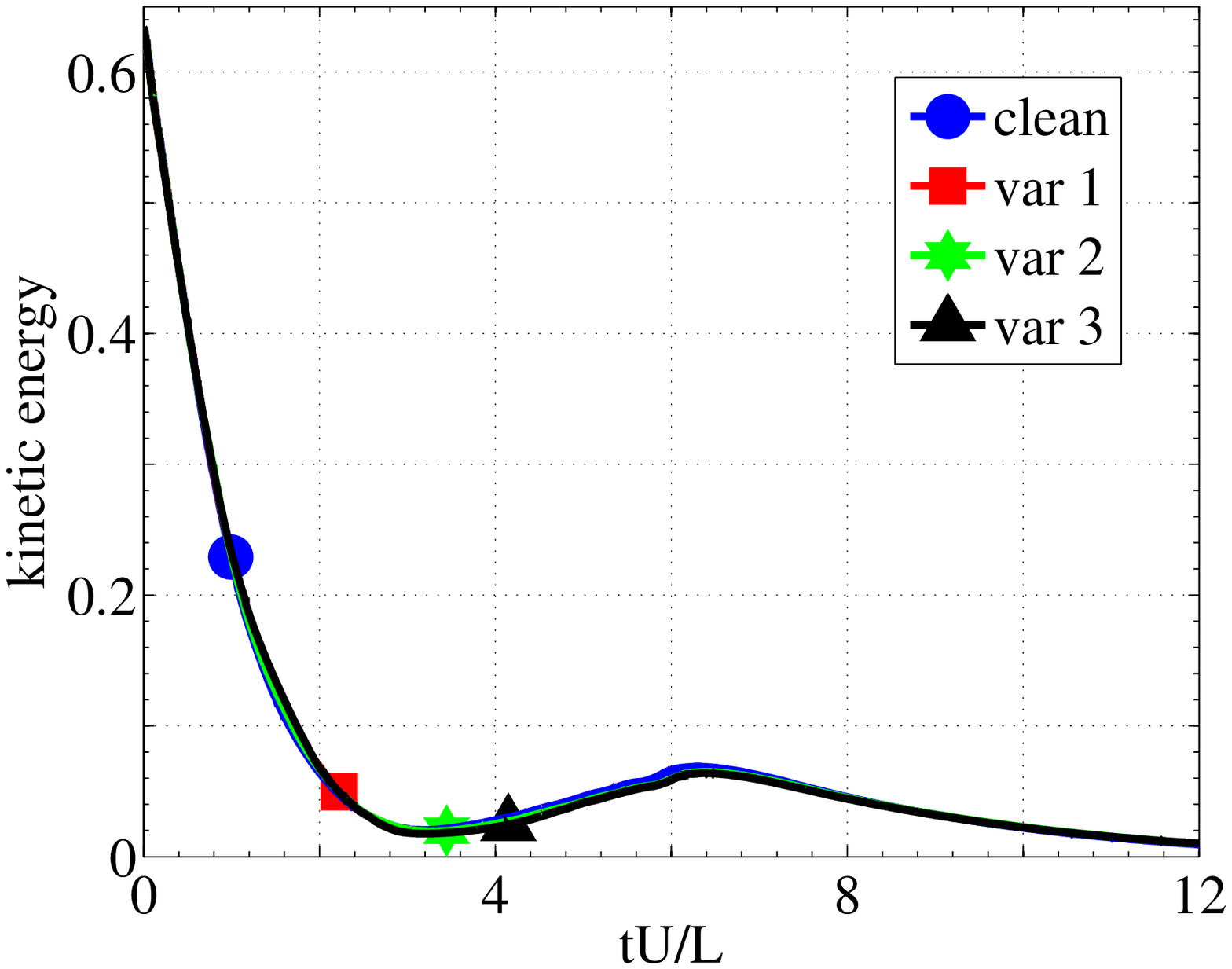}}}
\put(7.75, -0.5){{\includegraphics[width=7.5cm]{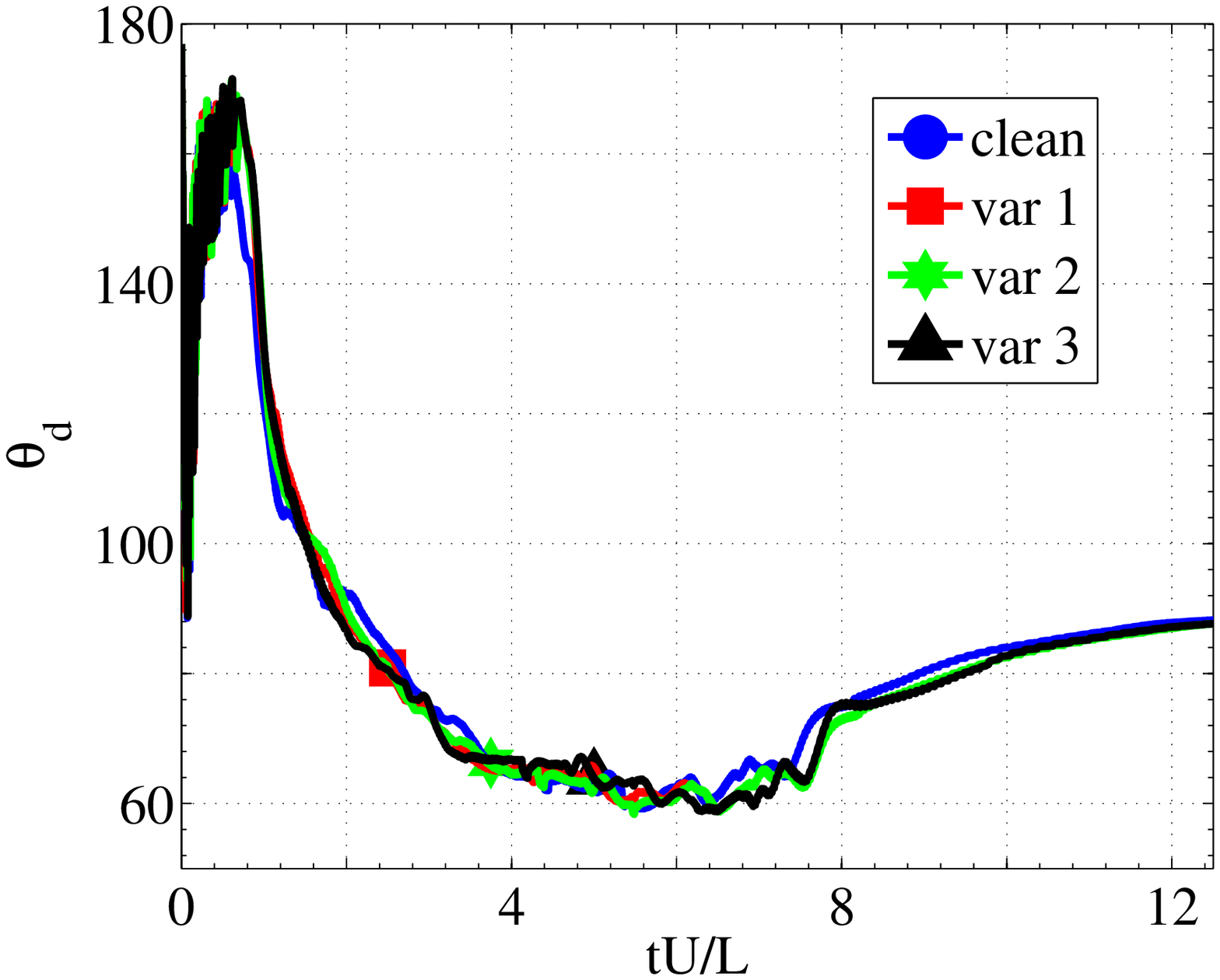}}}
\put(3.25,11.35){$(a)$}
\put(11.75,11.35){$(b)$}
\put(3.25, 5){$(c)$}
\put(11.75, 5){$(d)$}
\end{picture}
\end{center}
\caption{Effects of surfactant adsorption coefficient on the wetting diameter $(a)$, the sphericity $(b)$, the kinetic energy $(c)$, and the dynamic contact angle $(d)$  of an impinging droplet with $\Rey=4212$, $\Web$=$90$, $\Fro$=$92$, and  $\theta_e^0$=$90^\circ$. Var~1: $\alpha=1$,  Var~2: $\alpha=10$ and  Var~3: $\alpha=100$.
\label{case05}}
\end{figure}
\begin{table}
\caption{Comparison of flow and geometric  parameters  of droplet impingement simulations for    different     examples considered in section~\ref{seccase04}.} 
\begin{center}
\begin{tabular}{lccccccc} 
\hline \vspace*{-2mm}\\
Variants &  $\underset{\text{(I)}}{\text{End time}}$ &   $\displaystyle\max_{t\in(0,\text{I}]}|\delta_V(t)|$  & $\displaystyle\max_{t\in(0,\text{I}]}|\delta^\Gamma_c(t)|$ &    $\displaystyle\max_{t\in(0,\text{I}]}\frac{d(t)}{d_0}$ &   $\underset{\text{at } t=\text{I}}{\text{sphericity}}$ & $\underset{\text{at } t=\text{I}}{\text{kinetic energy}}$ & $\underset{\text{at } t=\text{I}}{ \theta_d(C_{\Gamma_1})}$ \vspace*{2mm}\\ 
\hline 
Clean     & 12.64 & 0.0502  & -       & 3.0721    & 0.5592  &   0.0069 & 88.24 \\
Var.~1   & 6.16 & 0.0461  & 0.0104  & 3.1041    & 0.4032  &   0.0627 & 62.69 \\
Var.~2  & 13.18 & 0.0567  & 0.0350  & 3.1122    & 0.5573  &   0.0059 & 87.93 \\
Var.~3 & 13.45 & 0.0469  & 0.0338  & 3.0849    & 0.5613  &   0.0053 & 88.59\\
\hline 
\end{tabular}
\end{center}
\label{tabcase05}
\end{table}

We next consider a surfactant droplet impingement with $\theta_e^0=90^\circ$ in which the surfactants have no effect on the dynamic contact angle when the model~\eqref{casurf3} is considered. However, the surfactants will increase the Weber number by reducing the surface tension, and consequently the maximum wetting diameter will be increased. We consider the same flow properties and the droplet configurations as in Section~\ref{seccase03} but with $\theta_e^0=90^\circ$. The computed variants are  (1)~$\alpha=1$,  (2)~$\alpha=10$ and  (3)~$\alpha=100$. Computationally obtained wetting diameters of these variants of surfactant droplets are compared with the clean droplet case. Note that the given initial distribution of the surfactant on the interface is uniform and the surface Peclet number is small, and therefore the effect of Marangoni convection will be negligible.

The numerical results, the wetting diameter, sphericity, kinetic energy and dynamic contact angle, of these surfactant droplet variants and the clean droplet case are presented in Fig.~\ref{case05}. Even though the wetting diameter cure of all variants are similar in~\eqref{case05} (a), a close-up view of the wetting diameter in~\eqref{case05} (a) reveals the difference in the maximum wetting diameter. Further, to quantify the influence of surfactants, different parameters are tabulated in Table~\ref{tabcase05}. Also, the time at which each variant of droplet breaks down (End time) is presented in Table~\ref{tabcase03}. In comparison to the droplet with $\theta_e^0=100^\circ$ (Section~\ref{seccase03}), the topological changes occur later, say around the dimensionless time $t=13$.


\subsection{Influence of adsorption in an impinging droplet   with $\theta_e^0=125^\circ$}\label{seccase05}
\begin{figure}[t!]
\begin{center}
\unitlength1cm
\begin{picture}(15, 11.5)
\put(-0.5,5.75){{\includegraphics[width=7.5cm]{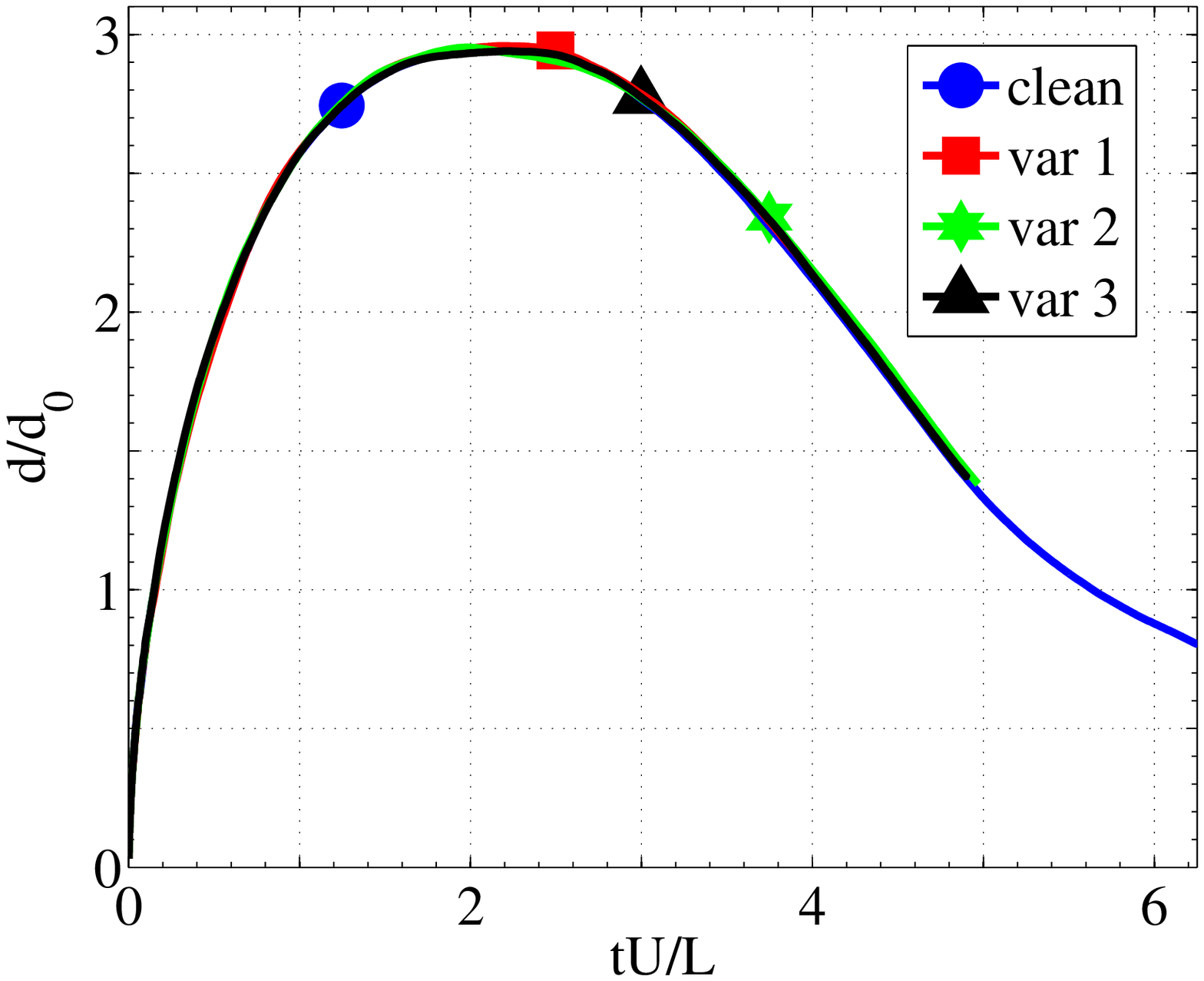}}}
\put(1.,6.75){{\includegraphics[width=3.25cm]{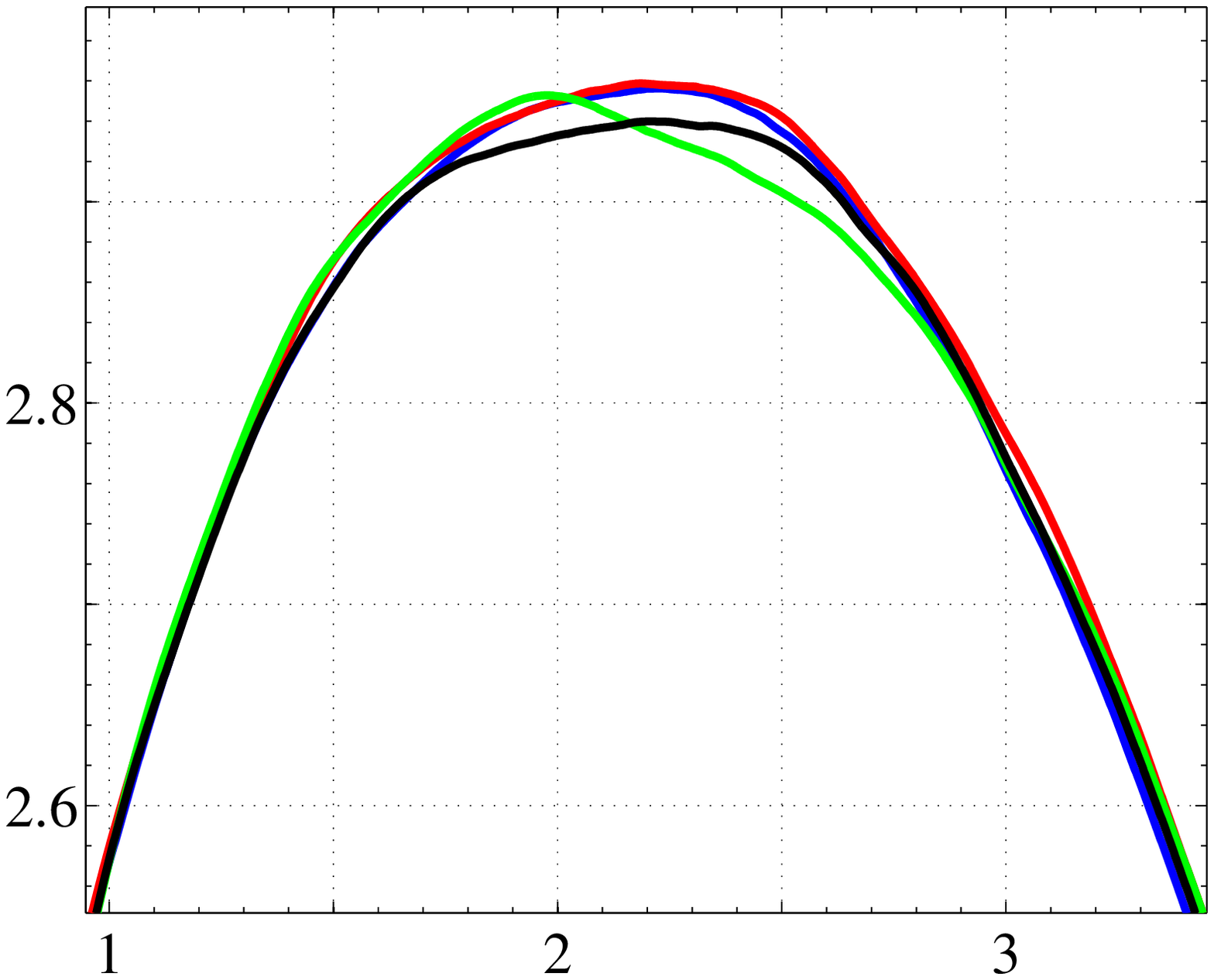}}}
\put(7.75,5.75){{\includegraphics[width=7.5cm]{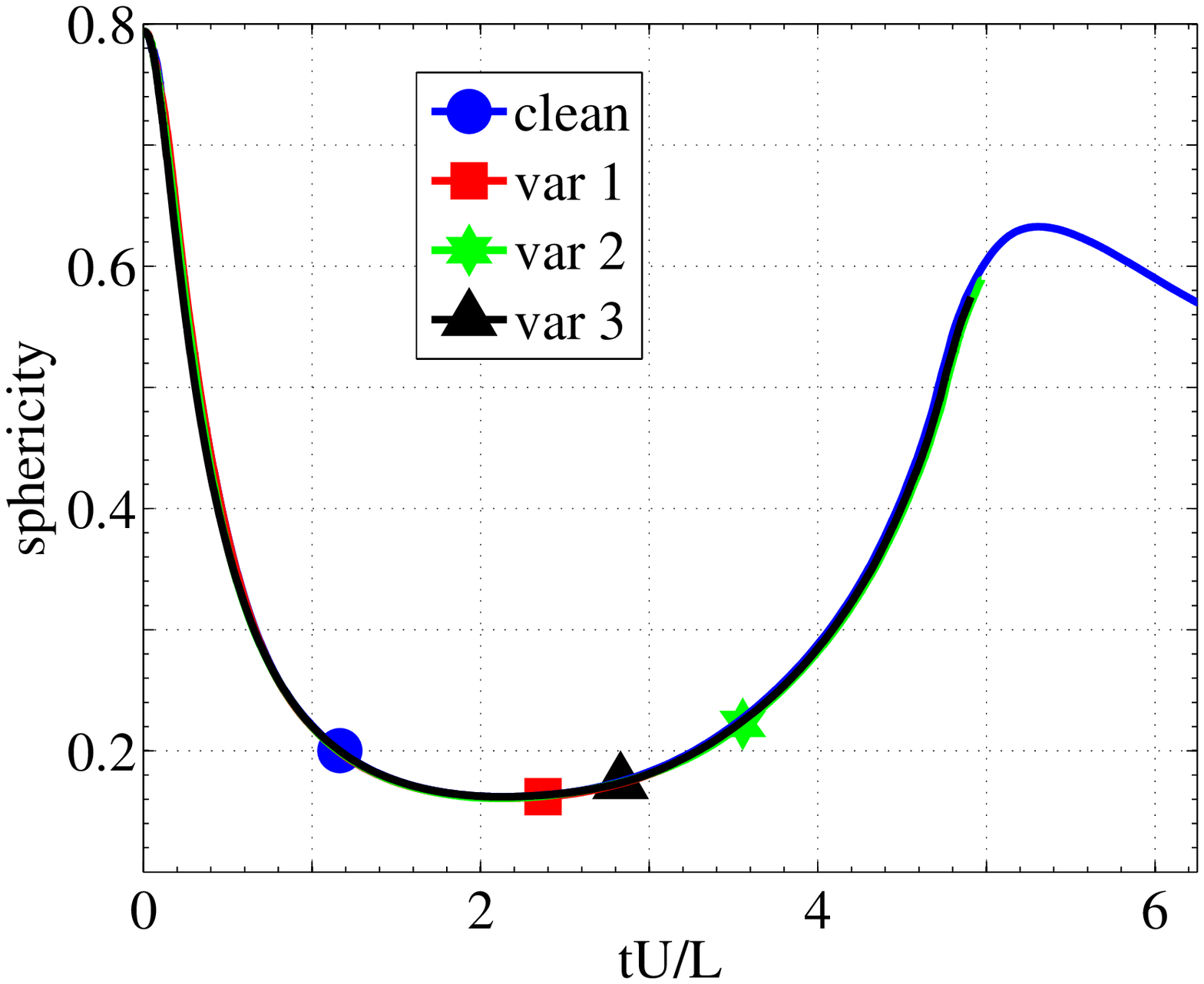}}}
\put(-0.5, -0.5){{\includegraphics[width=7.5cm]{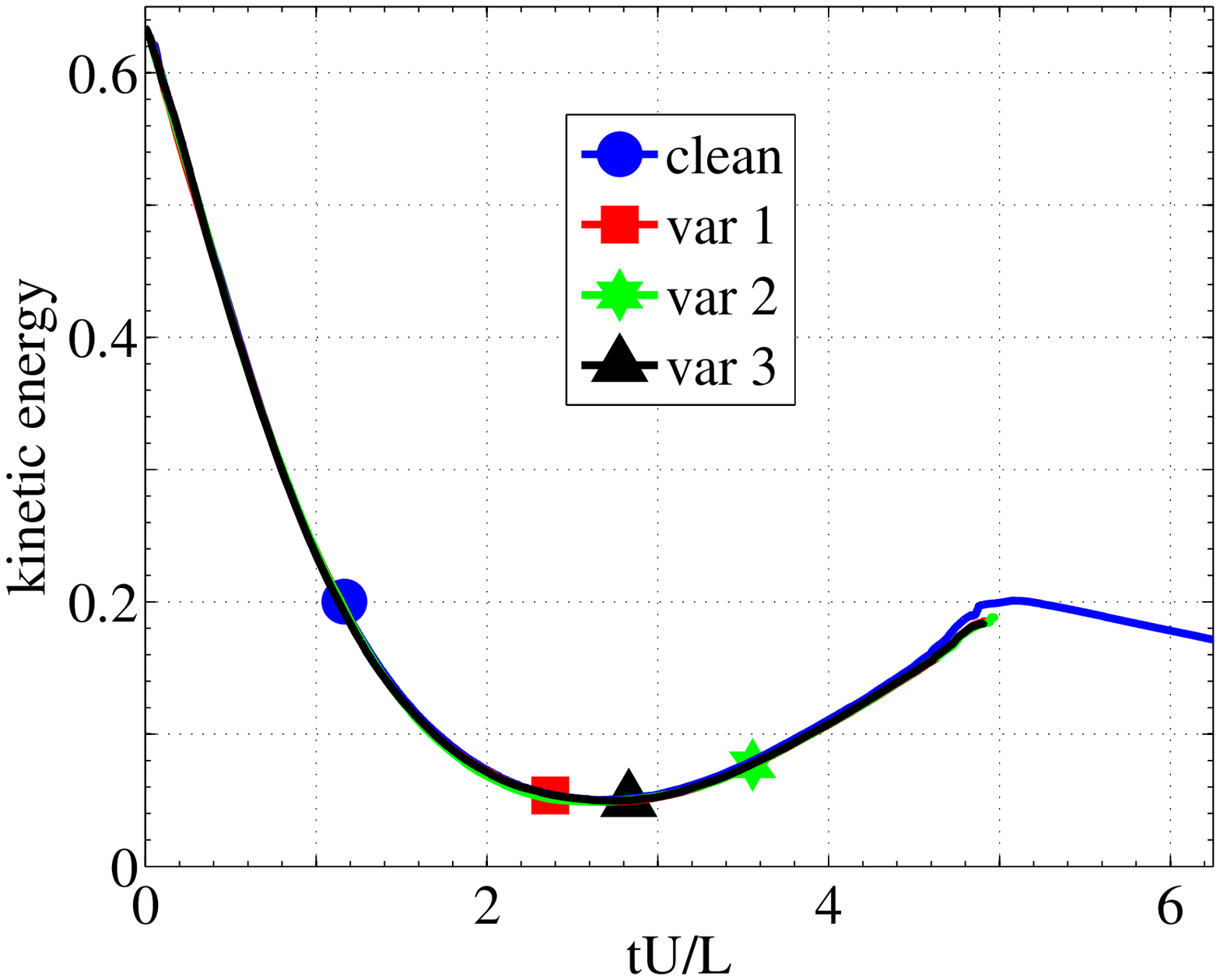}}}
\put(7.75, -0.5){{\includegraphics[width=7.5cm]{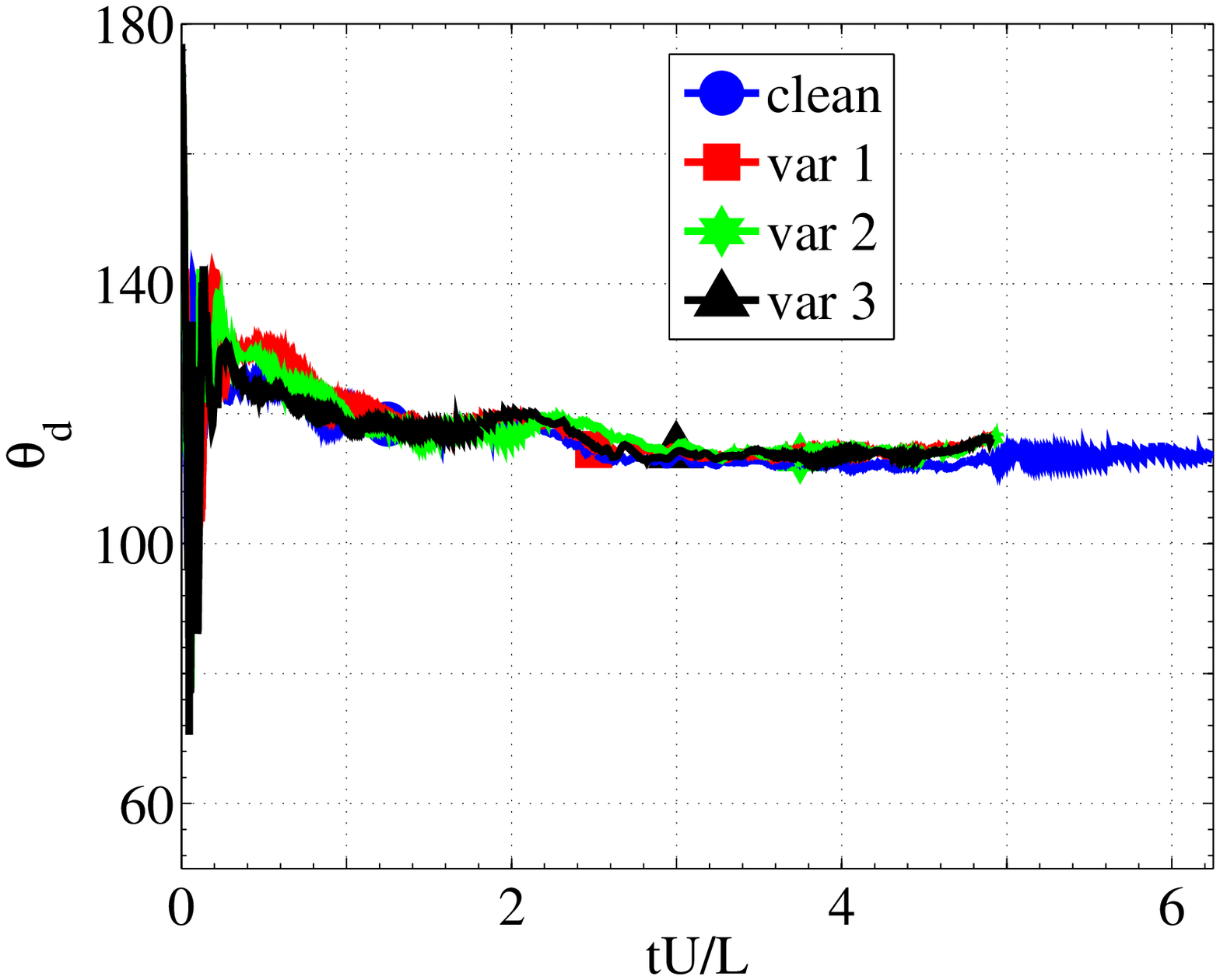}}}
\put(3.25,11.35){$(a)$}
\put(11.75,11.35){$(b)$}
\put(3.25, 5){$(c)$}
\put(11.75, 5){$(d)$}
\end{picture}
\end{center}
\caption{Effects of surfactant adsorption coefficient on the wetting diameter $(a)$, the sphericity $(b)$, the kinetic energy $(c)$, and the dynamic contact angle $(d)$  of an impinging droplet with $\Rey=4212$, $\Web$=$90$, $\Fro$=$92$, and  $\theta_e^0$=$125^\circ$. Var~1: $\alpha=1$,  Var~2: $\alpha=10$ and  Var~3: $\alpha=100$.
\label{case07}}
\end{figure}

We next consider a non-wetting water droplet with $\theta_e^0=125^\circ$. The flow properties and the droplet configurations are same as in Section~\ref{seccase03}, except $\theta_e^0=125^\circ$. Once again the computed variants are (1)~$\alpha=1$,  (2)~$\alpha=10$ and  (3)~$\alpha=100$. The numerical results (wetting diameter, sphericity, kinetic energy and dynamic contact angle) of these   variants and the clean droplet case are presented in Fig.~\ref{case07}. Since the equilibrium contact angle is large in the high Reynolds number example, the droplet spreads, recoils and bounce very quickly. In particular, the topological changes occur early in comparison with $\theta_e^0=100^\circ$  and $\theta_e^0=90^\circ$ cases (Section~\ref{seccase03} and \ref{seccase04}). Further, an interesting observation is that the topological changes in the surfactant droplet occurs   early than the clean droplet, see Table~\ref{tabcase07}. In the presence of surfactants, the surface tension will be less and eventually the topological changes are expected early. Other than this effect, the surfactant has almost no effects on the flow dynamics of the droplet, see Table~\ref{tabcase07}, where the tabulated values are almost identical. 

\begin{table}
\caption{Comparison of flow and geometric  parameters  of droplet impingement simulations for  different  examples considered in section~\ref{seccase05}.} 
\begin{center}
\begin{tabular}{lccccccc} 
\hline \vspace*{-2mm}\\
Variants &  $\underset{\text{(I)}}{\text{End time}}$ &  $\displaystyle\max_{t\in(0,\text{I}]}|\delta_V(t)|$  & $\displaystyle\max_{t\in(0,\text{I}]}|\delta^\Gamma_c(t)|$ &    $\displaystyle\max_{t\in(0,\text{I}]}\frac{d(t)}{d_0}$ &   $\underset{\text{at } t=\text{I}}{\text{sphericity}}$ & $\underset{\text{at } t=\text{I}}{\text{kinetic energy}}$ & $\underset{\text{at } t=\text{I}}{ \theta_d(C_{\Gamma_1})}$ \vspace*{2mm}\\ 
\hline 
Clean    & 6.40 &  0.0148  & -       & 2.9563   & 0.5593  &   0.1651 & 114.79 \\
Var.~1   & 4.91 & 0.0130  & 0.0246  & 2.9588   & 0.5735  &   0.1854 & 116.44 \\
Var.~2  & 4.97 & 0.0140  & 0.0234  & 2.9528   & 0.5887  &   0.1881 & 116.10 \\
Var.~3 & 4.90 & 0.0144  & 0.0164  & 2.9399   & 0.5746  &   0.1846 & 116.19\\
\hline 
\end{tabular}
\end{center}
\label{tabcase07}
\end{table}

\section{Summary and Observations}
A finite element scheme using the arbitrary Lagrangian--Eulerian approach is presented for computations of 3D-axisymmetric impinging droplets with soluble surfactants. The key ingredients of the scheme are the inclusion of the Marangoni effects without calculating the surface gradient of surfactant concentration on the free surface, a surfactant-dependent dynamic contact angle model, which is independent of numerical parameters and velocity at the contact line, and an accurate inclusion of surface forces with  isoparametric finite elements in a moving mesh. The numerical procedure includes the solution of the time-dependent Navier--Stokes equations,   the bulk surfactant concentration equation and the surface evaluation equations. Since the free surface resolved moving meshes are used, 
the discrete representation of the free surface is used as a computational mesh for the  surface evaluation equations. Further,   an iteration of  Gauss-Seidel type is employed for an implicit treatment of the adsorption/desorption balance condition for the surfactant mass transfer. A mesh convergence study and comparisons of  computationally obtained wetting diameter  with experimental results  are performed to validate the scheme. An excellent conservation of the fluid mass and  of the total surfactant mass is obtained with the proposed scheme. A number of computations for impinging droplets with soluble surfactants are performed, and the observations are summarized below.

\begin{itemize}
 \item An increase in the surfactant concentration, decreases the contact angle, when the equilibrium contact angle  value of the corresponding clean droplet is less than $90^\circ$.  Contrarily,  the contact angle increases further when the equilibrium contact angle   is greater than $90^\circ$. 
 \item The nonuniform surfactant concentration on the free surface induces the Marangoni effect.
 \item Apart from the Marangoni effect, the surfactant concentration decreases the surface tension force and thus increases the $\Web$ number. It eventually increases the maximum wetting diameter  during the deformation of the droplet.
 \item The effects of surfactants   are more on the wetting droplet droplet in comparison with the non-wetting droplets. 
 \item The presence of  surfactants at the contact line reduces the  contact angle and increase the surface force in wetting droplets. It eventually enforces the  wetting droplet to spread faster. Contrarily, the presence of  surfactants  at the contact line in non-wetting droplets  increases  the contact angle, and therefore the spreading is not affected much even though the surface force increases.
 \item Due to the increase in the surface force,  topological changes (breaking/splashing)   occur early in surfactant droplets in comparison with the clean droplet.
 \item Surfactants alter  the equilibrium contact angle and consequently surfactants change the equilibrium wetting diameter. 
\end{itemize}

\section*{Acknowledgment}
The author  would like to thank the Alexander von Humboldt Foundation for a partial support.

\bibliographystyle{elsart-num-sort}
\bibliography{../../masterlit}

\end{document}